\documentclass[10pt,twocolumn,amsmath,amssymb,aps,prb]{revtex4-2}

\usepackage{graphicx}
\usepackage{dcolumn}
\usepackage{bm}
\usepackage[utf8]{inputenc}
\usepackage[usenames,dvipsnames]{color}
\usepackage{tabularx}

\usepackage[english]{babel}
\usepackage[normalem]{ulem}
\usepackage{cancel}
\usepackage{xcolor}

\usepackage{hyperref}

\newcommand{\rmi}{{\rm i}}
\newcommand{\rmd}{{\rm d}}

\def\XXint#1#2#3{{\setbox0=\hbox{$#1{#2#3}{\int}$}
     \vcenter{\hbox{$#2#3$}}\kern-.5\wd0}}

\begin{document}

\title{Screening in Hubbard models with long-range interactions}
  
\author{Florian Gebhard$^1$}
\email{florian.gebhard@physik.uni-marburg.de}
\author{Kevin Bauerbach}
\email{kevin.bauerbach@physik.uni-marburg.de}
\affiliation{$^1$Fachbereich Physik, Philipps-Universit\"at Marburg,
  35032 Marburg, Germany}
\author{\"Ors Legeza$^{2,3,4}$}
\email{legeza.ors@wigner.hu}
\affiliation{$^2$Strongly Correlated Systems Lend\"ulet Research Group, 
Wigner Research Centre for
Physics, P.O.\ Box 49, 1525 Budapest, Hungary}
\affiliation{$^3$Institute for Advanced Study,
  Technical University of Munich, Lichtenbergstra\ss e 2a,
  85748 Garching, Germany}
\affiliation{$^4$Parmenides Foundation, Hindenburgstraße 15, 82343, P\"ocking, Germany}

\date{Polished version for PRB as of May 14, 2025}

\begin{abstract}%
We provide solid evidence for the long-standing presumption 
that model Hamiltonians with short-range interactions 
faithfully reproduce the physics of the long-range Coulomb interaction
in real materials. 
For this aim, we address a generic Hubbard model that captures
the quantum phase transitions between metal, Mott insulator, 
and charge-density-wave insulator, in the absence of Fermi-surface nesting.
By comparing the quantum phase diagrams for the $1/r$-Hubbard model  
on a half-filled chain with nearest-neighbor and
$1/r$-long-range interactions, we argue
that the inclusion of long-range interactions is not crucial for a proper description 
of interacting many-electron systems. 
To this end, we employ the Density Matrix Renormalization Group method on finite lattices and antiperiodic boundary conditions
to determine the quantum phase transitions between the metallic Luttinger liquid for
weak interactions, the Mott-Hubbard insulator for dominant on-site interactions,
and the charge-density wave insulator for dominant inter-site interactions. 
The two phase diagrams {\em qualitatively\/} agree inasmuch as the quantum phase transitions are continuous in both cases. 
Moreover, simple Hartree-Fock theory and the atomic limit provide renormalization factors
that allow us to {\em quantitatively\/} map the two phase diagrams onto each other.
As a practical advantage, our findings imply that computational efforts can be reduced tremendously by
using models with short-range interactions only. 
\end{abstract}



\maketitle

\section{Overview}
\label{sec:overview}

This overview introduces the reader to the fundamental issues,
summarizes and illustrates our central results, and gives an outline 
to our work.
After a brief introduction to the interacting-electron problem
in Sect.~\ref{sec:Introduction}, we
present the quantum phase diagrams for the $1/r$-Hubbard model with nearest-neighbor
interaction and with $1/r$-long-range interactions in Sect.~\ref{subsec:phasediagramscomp}.
As central results of this work we show 
that the two phase diagrams agree qualitatively and that they can be mapped onto each other 
using simple renormalization factors that can be obtained from
Hartree-Fock theory and the atomic limit. 
In the remainder of this work, 
as outlined in Sect.~\ref{sec:Outline},
we discuss the ground-state properties of the $1/r$-Hub\-bard model with $1/r$-long-range interactions
that we obtain from using the numerical Density-Matrix Renormalization Group (DMRG) method.

\subsection{Introduction}
\label{sec:Introduction}

Electrons in a perfectly ordered solid at zero temperature pose a formidable 
quantum many-particle problem~\cite{FetterWalecka,Mahan,NegeleOrland}.
Modern-day computers can solve effective single-particle problems
to arbitrary accuracy so that Hartree(-Fock) calculations
or density-functional theory provide relevant and reliable insight into the overall band structure.
However, the single-particle picture cannot capture correlation effects,
e.g., the correlation contribution to band gaps, or Mott insulating phases that arise
due to the electron-electron interaction~\cite{Mott1949,Mottbook,Gebhardbook,Solyom3}.

In the early 1960s, Hubbard~\cite{Hubbard1963} discussed
a series of approximations to derive a model that still
contains the essence of the electronic many-body problem.
The Hubbard model, independently also introduced by 
Gutzwiller~\cite{Gutzwiller1963} and Kanamori~\cite{Kanamori},
describes a single $s$-band of bandwidth~$W$ with nearest-neighbor electron transfers and
purely on-site interactions of strength~$U$ between
electrons of opposite spin orientation. Therefore, it
disregards the complexity of multi-band systems and ignores the
long-range nature of the Coulomb interaction. 
Due to these restrictions, the Hubbard model is often dismissed
as a simplistic toy-model that has little relevance for real materials, if any.

Despite its conceptual simplicity, the Hubbard model captures essential 
correlation effects. For example, in finite dimensions
it describes a continuous Mott transition 
at half band-filling as a function of $U/W$~\cite{LiebWu,GoehmannBuchHubbard},
and also itinerant anti-ferromagnetism in single-band metals~\cite{Anderson1959,Gebhardbook,Solyom3}. 
More specific features are not necessarily contained in the bare Hubbard model, 
such as excitons, i.e., 
bound electron-like and hole-like states in the optical excitation spectrum of an insulator.
For the description of excitons in Mott insulators~\cite{EsslerGebhardJeckelmann},
the inclusion of a nearest-neighbor 
interaction of strength~$V_{\rm NN}$ is mandatory (`extended Hubbard model'~\cite{Gebhardbook,PhysRevB.50.14016,PVD1994}).

The question remains whether or not the extended Hubbard model
captures the essence of the long-range Coulomb interaction. 
For an electron and a hole, both short-range and long-range interactions 
can give rise to the formation of an exciton. However, 
the nature of the Mott transition could be altered because 
electrons in the metallic state screen the Coulomb interaction~\cite{FetterWalecka,Mahan} 
whereas the Coulomb interaction
remains long-ranged at low energies in the insulating phase. 
Therefore, in the presence of long-range interactions 
the gap might open discontinuously as a function of 
the Coulomb parameters, in contrast to the case of the bare Hubbard model.
In this work, we argue that the extended Hubbard model reproduces
the quantum phase diagram of the Hubbard model with long-range Coulomb interactions.

It is very difficult to address these issues because the Hubbard model and its variants 
pose true many-particle problems that are unsolvable analytically, in general. 
An exception is one spatial dimension where the Bethe Ansatz 
provides exact results for the Hubbard model~\cite{GoehmannBuchHubbard}.
Unfortunately, in one spatial dimension the Mott transition generically occurs at $U=0^+$ due to the nesting of the 
Fermi points~\cite{Gebhardbook,LiebWu,Thierrybook}.
Nesting is not prevalent in higher dimensions~\cite{Fermisurfacenesting,Nesting}. Instead, 
the transition should generically occur when the Coulomb interaction
is of the order of the bandwidth, $U_{\rm c}\approx W$, 
in line with Mott's original arguments~\cite{Mott1949,Mottbook}.
Nesting can be avoided in one dimension when all electrons move in the same direction as
is the case for the $1/r$-Hubbard model with its linear dispersion 
relation, where the electron transfer amplitudes between sites decay proportional to the
inverse of their distance~$r$ (`$1/r$-Hubbard model').  
In the absence of nesting, Umklapp scattering processes do not occur
and the Mott transition is located at a finite interaction strength, $U_{\rm c}=W$ for
the $1/r$-Hubbard model~\cite{GebhardRuckenstein,GebhardGirndtRuckenstein,Gebhardbook}.

A number of exact results are known for the $1/r$-Hub\-bard model
that provide benchmark tests for 
the numerical density-matrix renormalization group (DMRG) method. 
It permits to study systems with up to $L=64$ lattice sites and periodic boundary conditions
so that reliable results can be obtained from extrapolations to the thermodynamic 
limit~\cite{1overRHubbard}.
Using DMRG we obtained the quantum phase diagram for the extended $1/r$-Hubbard 
model~\cite{1overRHubbard-NN}. In this work, we derive the 
zero-temperature phase diagram for the $1/r$-Hubbard model with
$1/r$-long-range interactions. Our comparison shows no 
qualitative differences. Moreover, we can map the two phase diagrams onto each other quantitatively
using simple renormalization factors that we derive from bare Hartree-Fock theory and the atomic limit, respectively.

\subsection{Phase diagrams}
\label{subsec:phasediagramscomp}

\begin{figure}[t]
    \begin{center}
    (a) \includegraphics[width=8cm]{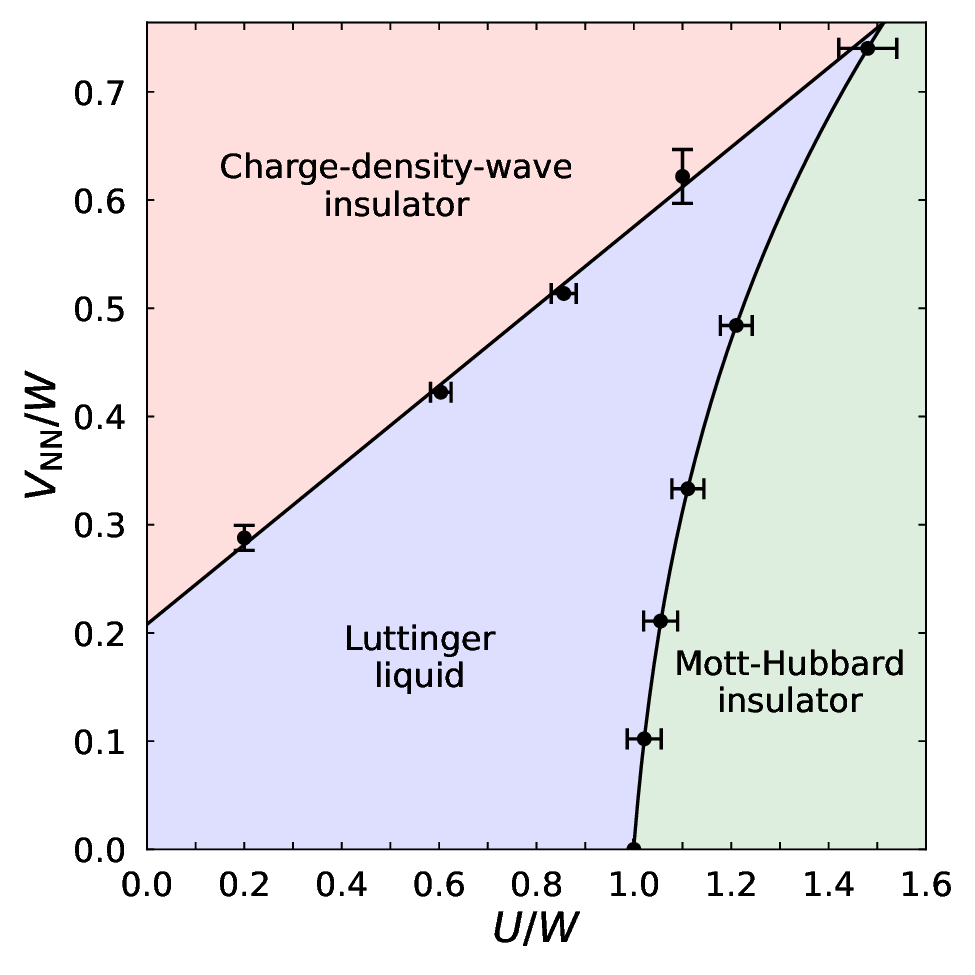}\\
    (b) \includegraphics[width=8cm]{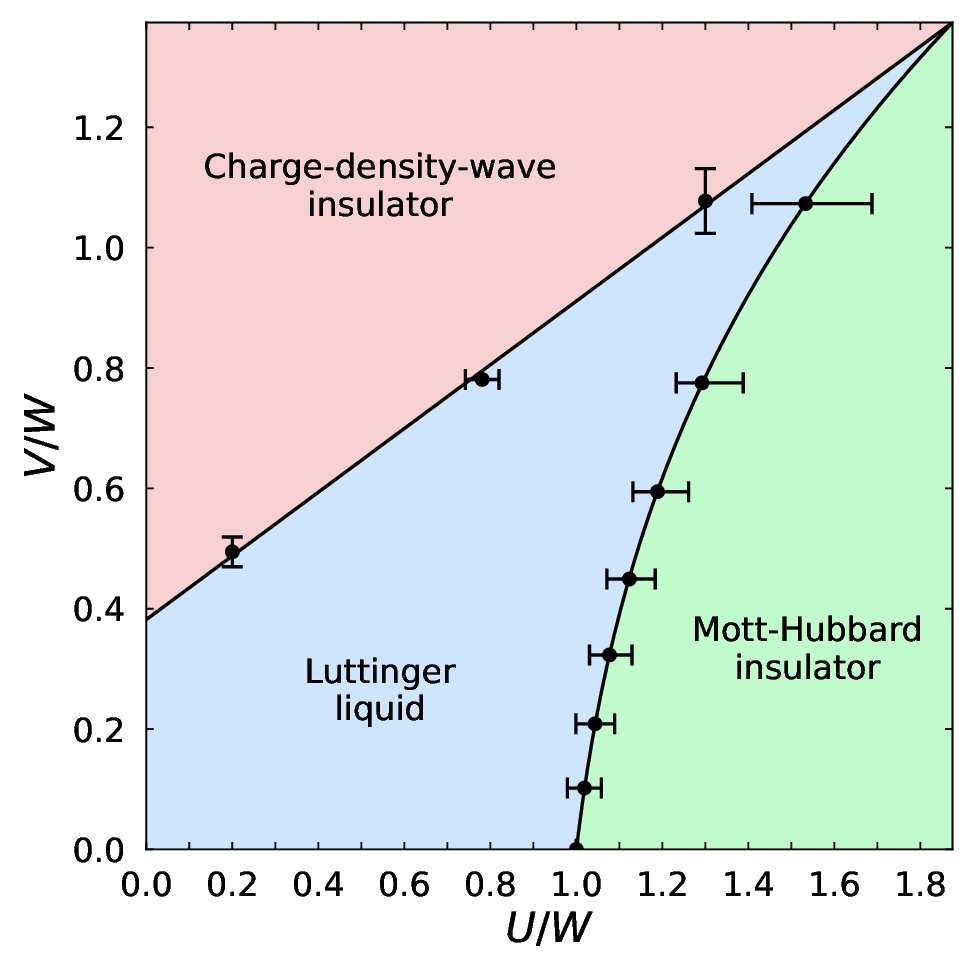}
    \end{center}
    \caption{(a) Phase diagram of the one-dimensional $1/r$-Hubbard model with nearest-neighbor interactions~$V_{\rm NN}$.
    (b) Phase diagram of the one-dimensional $1/r$-Hubbard model with $1/r$-long-range interactions. Dots: estimate for the critical interaction, $\overline{U}_{\rm c}$, with error bounds; 
      continuous lines: polynomial fits, see Sect.~\ref{sec:RenormalizationPhaseDiagram}.
      \label{QPDiagramAndComparison}}
\end{figure}

In Fig.~\ref{QPDiagramAndComparison}a we recall the quantum phase diagram of the $1/r$-Hubbard model with repulsive nearest-neighbor interactions~\cite{1overRHubbard-NN}. 
As can be argued using perturbation theory, the metallic Luttinger liquid appears in the region of weak interactions ($W \gg U, V_{\rm NN}$), 
the Mott-Hubbard insulator is found for dominant Hubbard interactions ($U \gg W, V_{\rm NN} $), and the charge-density-wave (CDW) phase occurs for 
dominant inter-site interactions ($V_{\rm NN} \gg W, U)$. 

For dominant Coulomb interactions, the separation line between the Mott-Hubbard insulator and the CDW insulator is given by
$V_{\rm NN} = U/2$ to leading order, with a stabilization tendency toward the Mott-Hubbard insulator in the next order in $1/U$.

In the absence of inter-site interactions, the Mott-Hubbard transition occurs
at $U_{\rm c}(V=0)=W$~\cite{GebhardRuckenstein,Gebhardbook,1overRHubbard}.
The inclusion of a repulsive nearest-neighbor interaction stabilizes the metallic phase
because the additional repulsive interaction softens the two-particle scattering potential in position space. 

For comparison, in Fig.~\ref{QPDiagramAndComparison}b, we show the quantum phase diagram of the $1/r$-Hubbard model with repulsive inter-site interactions 
that decay proportional to $V_{\rm LR}/r$ ($1/r$-long-range interactions).
As our first main result, we see that the quantum phase diagrams agree {\em qualitatively}. 
Both the Mott and CDW transitions remain {\em continuous\/} even in the presence of $1/r$-long-range interactions. 

\begin{figure}[t]
  \begin{center}
  \includegraphics[width=8cm]{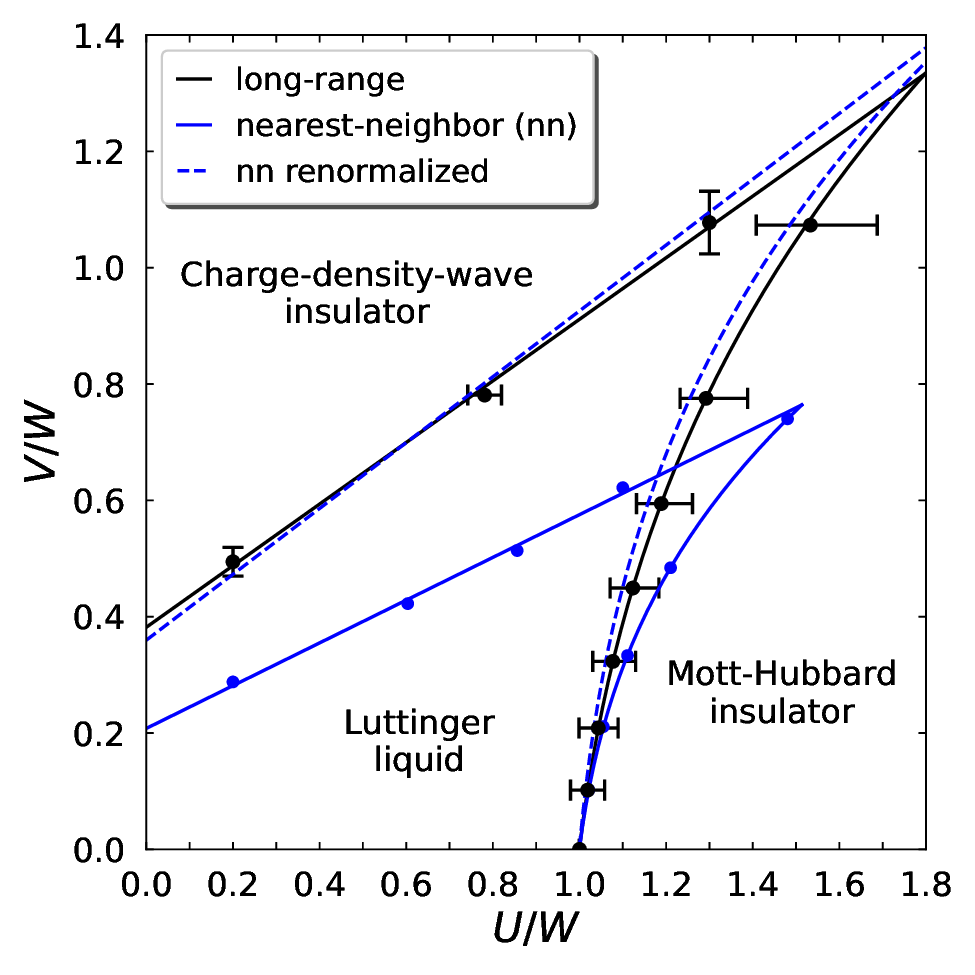}
  \end{center}
  \caption{Renormalized phase diagram. The phase separation lines of the $1/r$-Hubbard model with $1/r$-long-range interactions
 (black solid line) and nearest-neighbor interactions (blue solid line) can be mapped onto each other using $V_{\rm NN}=R_{\rm qp} V_{\rm LR}$ 
with $R_{\rm CDW} = 0.606$ and $R_{\rm Mott}=\ln 2\approx 0.693$ for the CDW and Mott quantum phase transitions.
The renormalized quantum phase transition lines for the extended $1/r$-Hubbard model (dashed blue lines) reproduce those of the
$1/r$-Hubbard model with $1/r$-long-range interactions (black solid lines) within error bars.\label{fig:RenormPhaseDiagram}}
\end{figure}

The agreement can be made {\em quantitative\/} within error bounds by using simple renormalization factors, see Fig.~\ref{fig:RenormPhaseDiagram}.
This approximate yet quantitatively satisfactory mapping procedure is our second main result. 

The phase diagrams are characterized by phase separation lines.
For $V>V_{\rm c}^{\rm CDW}(U)$ the CDW insulator is stable, whereas 
the Mott-Hubbard insulator prevails for $U>U_{\rm c}^{\rm Mott}(V) \geq 1$ when $0\leq V\leq V_{\rm c}^{\rm CDW}(U)$. 
For small interactions, Hartree-Fock theory predicts $V_{\rm c,NN}^{\rm CDW}(0)/V_{\rm c,LR}^{\rm CDW}(0)\approx R_{\rm HF}^{\rm CDW}(0)=0.606$,
see appendix, which we use to renormalize the CDW critical line by $V_{\rm NN}=R_{\rm CDW}V_{\rm LR}$, $R_{\rm CDW}=0.61$.

On the other hand, in the atomic limit, the CDW and Mott-Hubbard insulators are separated by
$V_{\rm c}^{\rm CDW}(U)=U/(2\ln(2))\approx 0.721U$ for the $1/r$-Hubbard model with $1/r$-long-range interactions,
see appendix.
Thus, the renormalization factor in the atomic limit is given by $R^{\rm al}_{\rm Mott}=(1/2)/[1/(2\ln 2)]=\ln 2\approx 0.693$.
In Fig.~\ref{fig:RenormPhaseDiagram}, the Mott critical lines $U_{\rm c}^{\rm Mott}(V)$ 
of the $1/r$-Hubbard models with nearest-neighbor 
and $1/r$-long-range interactions are then mapped onto each other using $V_{\rm NN}=R_{\rm Mott}V_{\rm LR}$, $R_{\rm Mott}=0.69$.

The successful mapping indicates that the $1/r$-long-range interaction of strength~$V_{\rm LR}$ in one dimension
can be replaced by a nearest-neighbor interaction~$V_{\rm NN}$ that is smaller by a factor $R\approx 2/3$,
$V_{\rm NN}\approx R V_{\rm LR}$.

Before we proceed, we caution the reader that our findings relate to the ground-state and low-energy physics
  of interacting quantum particles. It is well known that an attractive potential well can be used 
  to describe semi-quantitatively the $1s$ bound state of the hydrogen atom or of an exciton.
  However, such an approach fails to describe properly the fine structure of the wave functions
  or the energy of higher-excited states. Similarly, the long-range tail of the Coulomb interaction
  can give rise to exotic phenomena, particularly under non-equilibrium conditions.
  Long-range interactions are also more significant in dilute bosonic systems, see the example of polar bosonic
  gases in one-dimensional lattice~\cite{polargas},
  or in the case of bosonic many-body localization~\cite{manybodylocalization}, and references therein.

\subsection{Outline}
\label{sec:Outline}

For better comparability, we closely follow the structure of our previous
publication~\cite{1overRHubbard-NN}. Therefore, we start in Sect.~\ref{sec:modeldef} 
with the definition of the Hubbard model with $1/r$-long-range electron transfers as well as
on-site and $1/r$-long-range Coulomb interactions.
We introduce the ground-state properties of interest, namely,
the ground-state energy, the two-particle gap, the
momentum distribution, and the density-density correlation function
from which we determine the Luttinger parameter in the metallic phase
and the charge-density-wave (CDW) order parameter. 

In Sect.~\ref{sec:gsproperties} we present results for the ground-state properties. In addition,
we discuss their finite-size dependencies and extrapolations to the thermodynamic limit where appropriate. 

In Sect.~\ref{sec:MTnearest-neighbor}
we focus on the Mott transition in the presence of long-range interactions between electrons on different lattice sites. 
We propose and discuss several methods to extract the critical interaction
strength for the Mott-Hubbard transition based on the ground-state energy, the two-particle gap, the Luttinger parameter, 
and the structure factor, whereby we study the Mott transition at fixed
$v\equiv V_{\rm LR}/U$ in the range $0\leq v \leq 0.7$ with increment $\Delta v=0.1$, 
in units of the bandwidth, $W\equiv 1$.
Moreover, we address the transition to the CDW insulator
in the presence of long-range interactions, which can be determined using the CDW order parameter and the Luttinger parameter. 
We study this transition for a fixed ratio $v = V_{\rm LR}/U = 1$, as well as for fixed values of $U = 0.2$ and $U = 1.3$.

In Sect.~\ref{sec:RenormalizationPhaseDiagram}, we present and discuss 
the complete phase diagram for the model with long-range interactions. 
We compare it in detail with the phase diagram of the extended Hubbard model 
which contains only on-site and nearest-neighbor interactions. 
In addition, we introduce a method to quantitatively map the two quantum
phase diagrams onto each other. 

In our conclusions, Sect.~\ref{sec:conclusions}, we summarize our findings and discuss
their relevance in a broader context.
Hartree-Fock calculations for the CDW transition are collected in the appendix.

\section{\texorpdfstring{$\bm{1/r}$}{1/r}-Hubbard model}
\label{sec:modeldef}

\subsection{Hamiltonian}

In this work, we address the one-dimensional single-band $1/r$-Hubbard model~\cite{Gebhardbook}
with on-site ($U\hat{D}$) and long-range interactions ($V\hat{V}$)
\begin{equation}
  \hat{H}=\hat{T}+U\hat{D}+ V \hat{V} 
  \label{eq:fullHubbardmodel}
\end{equation}
on a ring with $L$~sites; $L$ is even. For convenience, we abbreviate $V\equiv V_{\rm LR}$.
We discuss the kinetic energy and the Coulomb interaction terms separately.

\subsubsection{Kinetic energy}

The kinetic energy describes the tunneling of electrons with
spin~$\sigma=\uparrow,\downarrow$ along a ring with $L$ sites,
\begin{equation}
      \hat{T} = \sum_{\substack{l,m=1\\
      l\neq m; \sigma}}^{L}t(l-m)
    \hat{c}_{l,\sigma}^+\hat{c}_{m,\sigma}^{\vphantom{+}} \; .
      \label{eq:defT}
\end{equation}
The creation and annihilation operators $\hat{c}_{l,\sigma}^{+}$,
  $\hat{c}_{l,\sigma}^{\vphantom{+}}$
  for an electron with spin
$\sigma=\uparrow,\downarrow$ on lattice site~$l$ obey the usual
anti-commutation relations for fermions.

In this work, we focus on the $1/r$-Hubbard model for which the electron transfer between lattice sites is defined by
\begin{eqnarray}
  t(r) &=& (-\rmi t ) \frac{(-1)^r}{d(r)} \; ,\nonumber \\
  d(r)& =& \frac{L}{\pi}\sin\left(\frac{\pi r}{L}\right) \; .
  \label{eq:defTconstituents}
\end{eqnarray}
Here, $d(l-m)$ is the chord distance between the sites~$l$ and $m$ on a ring.
In the thermodynamic limit and for $|l-m|\ll L $ fixed,
we have $d(l-m)= (l-m)+ \mathcal{O}(1/L^2)$, and the electron
transfer amplitude between two sites decays inversely proportional
to their distance (`$1/r$-Hubbard model').

Since $L$ is even,
we have anti-periodic electron transfer amplitudes
because $d(L+ r)= -d(r)$. Therefore, we choose
anti-periodic boundary conditions
\begin{equation}
\hat{c}_{L+l,\sigma}=-\hat{c}_{l,\sigma}
\end{equation}
for the operators, too.
Under these boundary conditions,
the kinetic energy operator becomes diagonal in Fourier space,
\begin{eqnarray}
  \hat{C}_{k,\sigma}^+
  &= & \frac{1}{\sqrt{L}}\sum_{l= 1}^{L}
  e^{\rmi  kl } \hat{c}_{l,\sigma}^+ \; , \nonumber \\
  \hat{c}_{l,\sigma}^{+}
    &= & \frac{1}{\sqrt{L}}\sum_k
    e^{-\rmi  kl } \hat{C}_{k,\sigma}^+ \; ,    \nonumber \\
    k&=& \frac{(2m+ 1)\pi}{L}\;, \; m= -\frac{L}{2}, \ldots, \frac{L}{2}-1\; ,
    \label{eq:FTofoperators}
\end{eqnarray}
so that
\begin{equation}
  \hat{T}=\sum_{k,\sigma}\epsilon(k) 
  \hat{C}_{k,\sigma}^+\hat{C}_{k,\sigma}^{\vphantom{+}}\; .
\label{eq:lineardispersion}
\end{equation}
The dispersion relation of the $1/r$-Hubbard model is linear,
$\epsilon(k)=t k$. We set $t=1/(2\pi)$ for a unit bandwidth,
$W\equiv 1$.

In  this work, we focus on the case of a paramagnetic half-filled ground state
where we have the same number of electrons per spin species, 
$N_{\uparrow}= N_{\downarrow}$,
that equals half the number of lattice sites, $N_{\sigma}=L/2$
($\sigma=\uparrow,\downarrow$).

\subsubsection{Coulomb interaction}

The Coulomb interaction is parameterized by two terms
in Eq.~(\ref{eq:fullHubbardmodel}).
The on-site (Hubbard) interaction~\cite{Hubbard1963,Gutzwiller1963,Kanamori}
acts locally between two electrons with opposite spins,
\begin{equation}
  \hat{D}= \sum_{l=1}^L \hat{n}_{l,\uparrow}\hat{n}_{l,\downarrow}
  \; ,  \quad
\hat{n}_{l,\sigma}=\hat{c}_{l,\sigma}^+\hat{c}_{l,\sigma}^{\vphantom{+}}
  \; ,
\end{equation}
where $\hat{n}_{l,\sigma}$
counts the number of electrons with spin $\sigma$ on site~$l$,
and $\hat{n}_l=\hat{n}_{l,\uparrow}+\hat{n}_{l,\downarrow}$
counts the number of electrons
on site~$l$. The corresponding operators for the total number
of electrons with spin~$\sigma=\uparrow,\downarrow$
are denoted by $\hat{N}_{\sigma}=\sum_l\hat{n}_{l,\sigma}$,
and $\hat{N}=\hat{N}_{\uparrow}+\hat{N}_{\downarrow}$.

To discuss the influence of the extended nature of the Coulomb interaction,
we consider 
\begin{eqnarray}
  \hat{V}&=&
  \sum_{r=1}^{L/2} V(r) \sum_{l=1}^{L} (\hat{n}_l-1)(\hat{n}_{l+r}-1)\;,\nonumber\\
  V(r) &=& \frac{1}{d(r)}
  \label{eq:NNinteraction}
\end{eqnarray}
with $d(r)$ from Eq.~(\ref{eq:defTconstituents}). 
Due to the ring geometry, we limit ourselves to distances $1\leq r\leq L/2$.

For comparison, we shall also show results for the extended $1/r$-Hubbard model,
$V(r)=\delta_{r,1}$, as collected in Ref.~\cite{1overRHubbard-NN}.

\subsubsection{Particle-hole symmetry}

We study the case of an anti-symmetric dispersion relation
where $t(-r)=-t(r)$. 
Under the particle-hole transformation
\begin{equation}
  \hat{c}_{l,\sigma}^{\vphantom{+}} \mapsto \hat{c}_{l,\sigma}^+ \quad
  , \quad \hat{n}_{l,\sigma} \mapsto 1-\hat{n}_{l,\sigma} \; ,
\end{equation}
the kinetic energy remains unchanged,
\begin{eqnarray}
\hat{T} &\mapsto& \sum_{\substack{l,m=1\\      l\neq m; \sigma}}^{L}
t(l-m)\hat{c}_{l,\sigma}^{\vphantom{+}}\hat{c}_{m,\sigma}^+
\nonumber \\
&=&
 \sum_{\substack{l,m=1\\      l\neq m; \sigma}}^{L}
  \left[ -t(m-l)\right]
  \hat{c}_{l,\sigma}^+\hat{c}_{m,\sigma}^{\vphantom{+}}
  = \hat{T}
\end{eqnarray}
  because $t(-r)=-t(r)$.
Furthermore,
\begin{equation}
  \hat{D}\mapsto \sum_{l=1}^L
  (1-\hat{n}_{l,\uparrow})(1-\hat{n}_{l,\downarrow})
  =\hat{D}-\hat{N}+L \; ,
\end{equation}
and
\begin{equation}
  \hat{V}\mapsto \hat{V} \;. 
\end{equation}
Therefore, $\hat{H}(N_{\uparrow},N_{\downarrow})$ has the same spectrum as
$\hat{H}(L-N_{\uparrow},L-N_{\downarrow})-U(2L-N)+LU$,
where $N=N_{\uparrow}+N_{\downarrow}$ is the particle number.

\subsection{Ground-state properties}
\label{subsec:gsprop}

The interplay between long-range interactions $V\hat{V}$, on-site (Hubbard) interactions $U\hat{D}$, 
and the fermions' kinetic energy~$\hat{T}$, characterized by the bandwidth~$W$,
gives rise to a complex quantum phase diagram at half band-filling, $n=N/L=1$.
We are interested in the metal-insulator transition where
the metallic Luttinger liquid for weak interactions, $U,V\ll W$, turns
into a paramagnetic Mott-Hubbard insulator
for large on-site interactions at some finite value $U_{\rm c}^{\rm Mott}(V)$ for $U \gg W,V$, 
or to a charge-density wave insulator at $V_{\rm c}^{\rm CDW}(U)$ for strong inter-site interactions, $V\gg U,W$.

For the $1/r$-Hubbard model, metal-insulator transitions can be determined through finite-size extrapolation of both 
the ground-state energy and the two-particle gap~\cite{1overRHubbard,1overRHubbard-NN}. 
Alternatively, the Luttinger parameter~\cite{PhysRevB.106.205133} and the finite-size extrapolation of the structure factor 
at the Brillouin zone boundary provide a way to identify the critical interaction strength. 
Additionally, the charge-density-wave state can be characterized using the CDW order parameter. 

In this section, we also present results for the momentum distribution for finite systems, which can be obtained via 
DMRG~\cite{White-1992b,Schollwock-2005}.

\subsubsection{Ground-state energy and two-particle gap}

We denote the ground-state energy by
\begin{equation}
E_0(N,L;U,V)= \langle \Psi_0 | \hat{H} |\Psi_0 \rangle
\end{equation}
for given particle number~$N$, system size~$L$, and interaction
parameters $U,V$.
Here, $|\Psi_0\rangle$ is the normalized ground state of the
Hamiltonian~(\ref{eq:fullHubbardmodel}).
We are interested in the thermodynamic limit,
$N, L\to \infty$ with $n=N/L$ fixed.
We denote the ground-state energy
per site and its extrapolated value by
\begin{eqnarray}
  e_0(N,L;U,V)&=& \frac{1}{L} E_0(N,L;U,V) \;, \nonumber \\
  e_0(n;U,V)&= & \lim_{L\to \infty} e_0(N,L;U,V) \; ,
\end{eqnarray}
respectively.

The two-particle gap is defined by
\begin{equation}
  \Delta_2(L;U,V) = \mu_2^+(L;U,V)-\mu_2^-(L;U,V) \; ,
  \label{eq:tpgapdef}
\end{equation}
where
\begin{eqnarray}
  \mu_2^-(L;U,V)&=& E_0(L,L;U,V)- E_0(L-2,L;U,V)
  \nonumber \; , \\
  \mu_2^+(L;U,V)  &=& E_0(L+ 2,L;U,V)- E_0(L,L;U,V)
    \nonumber \\
  \label{eq:defmu2plus}
\end{eqnarray}
are the chemical potentials for adding the last two particles to achieve half filling
and the first two particles beyond half filling, respectively.

Due to particle-hole symmetry, we have
\begin{equation}
  \mu_2^-(L;U,V)= 2U-\mu_2^+(L;U,V)
\end{equation}
so that
\begin{equation}
  \Delta_2(L;U,V) = 2\mu_2^+(L;U,V)-2U
  \label{eq:defDelta2}
\end{equation}
and
\begin{equation}
  \Delta_2(U,V) = \lim_{L\to\infty} \Delta_2(L;U,V)
  \label{eq:defDeltaTDlim2}
\end{equation}
in the thermodynamic limit.
We always consider the spin symmetry sector $S=S^z=0$. For this reason,
we study the two-particle gap rather than the single-particle gap.

The two added particles repel each other so that, in the thermodynamic limit,
they are infinitely separated from each other. Therefore, we have
\begin{equation}
\Delta_2(U,V)=2 \Delta_1 (U,V) \; ,
\end{equation}
where $\Delta_1(U,V)$ is the gap for single-particle excitations.
For finite systems, we expect the interaction energy
\begin{equation}
  e_{\rm R}(L;U,V) = \Delta_2(L;U,V)-2 \Delta_1 (L;U,V)=\mathcal{O}(1/L)>0
  \label{eq:defdeltaeR}
\end{equation}
to be positive, of the order $1/L$.
We verified that the interaction energy vanishes in the thermodynamic limit
for the case $V=0$~\cite{1overRHubbard}.
  
\subsubsection{Momentum distribution}

We also study the spin-summed momentum distribution in the ground state
at half band-filling, $N=L$,
\begin{eqnarray}
  n_k(L;U,V) &=& \langle \Psi_0 | \hat{n}_{k,\uparrow} + \hat{n}_{k,\downarrow}
  |\Psi_0\rangle \nonumber \\
  &= & \sum_{l,m;\sigma} e^{\rmi k (l-m)} P_{l,m;\sigma}
  \label{eq:defmomentumdistribution}
\end{eqnarray}
with $\hat{n}_{k,\sigma}=\hat{C}_{k,\sigma}^+\hat{C}_{k,\sigma}^{\vphantom{+}}$
and the single-particle density matrix $P_{l,m;\sigma}=
\langle \Psi_0 |\hat{c}_{l,\sigma}^+ \hat{c}_{m,\sigma}^{\vphantom{+ }} |\Psi_0\rangle$.
Due to particle-hole symmetry, we have
\begin{equation}
n_k(L;U,V)= 1-n_{-k}(L;U,V)
\end{equation}
at half band-filling. Therefore, it is sufficient to study wave numbers from
the interval $-\pi<k<0$.

\subsubsection{Density-density correlation function and Luttinger parameter}
\label{subsec:CNNexact}

Lastly, we address the density-density correlation function
at half band-filling, $N= L$,
\begin{equation}
  C^{\rm NN}(r,L;U,V) = \frac{1}{L} \sum_{l=1}^L
  \bigl(\langle\hat{n}_{l+r}\hat{n}_l \rangle
  - \langle \hat{n}_{l+r}\rangle \langle \hat{n}_l \rangle\bigr) \; ,
  \label{eq:CNNdef}
\end{equation}
where $\langle \ldots \rangle \equiv \langle \Psi_0| \ldots | \Psi_0\rangle$.
The limit $L\gg r\gg 1$ for $U,V \ll W$
is also accessible
from field theory~\cite{Thierrybook,PhysRevLett.64.2831,PhysRevB.39.4620},
\begin{equation}
  C^{\rm NN}(r\gg 1;U,V) \sim - \frac{K(U,V)}{(\pi r)^2}
  +\frac{A(U,V) (-1)^r}{r^{1+K}[\ln(r)]^{3/2}} + \ldots \, ,
  \label{eq:CNNfieldtheory}
\end{equation}
where $A(U,V)$ is a constant that depends on the interaction but not
on the distance~$r$.

We extract the Luttinger parameter~$K(U,V)$ from the structure factor,
\begin{equation}
  \tilde{C}^{\rm NN}(q,L;U,V)
  = \sum_{r=0}^{L-1}e^{-{\rm i}q r} C^{\rm NN}(r,L;U,V) 
  \;,
  \label{eq:CNNtildedef}
\end{equation}
where the wave numbers are from momentum space,
$q=(2\pi/L)m_q$, $m_q=-L/2,-L/2+1,\ldots,L/2-1$.
By construction, $\tilde{C}^{\rm NN}(q=0,L;U,V)=0$
because the particle number is fixed,
$N=L$ in the half-filled ground state. 
In the thermodynamic limit, the structure factor $\tilde{C}^{\rm NN}(q,L;U,V)$ remains of the order unity even in the CDW phase because we subtract
the contributions of the long-range order in the definition~(\ref{eq:CNNdef}).

The transition to a charge-density-wave insulator can be monitored from the CDW order parameter.
In this work, we do not study the standard CDW order parameter,
\begin{equation}
    D(L;U,V)=  \frac{1}{L} \left| \sum_{r=0}^{L-1} (-1)^r \left(\langle \hat{n}_{r} \rangle -1 \right)\right|\leq 1\; .
    \label{eq:CDWorderparameterdef}
\end{equation}
Instead, we include all short-range contributions and address
\begin{equation}
    N_{\pi}(L;U,V)=  \frac{1}{L} \sum_{r=0}^{L-1} (-1)^r \frac{1}{L} \sum_{l=0}^{L-1} \left(\langle \hat{n}_{r+l} \hat{n}_{l}\rangle -1 \right) \; .
    \label{eq:ourCDWorderparameterdef}
\end{equation}
When the charges are homogeneously distributed, $\langle \hat{n}_l\rangle=1$, we have
$N_{\pi}(L;U,V)=\tilde{C}^{\rm NN}(\pi,L;U,V)/L$, and the order parameter vanishes in the metallic phase.
More generally, in the thermodynamic limit we have $N_{\pi}(U,V)=(D(U,V))^2$. In the $1/r$-Hubbard model
with its long-range electron transfer, it is advantageous to analyze $N_{\pi}(L;U,V)$
to facilitate a reliable finite-size analysis. 

When Eq.~(\ref{eq:CNNfieldtheory}) is employed, it follows
that the Luttinger parameter for finite systems,
\begin{equation}
  K(L;U,V)
  = \frac{L}{2} \tilde{C}^{\rm NN}(2\pi/L,L;U,V) \; ,
  \label{eq:KfromCNNfinite}
\end{equation}
can be used to calculate the Luttinger parameter in the thermodynamic limit,
\begin{eqnarray}
  K(U,V)&=& \lim_{L\to\infty} K(L;U,V)\nonumber \\
  &=& \pi \lim_{q\to 0}\frac{\tilde{C}^{\rm NN}(q;U,V)}{q}\; ,
  \label{eq:KfromCNN}
\end{eqnarray}
where we denote the structure factor
in the thermodynamic limit by $\tilde{C}^{\rm NN}(q;U,V)$.
Using Eq.~(\ref{eq:KfromCNN}), the Luttinger parameter can be
calculated numerically
with very good accuracy~\cite{Ejima2005}.
The Luttinger parameter can be used to locate the metal-insulator transition
in one dimension. 

\section{Ground-state properties}
\label{sec:gsproperties}

Before we investigate the Mott transition for the half-filled $1/r$-Hubbard model
with $1/r$-long-range interactions 
in more detail in the next section,
we present DMRG results for the ground-state energy, the two-particle gap, the momentum distribution,
the structure factor, and the CDW order parameter. 
For the numerical calculations we employ the same high accuracy SU(2) spin adapted DMRG code as
for our previous work~\cite{1overRHubbard,1overRHubbard-NN} with bond dimension 
$D_{\rm SU(2)}$ up to $D_{\rm SU(2)}^{\rm max}=6000$.

\subsection{Ground-state energy}

For $V= 0$, the ground-state energy per site for finite system sizes 
is given by ($n=N/L$,
$N$: even)~\cite{GebhardRuckenstein,Gebhardbook,1overRHubbard}
\begin{eqnarray}
  e_0 &=&\frac{1}{4} n(n-1) +\frac{U}{4} n   \label{eq:gsenergyNeven} \\
  && -\frac{1}{2L} \sum_{r=0}^{(N/2)-1}\sqrt{1+U^2-4U(2r+1-L/2)/L} \nonumber 
\end{eqnarray}
with the abbreviation $e_0\equiv e_0(N,L;U,V= 0)$.

\begin{figure}[t]
 \begin{flushleft}
(a)  \includegraphics[width=7cm]{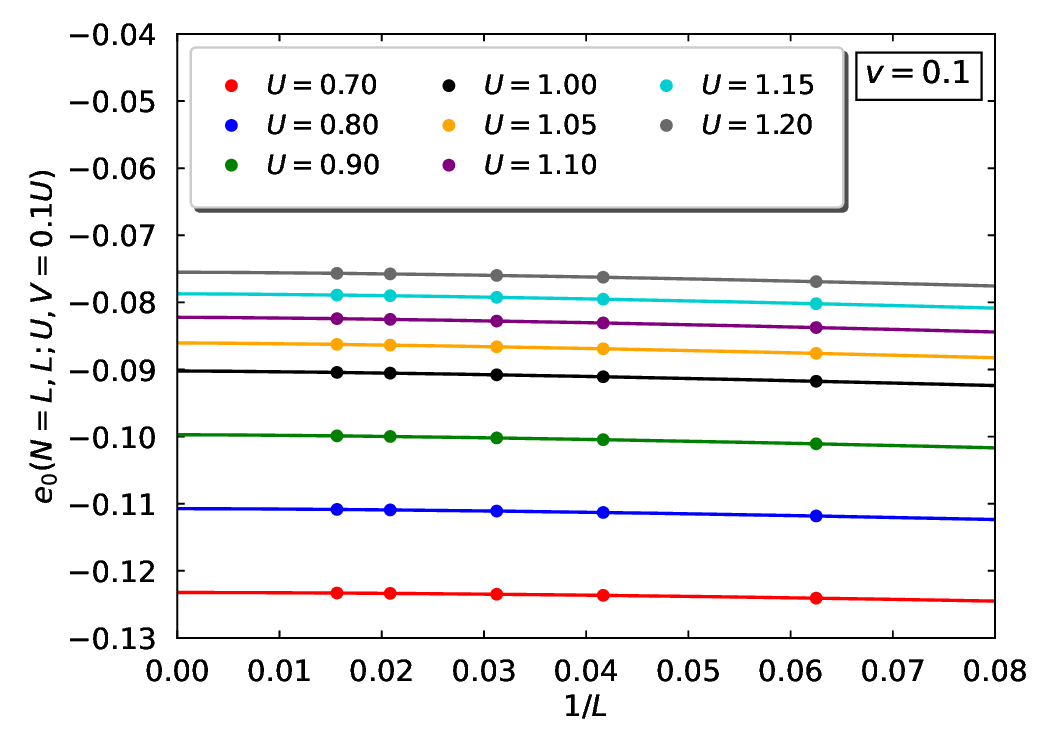}\\[12pt]
(b)  \includegraphics[width=7cm]{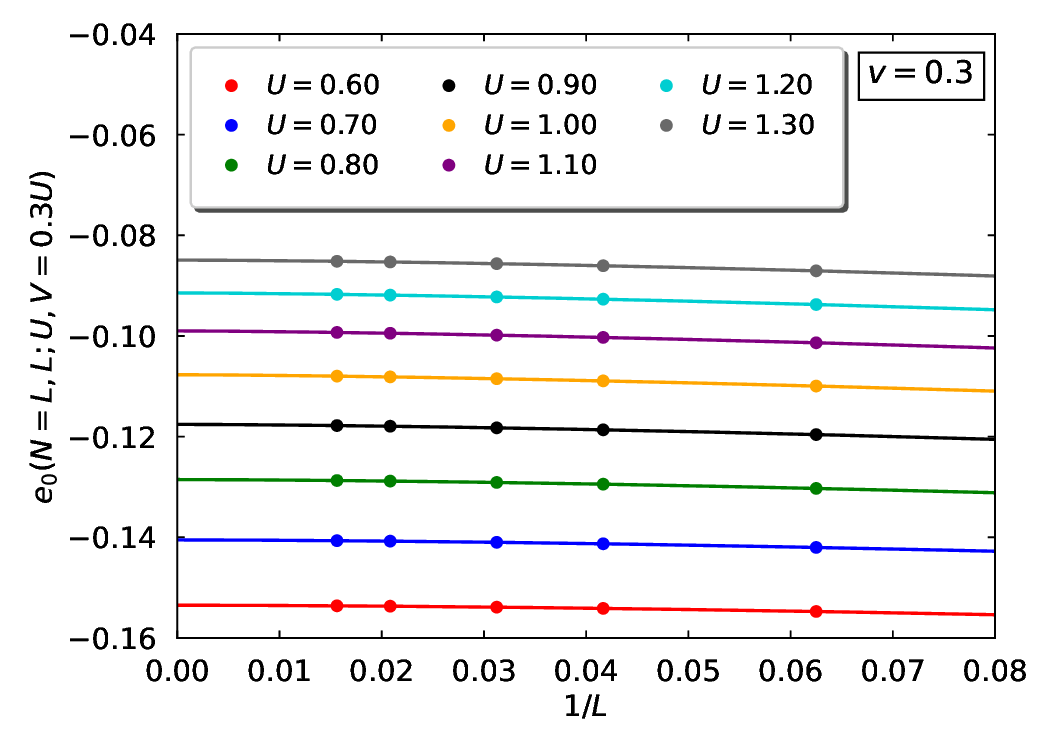}\\[12pt]
(c)  \includegraphics[width=7cm]{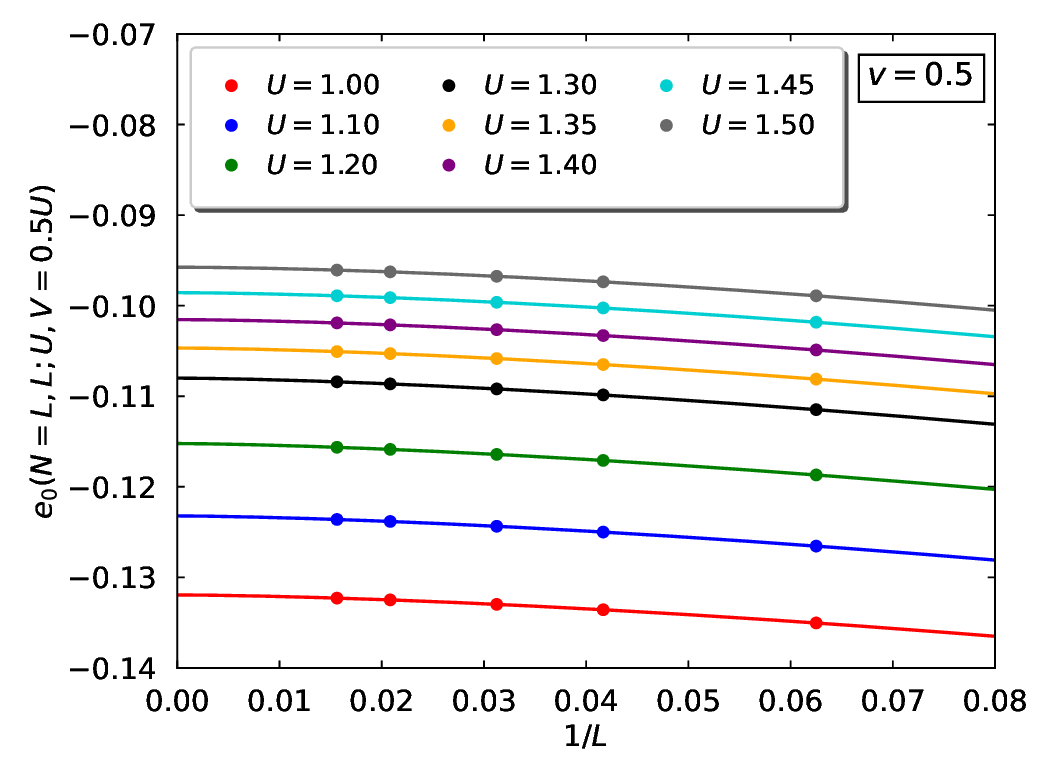}
 \end{flushleft}
 \caption{Ground-state energy per lattice site
    at half band-filling, $e_0(N=L,L;U,V)$,
    for the $1/r$-Hubbard model with $1/r$-long-range interactions
    as a function of $1/L$ for $L=8,16,24,32,48,64$
    and various values for $U$
    for (a) $v=0.1$, (b) $v=0.3$, (c) $v=0.5$.
    The continuous lines are fits to the algebraic
    fit function~(\protect\ref{eq:ezeroextrapolation}).
    The intercept of the extrapolation curves with the ordinate
    defines the extrapolation estimate $e_0(n=1;U,V)$
    in the thermodynamic limit.\label{fig:gsenergy}}
\end{figure}

In  the thermodynamic limit at $n=1$,
the ground-state energy per site becomes particularly simple,
\begin{eqnarray}
  e_0(n= 1;U\leq 1,V= 0)
  &=& -\frac{1}{4} + \frac{U}{4}-\frac{U^2}{12} \; ,\nonumber \\
  e_0(n= 1;U\geq 1,V= 0) &=& -\frac{1}{12 U} \; .
  \label{eq:e0V0TDLhalffilling}
\end{eqnarray}
The analytic expressions~(\ref{eq:gsenergyNeven})
and~(\ref{eq:e0V0TDLhalffilling}) are useful for a
comparison with numerical data at $V=0$.

For finite~$V$, we can use first-order perturbation theory for weak interactions, $U,V\ll1$, 
and at half band-filling to find
\begin{equation}
    e_0^{\rm PT}(U,V)=-\frac{1}{4}+\frac{U}{4}\left(1-\frac{8v}{\pi^2}\frac{7\zeta(3)}{8}\right)+\mathcal{O}(U^2)
    \label{eq:e01storderpt}
\end{equation}
with $v=V/U$ in the thermodynamic limit and $\zeta(x)$ is the Riemann zeta function. 
Note that Eq.~(\ref{eq:e01storderpt}) holds for all $v$, as long as $U,V \ll 1$.

We display the ground-state energy per site
at half band-filling, $e_0(L,L;U,V)$, 
as a function of the inverse system size ($L= 8,16,24,32,48,64$)
and various values of $U$ in Fig.~\ref{fig:gsenergy}a ($v=0.1$),
Fig.~\ref{fig:gsenergy}b ($v=0.3$), and Fig.~\ref{fig:gsenergy}c ($v=0.5$).
For the extrapolation to the thermodynamic limit, we use the
algebraic fit function
\begin{equation}
  e_0(L,L;U,V) = e_0(n=1;U,V) + a_0(U,V) \left(\frac{1}{L}\right)^{\gamma_0(U,V)} \; ,
  \label{eq:ezeroextrapolation}
  \end{equation}
where $e_0(n=1;U,V)$ denotes the 
numerical estimate for the ground-state energy density
in the thermodynamic limit; $a_0(U,V)$ and $\gamma_0(U,V)$ are the two other
fit parameters.
This extrapolation scheme is appropriate
for $V= 0$~\cite{1overRHubbard} because the ground-state energy
per site scales with $(1/L)^2$ for $U\neq 1$ and with $(1/L)^{3/2}$
for $U= U_{\rm c}(V= 0)= 1$, as follows from Eq.~(\ref{eq:gsenergyNeven}).

More generally, we {\em assume\/} for all $(U,V)$
\begin{equation}
    \gamma_0(U,V) = \left\{ \begin{array}{ccl}
    2 & \hbox{for} & U \neq U_{\rm c}(V) \\[6pt]
    \displaystyle \frac{3}{2} & \hbox{for} & U = U_{\rm c}(V) 
    \end{array}
    \right. \; .
    \label{eq:gamma0TDL}
\end{equation}
Eq.~(\ref{eq:gamma0TDL}) follows from the fact that
the ground-state energy density displays generic $(1/L)^2$ finite-size corrections when the
model is not critical. 
When the holon dispersion displays a square-root divergence at low energies for $U_{\rm c}(V)$, the exponent is reduced to $\gamma_0(U_{\rm c}(V),V)=3/2$,
as in the case of the bare $1/r$-Hubbard model.
We shall discuss the finite-size modifications
in detail in Sec.~\ref{sec:MTnearest-neighbor}.

\begin{figure}[tb]
    \includegraphics[width=7.7cm]{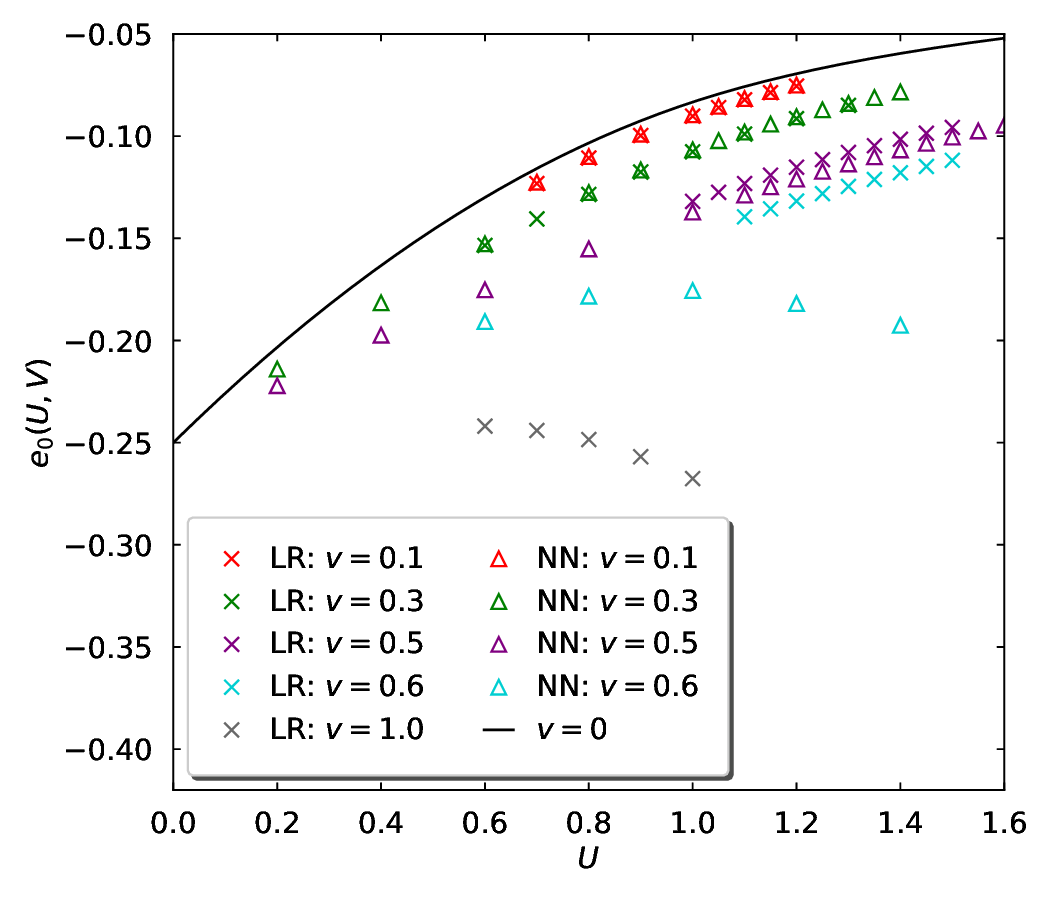}
 \caption{Ground-state energy density
   at half band-filling, $e_0(n=1;U,V)$,
   for the $1/r$-Hubbard model with $1/r$-long-range interactions (crosses) from the extrapolation
   to the thermodynamic limit in Fig.~\protect\ref{fig:gsenergy}, and for the extended $1/r$-Hubbard model
   (open triangles, see Ref.~\cite{1overRHubbard-NN}).
   The continuous line is the exact result
   for $V=0$, $e_0(n=1;U,V=0)$.
   \label{fig:gsenergyextrapolate}}
\end{figure}

The extrapolated ground-state energies are shown in Fig.~\ref{fig:gsenergyextrapolate} 
for $1/r$-long-range interactions (crosses) and for nearest-neighbor interactions
(open triangles), together with the exact result for $V=0$. 
For small interactions,
the long-range interactions in the particle-hole symmetric form
decrease the ground-state energy because
the Hartree contribution at half band-filling is subtracted
in the definition of the interaction, and the Fock contribution
is negative because of the exchange hole.
Therefore, the linear term in the interaction $(U/4)(1-7v\zeta(3)/\pi^2)$, see Eq.~(\ref{eq:e01storderpt}),
is smaller in the presence of repulsive long-range interactions.

At large interactions,
the ground-state energy approaches zero,  $\lim_{U\to \infty} e_0(n=1;U,V= vU)= 0$, as long as the charge-density wave is absent.
In the presence of a CDW, the ground-state energy
is negative and proportional to $U$ for large interactions, $e_0(U\gg 1,V)=U(1/2-v\ln 2)$, see appendix.
Therefore, the ground-state energy displays a maximum for $v\gtrsim 1$
because its slope as a function of~$U$ is positive for small interactions  
and negative for large interaction strengths in the CDW phase. This result further demonstrates that as soon as the condition $v\ln 2 > 1/2$ is satisfied, 
the CDW phase emerges in the strong-coupling regime. This allows us to infer the existence of a tri-critical point,
where Luttinger liquid, CDW insulator and Mott-Hubbard insulator coexist.
Strong-coupling theory to first order indicates that the tri-critical point 
lies on the line $v_{\rm tp}^{\rm LR} = 1/(2\ln 2)$, i.e., $V \approx  0.72134 U$.
For the extended $1/r$-Hubbard model, the same considerations give $v_{\rm tp}^{\rm NN} = 1/2$.

The comparison between the $1/r$-Hubbard models with $1/r$-long-range and nearest-neighbor interactions 
shows that the ground-state energy density remains similar for small interaction strengths. 
Discernible deviations emerge at moderate values of~$v$, e.g., for $v = 0.5$, and discrepancies become clearly visible for $v \gtrsim 0.6$. 
This behavior arises from the fact that the tri-critical point is reached for smaller $v$ in the case
of nearest-neighbor interactions, $v_{\rm tp}^{\rm NN}<v_{\rm tp}^{\rm LR}$. 
In comparison with the extended $1/r$-Hubbard model,
the CDW phase requires larger $v$ for $1/r$-long-range interactions.

\subsection{Two-particle gap}

For $V=0$ the two-particle gap is known for all system sizes~\cite{GebhardRuckenstein,Gebhardbook,1overRHubbard},
\begin{equation}
    \Delta_2(L;U\geq 1,V=0)=  U-1+\frac{2}{L} +\sqrt{(U-1)^2+\frac{4U}{L}}\; .
    \label{eq:Delta2analyt}
\end{equation}
\begin{figure}[t]
 \begin{center}
(a)  \includegraphics[width=7cm]{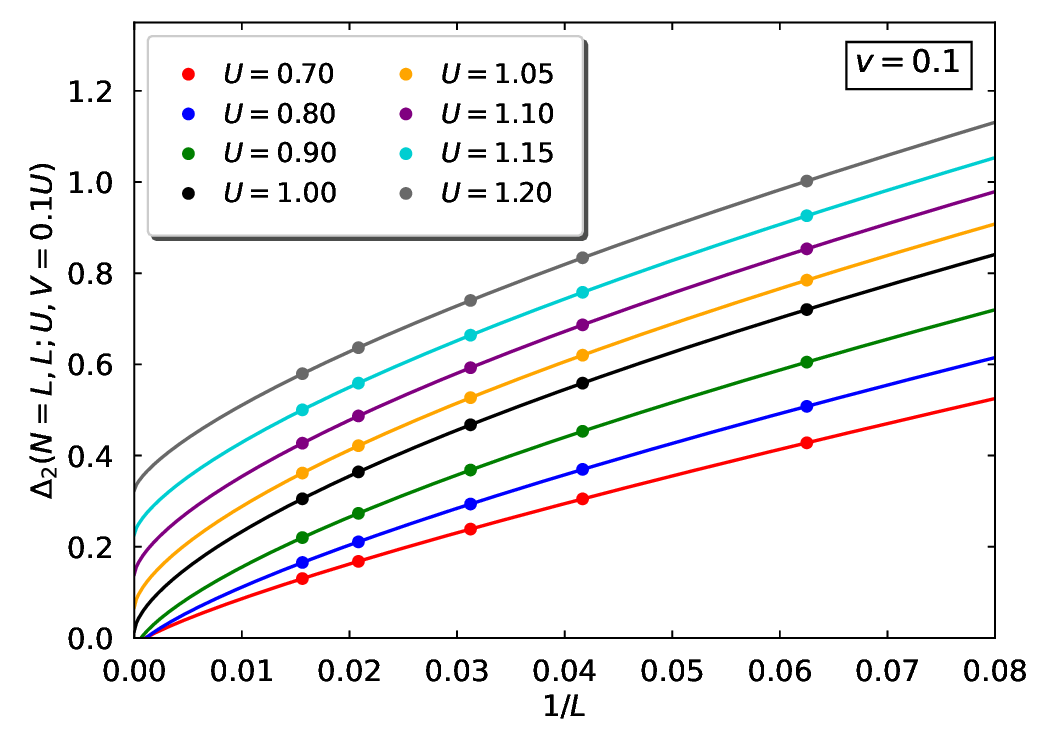}\\[6pt]
(b)  \includegraphics[width=7cm]{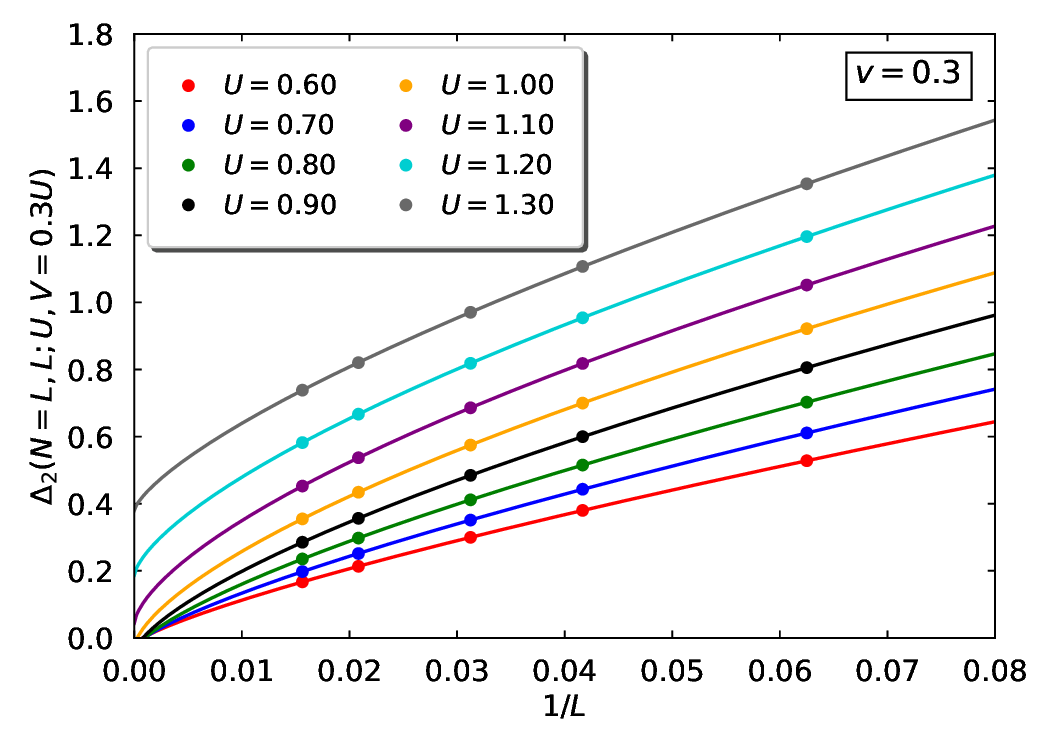}\\[6pt]
(c)  \includegraphics[width=7cm]{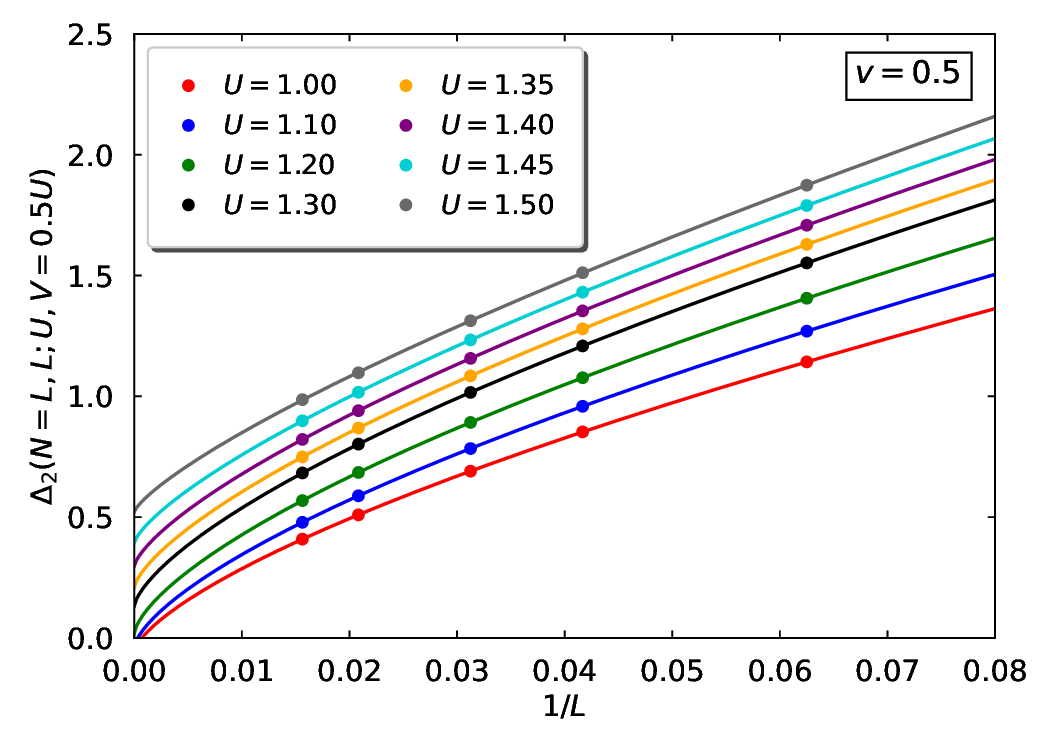}
 \end{center}
 \caption{Two-particle gap $\Delta_2(L;U,V)$
   for the extended $1/r$-Hubbard model with $1/r$-long-range interactions
    as a function of inverse system size for $L=8,16,24,32,48,64$ and various values for $U$
    for (a) $v=0.1$, (b) $v=0.3$, (c) $v=0.5$.
    The continuous lines are fits to the algebraic
    fit function~(\protect\ref{eq:gapextrapolationscheme}).
The intercept of the extrapolation curves with the ordinate
    defines the extrapolation estimate $\Delta_2(U,V)$
    for the two-particle gap.\label{fig:Delta2finite}}
\end{figure}
In the thermodynamic limit, we find
\begin{equation}
    \Delta_2(U\geq 1,V=0) = 2(U-1) \; .
    \label{eq:Delta2TDLVzero}
\end{equation}
The gap opens linearly above the critical interaction strength, $U_{\rm c}(V=0)=1$.
Eq.~(\ref{eq:Delta2analyt}) shows that the finite-size data approach the value
in the thermodynamic limit
\begin{equation}
  \Delta_2(L;U,V) =\Delta_2(U,V) + a_2(U,V) \left(\frac{1}{L}\right)^{\gamma_2(U,V)}
  \label{eq:gapextrapolationscheme}
\end{equation}
with $\gamma_2(U\neq U_{\rm c},V=0)=1$, $\gamma_2(U=U_{\rm c},V=0)=1/2$.

More generally, we {\em assume\/} for all $(U,V)$
\begin{equation}
    \gamma_2(U,V) = \left\{ \begin{array}{ccl}
    1 & \hbox{for} & U \neq U_{\rm c}(V) \\[6pt]
    \displaystyle \frac{1}{2} & \hbox{for} & U = U_{\rm c}(V) 
    \end{array}
    \right. \; .
    \label{eq:gamma2TDL}
\end{equation}
As for the ground-state energy, these exponents apply for very large system sizes. 
We shall discuss the finite-size modifications
in more detail in Sec.~\ref{sec:MTnearest-neighbor}.

\begin{figure}[b]
  \includegraphics[width=8cm]{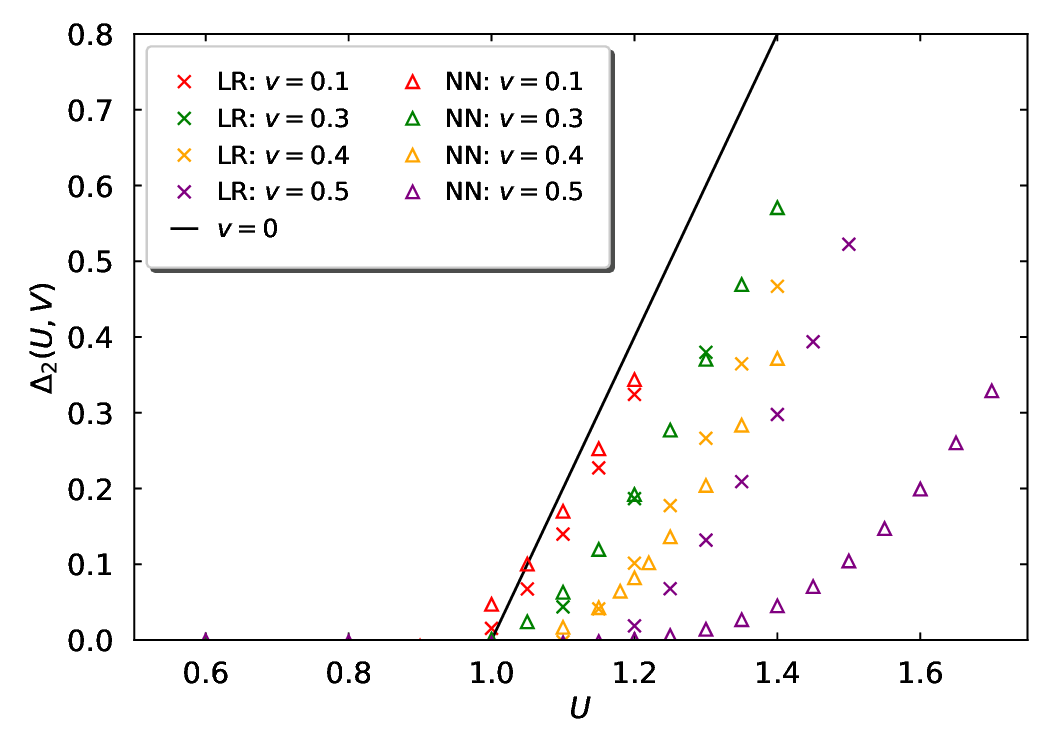}
  \caption{Two-particle gap $\Delta_2(U,V)$ for the $1/r$-Hubbard model
    as a function of $U$  for $v=V/U=0.1$ (red),
    $v=0.3$ (green), $v=0.4$ (orange), $v=0.5$ (purple),
    extrapolated from finite-size data with up to $L=64$ sites. 
The crosses represent the model with $1/r$-long-range interactions, the open triangles correspond to the model with only nearest-neighbor interactions~\cite{1overRHubbard-NN}.
    The continuous line is the exact result in the thermodynamic limit
    for $V=0$, $\Delta_2(U,V= 0)= 2(U-1)$, see Eq.~(\ref{eq:Delta2TDLVzero}).\label{fig:Delta2}}
\end{figure}

In Fig.~\ref{fig:Delta2finite}
we show the DMRG results for $\Delta_2(L;U,V)$ as a function of $1/L$ for
various values for $U$ as a function of $1/L$ for $L=8,16,24,32,48,64$ 
for (a) $v=0.1$, (b) $v=0.3$, (c) $v=0.5$.
The lines are fits to the algebraic function in Eq.~(\ref{eq:gapextrapolationscheme}).
The fits in Fig.~\ref{fig:Delta2finite}
are seen to agree very well with the data, showing a steep decrease
of the finite-size gap as a function of inverse system size. This indicates that
large system sizes are required to obtain reasonable gap extrapolations.

The extrapolated gaps $\Delta_2(U,V)$ are shown in Fig.~\ref{fig:Delta2}
as a function of~$U$ for $v=0$, $v=0.1$, $v=0.3$, $v=0.4$ and $v=0.5$ for the $1/r$-Hubbard model with $1/r$-long-range interactions (crosses) 
and for the $1/r$-Hubbard model with nearest-neighbor interactions (open triangles).
Finite-size effects smoothen the phase transition which makes it difficult 
to extract the critical point and the critical exponent for the opening of the Mott-Hubbard gap. 

In the bare $1/r$-Hubbard model, the Mott transition is solely governed by the on-site interaction. 
Once the Hubbard interaction exceeds the bandwidth, $U>W$, electrons become `localized', and the metallic state goes over to a Mott insulator. 
In the presence of inter-site interactions, the Mott transition is shifted to larger values which
implies a more stable metallic Luttinger liquid.

At first glance, the shift of the Mott transition to higher values in models with repulsive inter-site interactions  
seems counterintuitive. From a wave-mechanical viewpoint, however, the repulsive interaction between lattice sites softens
the two-particle scattering potential. 
Figuratively speaking, particles scattered by the weaker interaction $V$ do not fully experience the stronger on-site interaction $U$.
This ties in with the observation that the gap in the $1/r$-Hubbard model with $1/r$-long-range interactions
is smaller than that for the extended $1/r$-Hubbard model for the same $v$ because the long-range parts of the inter-site interaction further smoothen 
the two-particle scattering potential in position space.

\begin{figure}[t]
(a)  \includegraphics[width=7cm]{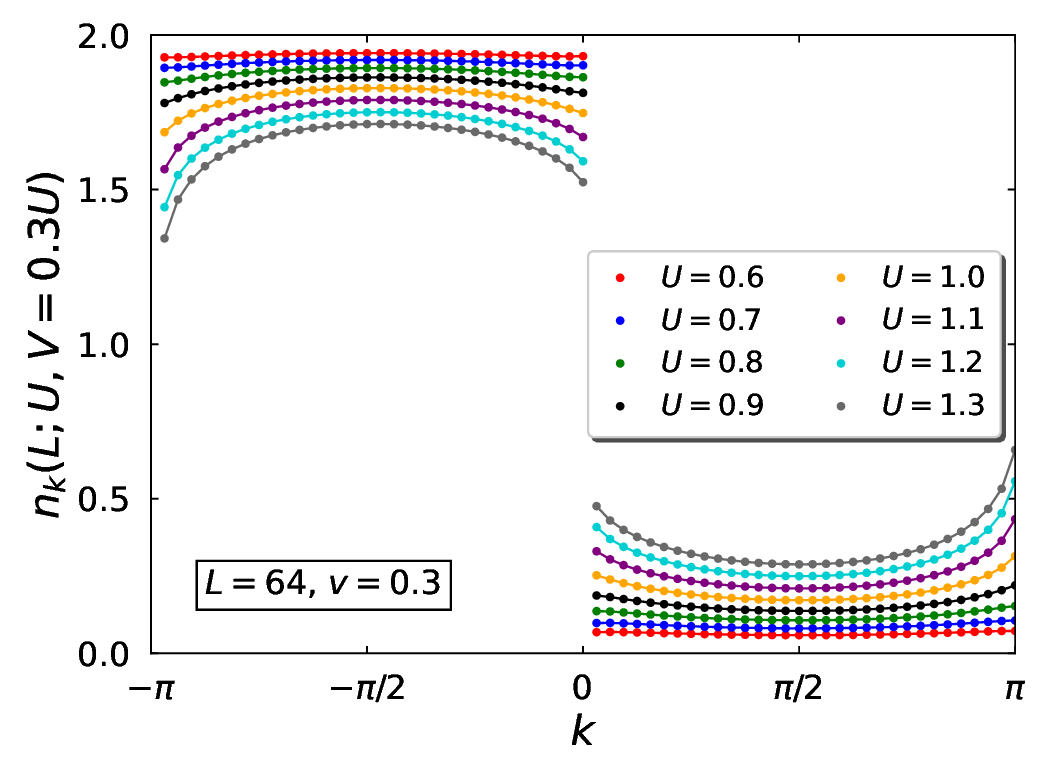}\\
(b)  \includegraphics[width=7cm]{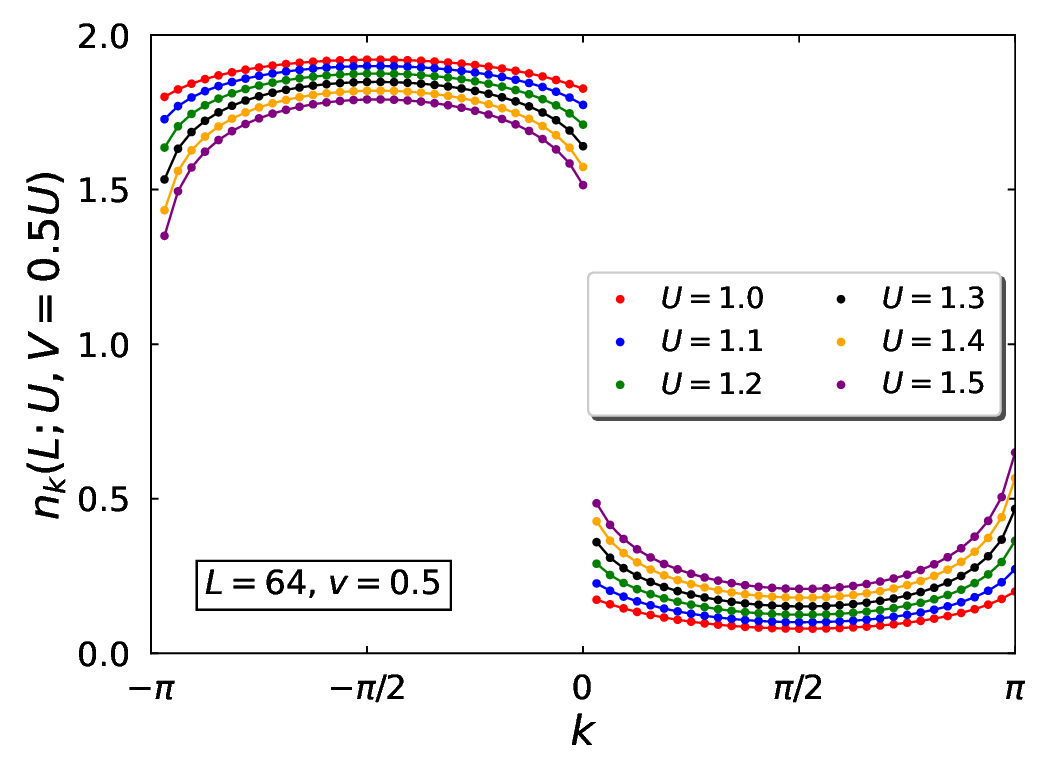}\\
(c)  \includegraphics[width=7cm]{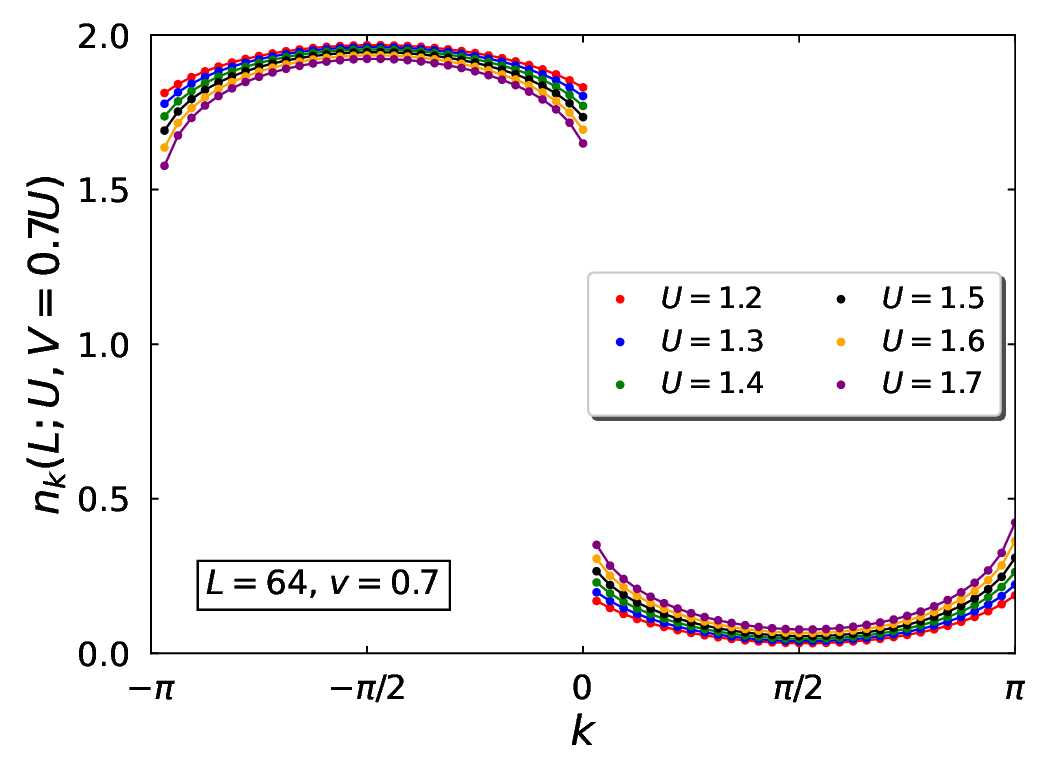}
 \caption{Momentum distribution $n_k(L;U,V)$ at half band-filling
   for the extended $1/r$-Hubbard model with $1/r$-long-range interactions
   for $L=64$ sites
   and  for various values for $U$
   for $v=0.3$, $v=0.5$, and $v=0.7$
   (from top to bottom). Continuous lines are a guide to the eye only.
   \label{fig:momdis}}
\end{figure}

\subsection{Momentum distribution}

In Fig.~\ref{fig:momdis} we show the momentum distribution from DMRG
at half band-filling for $L= 64$
sites for various values of~$U$ and $v= 0.3$, $v= 0.5$,
and $v= 0.7$ (from top to bottom). The momentum distributions resemble those 
obtained for the extended $1/r$-Hubbard model. 
For small interactions, the
momentum distribution looks like that of a Fermi liquid with
all states $-\pi <k<0$ occupied and all states $0<k<\pi$ empty.
For small~$U$, 
low-energy scattering processes are limited to 
the vicinity of the sole Fermi point $k_{\rm F}= 0$.
Indeed, in the field-theoretical limit, $U,V \ll 1$, the model reduces
to a bare $g_4$-model of only right-moving particles~\cite{Thierrybook}.
This `non-interacting Luttinger liquid' displays a jump discontinuity at
$k_{\rm F}$.

However, 
the $1/r$-Hubbard model is defined on a lattice
and the bandwidth is finite. Consequently, there is a left Fermi point at
$k_{{\rm F},2}= -\pi$ at the lower Brillouin zone boundary where the bare dispersion relation jumps by~$W$.
The left Fermi point starts to play a role when $U$ becomes
of the order of half the bandwidth.
States near $k_{{\rm F},2}$
are depleted more quickly as a function of~$U$ than those deeper in the
Brillouin zone. Therefore, as seen in Fig.~\ref{fig:momdis},
the momentum distribution develops a maximum (minimum) around $k= -\pi/2$ ($k=\pi/2$).

These considerations show that the Luttinger parameter must deviate from
unity, $K(U,V)< 1$, for all $(U,V)$, even though corrections to unity
are (exponentially) small for $U,V \ll 1$.
Therefore, $n(k)$ 
is a continuous function
in the 
$1/r$-Hubbard model for all $U,V>0$.

\subsection{Structure factor and CDW order parameter}

\begin{figure}[t]
(a)  \includegraphics[width=7.5cm]{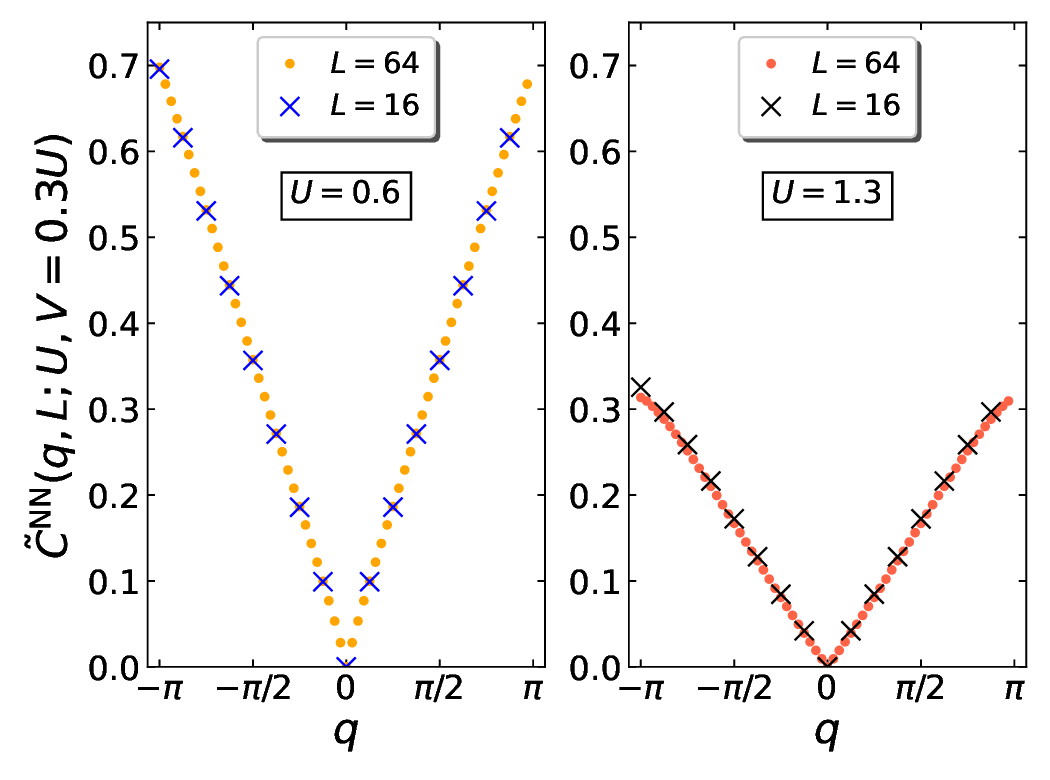}\\[12pt]
(b)  \includegraphics[width=7.5cm]{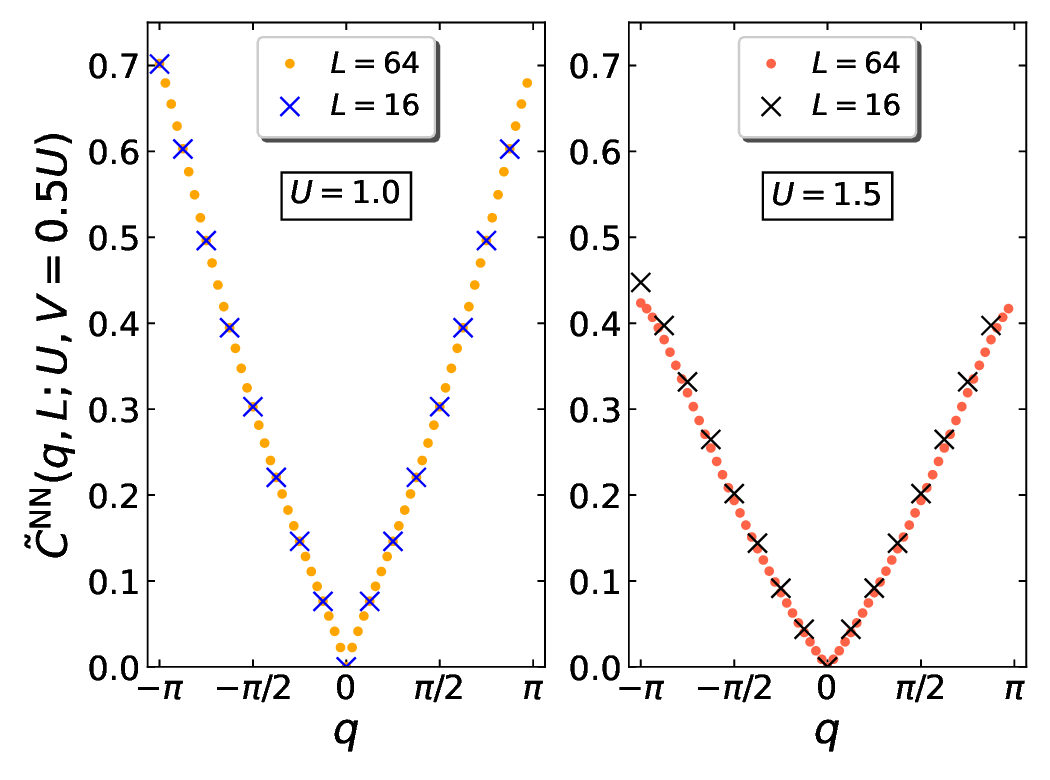}\\[12pt]
(c)  \includegraphics[width=7.5cm]{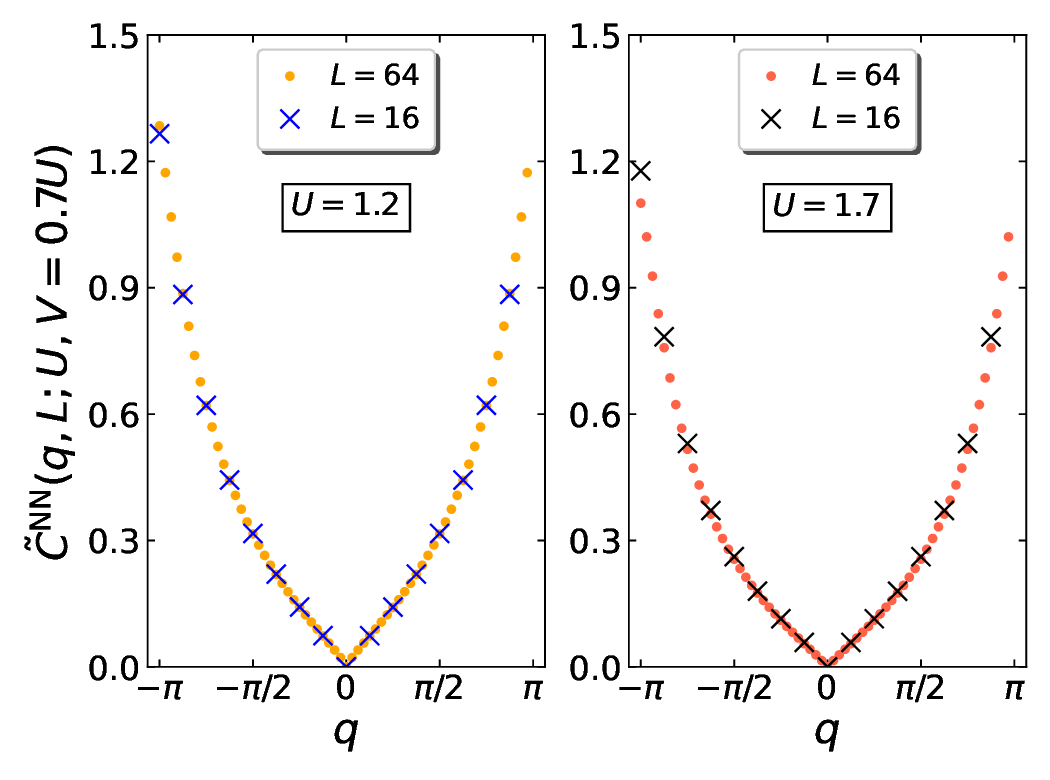}
 \caption{Structure factor $\tilde{C}^{\rm NN}(q,L;U,V)$
   for the extended $1/r$-Hubbard model with long-range interactions
    for $L=16, 64$ below (left) and above (right) the Mott transition
    for (a) $v=0.3$, (b) $v=0.5$, (c) $v=0.7$.
    Dotted lines are a guide to the eye only.}
    \label{fig:CNNq}
\end{figure}

Lastly, we show the structure factor from DMRG in Fig.~\ref{fig:CNNq}
for $v=0.3$, $v=0.5$, and $v=0.7$ (from top to bottom)
for the extended $1/r$-Hubbard model with $1/r$-long-range interactions at system sizes $L=16,64$
below (left) and above (right) the Mott transition. As for the momentum distribution,
the $1/r$-Hubbard model with nearest-neighbor interactions and with $1/r$-long-range interactions lead to 
the same qualitative behavior.

In general, finite-size effects are fairly small but larger
systems permit a much better resolution in momentum space.
In comparison with the exact result for the non-interacting system,
\begin{equation}
\tilde{C}^{\rm NN}(q,n=1;U=0,V=0)=\frac{|q|}{\pi} \; ,
\end{equation}
we see that the Hubbard interaction reduces the charge fluctuations.
This is expected because the suppression of double occupancies likewise
reduces the number of holes and the charges are more homogeneously distributed 
in the system. Therefore, the charge correlations become smaller when
we compare the left and right figure in the same row.

The repulsive inter-site interactions counter the effect of the Hubbard interaction
because pairs of a double occupancy
and a hole are energetically
favorable.
Therefore, the charge correlations increase when we
go from top to bottom in the left/right row, even though~$U$ also increases
from top to bottom.

\begin{figure}[t]
  \begin{center}
   (a) \includegraphics[width=8cm]{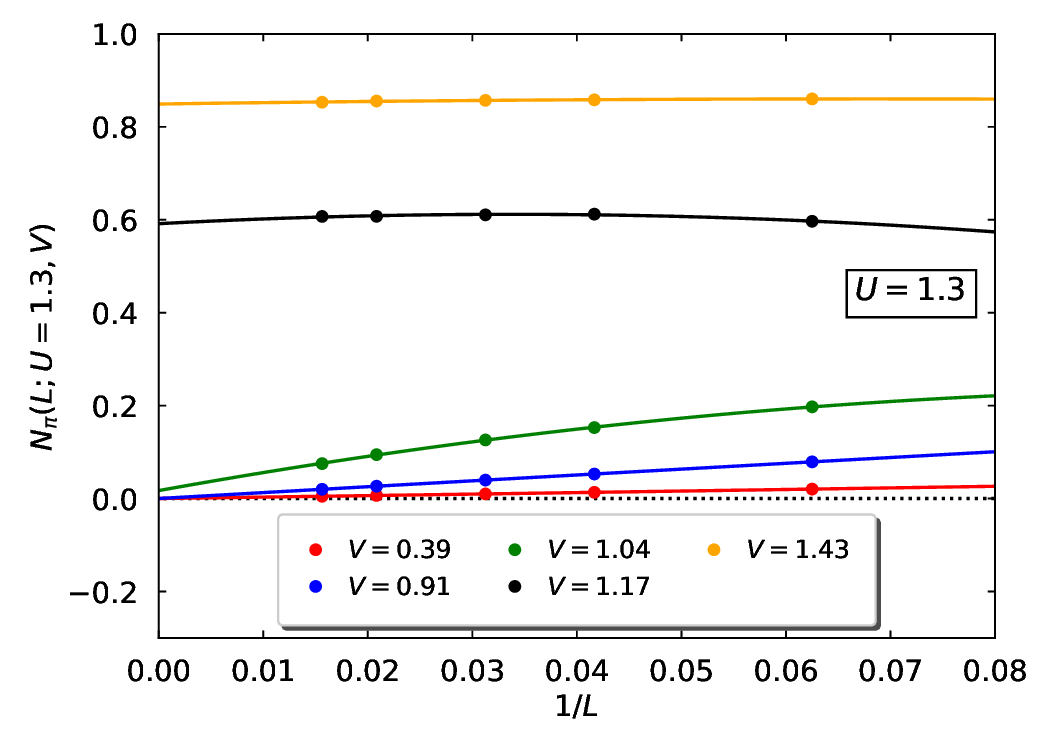}\\
   (b) \includegraphics[width=8cm]{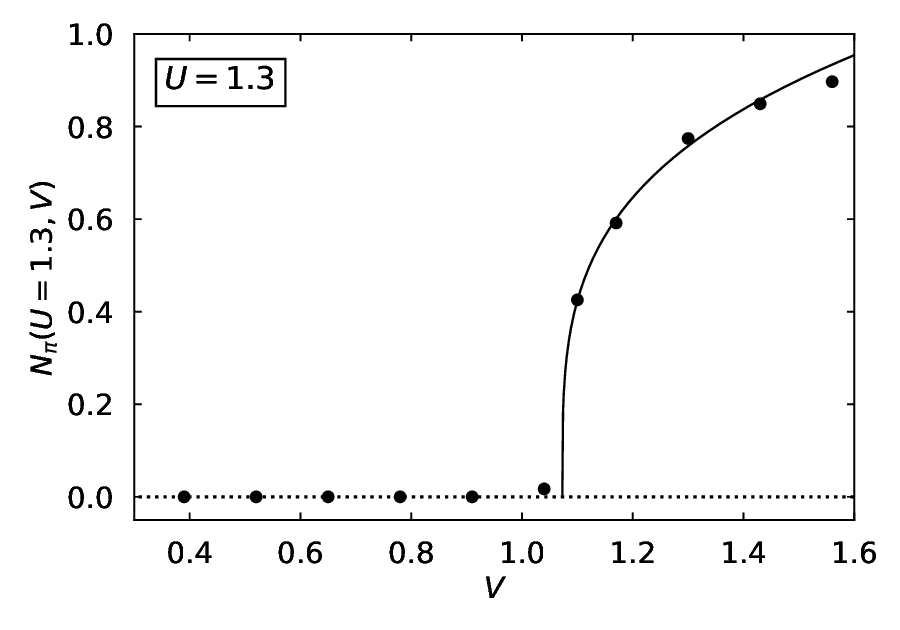}
  \end{center}
 \caption{(a) CDW order parameter $N_{\pi}(L;U=1.3,V)$ for the half-filled $1/r$-Hubbard model with $1/r$-long-range interactions
    as a function of $1/L$ ($L\leq 64$) for fixed $U=1.3$ and various $V$-values. Lines are a second-order polynomial fit in $1/L$,
    see Eq.~(\ref{eq:CDWfit2nd});
    (b) Extrapolated CDW order parameter $N_{\pi}(U=1.3,V)$ as a function of $V$. The line is an algebraic fit to the data in the vicinity
    of the CDW transition, see Eq.~(\ref{eq:CDWexponent}), with $V_{\rm c}(U=1.3)=1.07$, $N_0=1.14$ and $2\nu=0.27$.\label{fig:CDWorderparameter}}
\end{figure}

When the interactions between the electrons on different lattice sites increase beyond a certain threshold value
$V_{\rm c}^{\rm CDW}(U)$, the ground state displays charge-density-wave order.
In Fig.~\ref{fig:CDWorderparameter}a we show the charge-density-wave order parameter $N_{\pi}(L;U=1.3,V)$, see Eq.~(\ref{eq:ourCDWorderparameterdef}), 
as a function of $1/L$ for various values of~$V$, and 
the extrapolated result $N_{\pi}(U=1.3,V)$ into the thermodynamic limit using a second-order polynomial fit
in Fig.~\ref{fig:CDWorderparameter}b,
\begin{equation}
    N_{\pi}(L;U,V) = N_{\pi}(U,V) +\frac{N_1(U,V)}{L} + \frac{N_2(U,V)}{L^2} \; .
    \label{eq:CDWfit2nd}
\end{equation}
Apparently, the CDW order parameter is continuous over the CDW transition.
Close to the transition, $V\gtrsim V_{\rm c}(U)$,
\begin{equation}
N_{\pi}(U,V) =N_0\left[V-V_{\rm c}(U)\right]^{2\nu} \; ,
\label{eq:CDWexponent}
\end{equation}
where $\nu$ is the critical exponent for the CDW order parameter $D(U,V)$. 
Note that, in our previous article on the extended $1/r$-Hubbard model~\cite{1overRHubbard-NN}, 
we used data for $v=0.7$ to compute the order parameter.
Here, we trace the CDW transition as a function of $V$ for fixed $U$.

To make use of Eq.~(\ref{eq:CDWexponent}), 
the critical interaction $V_{\rm c}(U)$ must be known.
In addition, the region of validity of Eq.~(\ref{eq:CDWexponent}) is unknown {\sl a priori}.
Typically, one has to study system parameters close to the transition to obtain
a reliable estimate for $\nu$. Therefore, very large system sizes might be necessary 
to reach the scaling limit, and we have to be satisfied with the result from Fig.~\ref{fig:CDWorderparameter}b
that the CDW transition at $U=1.3$ is continuous with exponent $\nu \leq 1/2$. 

\section{Mott transitions}
\label{sec:MTnearest-neighbor}

In this section, we determine the critical value for the Mott-Hubbard and CDW transitions in the $1/r$-Hubbard model with $1/r$-long-range interactions.
We investigate the two-particle gap, the ground-state energy, the Luttinger parameter, and
the structure factor at the Brillouin zone boundary to locate the critical interaction strengths $U_{\rm c}^{\rm Mott}(V)$ and $U_{\rm c}^{\rm CDW}(V)$.

\subsection{Two-particle gap}
\label{subsec:tpgapgamma2}

To simplify the notation, we drop the superscript `Mott' 
when quantities cannot be confused with those for the CDW transition.

In our previous work~\cite{1overRHubbard}, we showed that the
exponent $\gamma_2(U)=\gamma_2(U,V=0)$ sensitively depends on~$U$ in the vicinity
of the Mott-Hubbard transition, and  the
critical interaction for the bare $1/r$-Hubbard model, $U_{\rm c}(V=0)=1$,
was obtained with an accuracy of one per mil. Moreover, we demonstrated in section~IV~A of~\cite{1overRHubbard-NN} that the minimum of the $\gamma$-parameter depends only marginally on the system size, especially for $L \geq 16$.

\begin{figure}[t]
  \begin{center}
    \includegraphics[width=8cm]{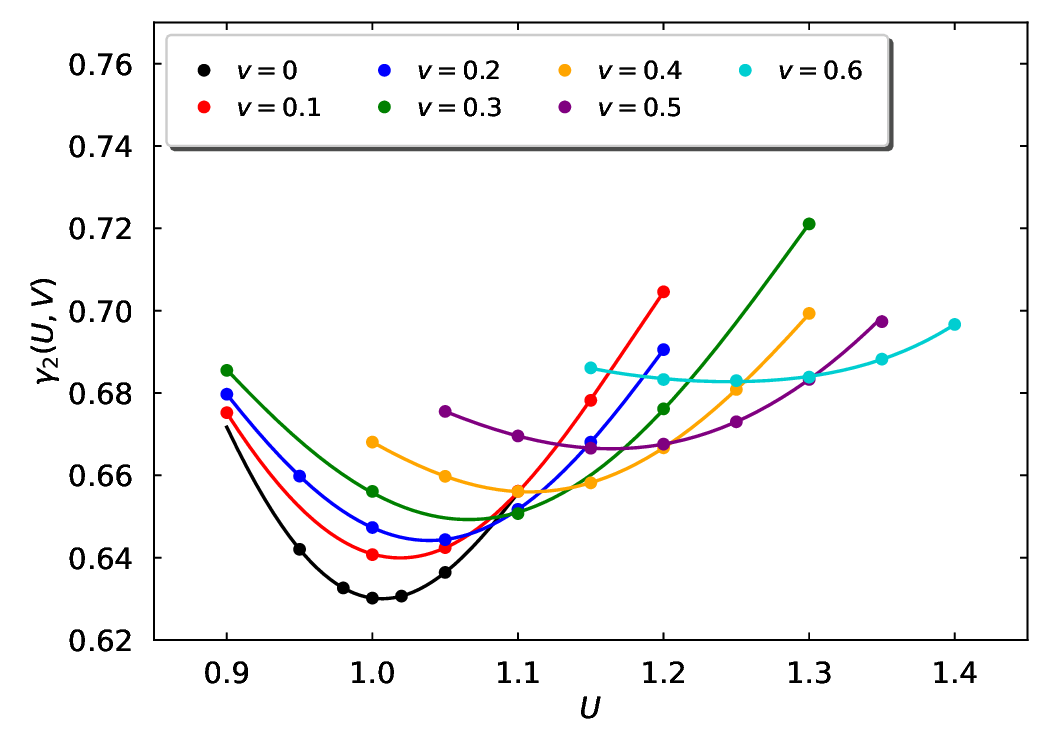}
  \end{center}
  \caption{Exponent $\gamma_2(U,V)$ 
    for the two-particle gap in the extended $1/r$-Hub\-bard model with $1/r$-long-range interactions as a function of $U$ for various values of $v=V/U$,
    based on system sizes $16\leq L \leq 64$.
The minimum of the curve determines $U_{\rm c,gap}(V)$.\label{fig:gammaofUinHM}}
\end{figure}

In Fig.~\ref{fig:gammaofUinHM} we display the exponent
$\gamma_2(U,V)$, as obtained from the fit of the finite-size data
in the range $16\leq L \leq 64$ to the algebraic function in Eq.~(\ref{eq:gapextrapolationscheme}).
Also shown are the quartic fits around the minima which lead to
the critical interactions $U_{\rm c,gap}(V)$, listed below in Table~\ref{tab:one}.
Note that the curves flatten out for increasing $v$.
Therefore, it is not possible to determine quantum-phase transitions for $v\gtrsim 0.7$ 
from the parameter $\gamma_2(U,V)$ for system sizes $L\leq 64$ and step size $\Delta U=0.05$.

The comparison with the exact value for $V=0$ shows that the gap exponent $\gamma_2(U,V)$ provides a fairly accurate estimate for
the critical interaction. The same accuracy can be obtained when using the ground-state energy exponent $\gamma_0(U,V)$,
as we shall show next.

\begin{figure}[b]
  \begin{center}
    \includegraphics[width=8cm]{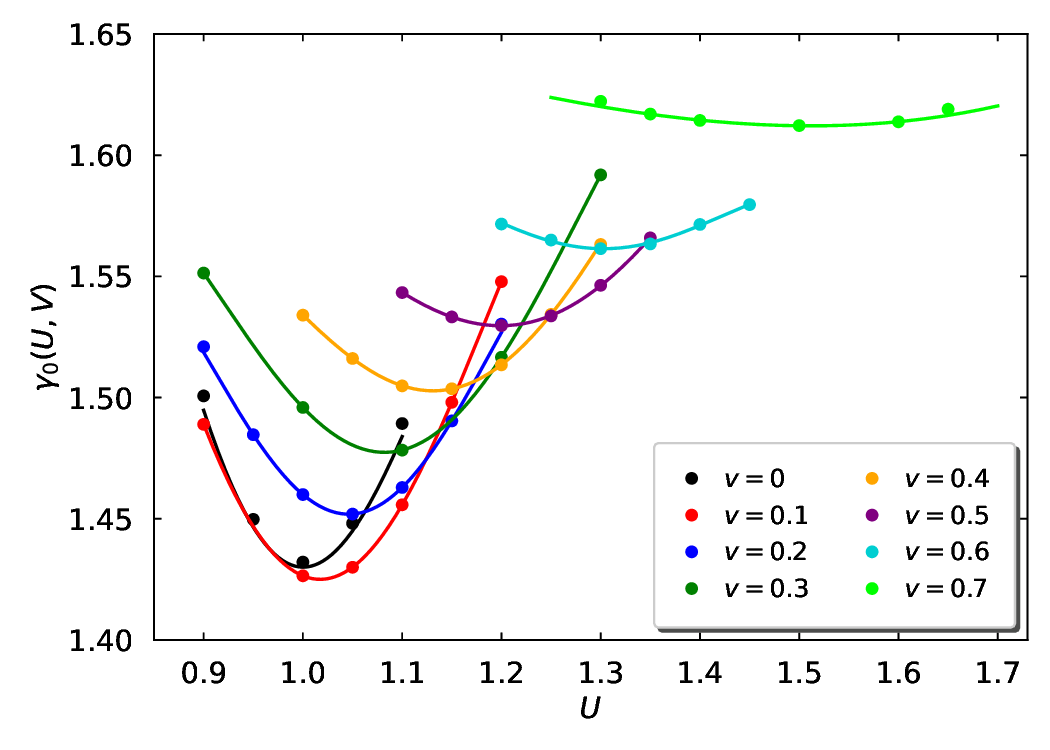}
  \end{center}
  \caption{Exponent $\gamma_0(U,V)$ 
    for the ground-state energy of the extended $1/r$-Hub\-bard model with $1/r$-long-range interactions as a function of $U$ for various values of $v=V/U$,
    based on system sizes $16\leq L \leq 64$.
The minimum of the curve determines $U_{\rm c,gs}(V)$.
    \label{fig:groundstategammaofUinHM}}
\end{figure}

\subsection{Ground-state energy}
\label{subsec:gsenergygamma0}

As seen from Eq.~(\ref{eq:gamma0TDL}), the $1/L$ corrections to the ground-state energy density also permit to locate the Mott transition in the extended $1/r$-Hubbard model,
in the same way as the two-particle gap. 
In Fig.~\ref{fig:groundstategammaofUinHM} we show the exponent
$\gamma_0(U,V)$, as obtained from the fit of the finite-size data
in the range $16\leq L \leq 64$ to the algebraic function in Eq.~(\ref{eq:ezeroextrapolation}).
Also shown in the figure are the quartic fits around the minima which lead to
the critical interactions $U_{\rm c,gs}(V)$ listed in Table~\ref{tab:one}.

\begin{table}[t]
  \begin{ruledtabular}
    \begin{tabular}[t]{rrrrrr}
      $V/U$ & $U_{\rm c,gap} (V)$ & $U_{\rm c,gs}(V)$ & $U_{\rm c,LL}(V)$ & $U_{\rm c,sf}(V)$ & $\overline{U}_{\rm c}(V)$ \\
      \hline \\[-7pt]
      $0$   & $1.009$ & $1.000$ & $1.033$ & $0.965$ & $1.002$ \\
      $0.1$ & $1.019$ & $1.017$ & $1.058$ & $0.979$ & $1.018$ \\
      $0.2$ & $1.039$ & $1.046$ & $1.089$ & $0.999$ & $1.043$ \\
      $0.3$ & $1.066$ & $1.082$ & $1.129$ & $1.031$ & $1.077$ \\
      $0.4$ & $1.108$ & $1.131$ & $1.183$ & $1.071$ & $1.123$ \\
      $0.5$ & $1.164$ & $1.199$ & $1.261$ & $1.132$ & $1.189$\\
      $0.6$ & $1.245$ & $1.303$ & $1.388$ & $1.232$ & $1.292$ \\
      $0.7$ & $-$     & $1.512$ & $1.684$ & $1.404$ & $1.533$ \\
    \end{tabular}
          \end{ruledtabular}
  \caption{Critical interaction strengths for the extended $1/r$-Hubbard model with $1/r$-long-range interactions,
as obtained from the two-particle gap, the ground-state energy, the Luttinger parameter,
and the structure factor for systems with $16\leq L\leq 64$ lattice sites.
    For $V=0$, the exact
    result in the
    thermodynamic limit is known~\cite{GebhardRuckenstein},
    $U_{\rm c}(V= 0)= 1$.
    \label{tab:one}}
  \end{table}

As seen from Table~\ref{tab:one},
the critical interaction strengths obtained from the minima of $\gamma_0(U,V)$ very well agree with the exact result at $V=0$ and also with the values obtained 
from the gap exponent $\gamma_2(U,V)$, with deviations in the low percentage range. 
Therefore, we find reliable estimates for the critical interaction strength for the Mott transition from $\gamma_0(U,V)$ and $\gamma_2(U,V)$.

\begin{figure}[t]
  \begin{center}
   (a) \includegraphics[width=8cm]{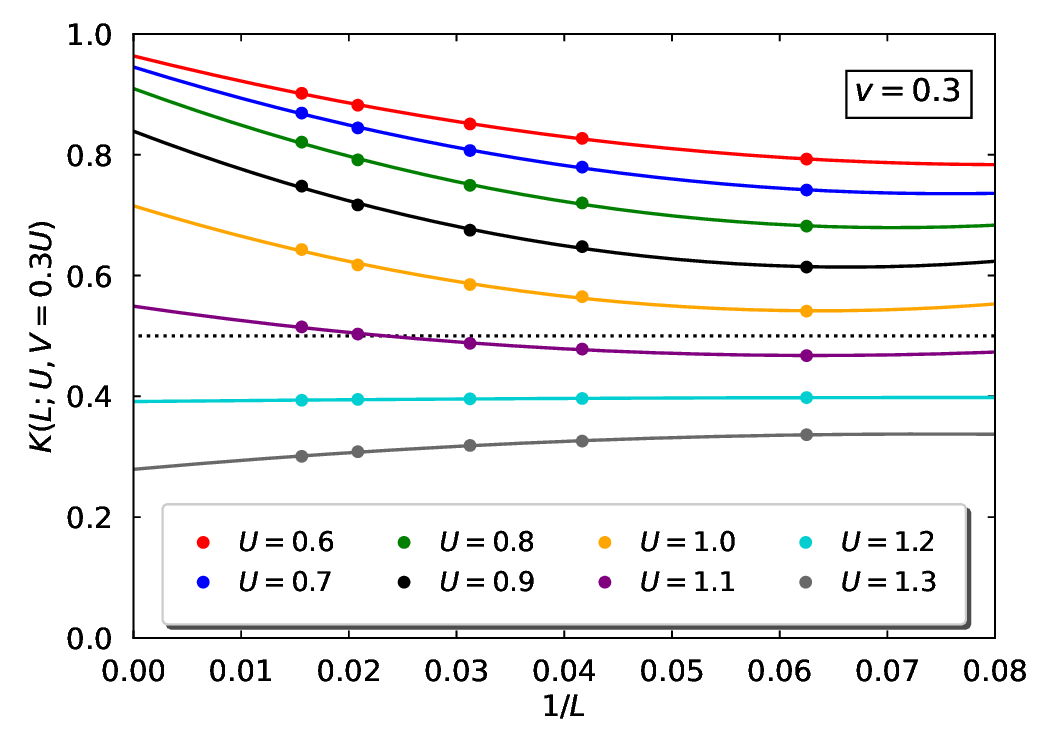}\\
   (b) \includegraphics[width=8cm]{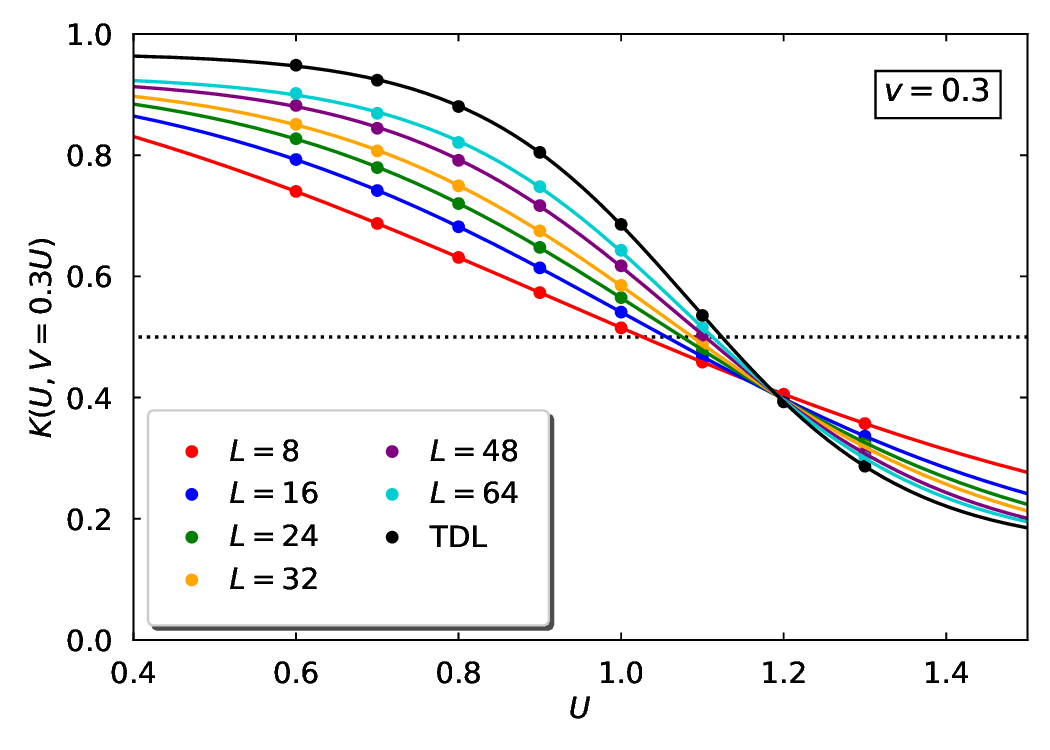}
  \end{center}
  \caption{Luttinger parameter $K(L;U,V)$ from DMRG
    for the $1/r$-Hubbard model with $1/r$-long-range interactions for $v=0.3$. (a) $K$ as a function of inverse system size for $L=16,24,32,48,64$ for various values for $U$ including a second order polynomial fit to the thermodynamic limit. (b) $K$
    as a function of $U$ for system sizes $L=8,16,24,32,48,64$. The lines correspond to the fit from Eq.~(\ref{eq:fitMTDLforallU}).
    The intersection of the extrapolation with $K_{\rm c}= 1/2$ determines $U_{\rm c,LL}(V)$. 
    The black line represents the thermodynamic limit (TDL), obtained from the fits in 
    Fig.~\ref{fig:LuttParameter03}a.
 \label{fig:LuttParameter03}}
\end{figure}

\subsection{Luttinger parameter}
\label{subsec:Luttingerparameter}

As an alternative to locate the Mott transition, we monitor the Luttinger
parameter and, for fixed ratios $v= V/U$~\cite{1overRHubbard}, we determine $U_{\rm c}(U,V)$ from the condition~\cite{Thierrybook}
\begin{equation}
  K(U_{\rm c}(V),v)= 1/2
  \; .
  \label{eq:Kmustbeonehalf}
\end{equation}

As an example, in Fig.~\ref{fig:LuttParameter03}
we show the Luttinger parameter
$K(L;U,V)$ from DMRG
for the $1/r$-Hubbard model with $1/r$-long-range interaction $V= 0.3U$
as a function of $U$ for system sizes $L= 8,16,24,32,48,64$
including a second-order polynomial extrapolation
to the thermodynamic limit.
The intersection of the extrapolation into the thermodynamic limit
with $K_{\rm c}= 1/2$ determines
$U_{\rm c}(V)$. 

To obtain a reliable estimate
for the intersection we can either
use the two data points closest to the transition
and perform a linear interpolation, in this case from $U= 1.1$
and $U= 1.2$. Alternatively, we use a four-parameter fit of the whole
data set that employs the information that the Luttinger parameter
deviates from unity by exponentially small terms for $U,V\to 0$,
\begin{equation}
    K(U,V)= a + b\tanh(c+ dU)
    \label{eq:fitMTDLforallU}
\end{equation}
to fit the extrapolated data for finite values of $U$ to a continuous
curve which is parameterized by $a,b,c,d$ that depends on~$v$.
Then, we solve Eq.~(\ref{eq:Kmustbeonehalf}) for $U_{\rm c,LL}(V)$.

Alternatively, we could have solved Eq.~(\ref{eq:Kmustbeonehalf}) for each system size, and extrapolated the resulting
system-size dependent critical interactions strengths to the thermodynamic limit.
Since the results deviate more strongly from the exact value for $V=0$, we refrain from pursuing this approach further.

As seen from Table~\ref{tab:one}, the critical value from the Luttinger parameter systematically
overestimates the correct interaction strengths by some five percent. The same overshooting was found for
the CDW transition in a one-dimensional model for
spinless fermions with nearest-neighbor interactions~\cite{PhysRevB.106.205133}.

Apparently, much larger systems are required to overcome this systematic error.
As in Ref.~\cite{1overRHubbard-NN}, we use the critical interaction strengths
$U_{\rm c,LL}(V)$ as an upper bound to the exact value $U_{\rm c}(V)$.

\begin{figure}[t]
  \begin{center}
   (a) \includegraphics[width=8cm]{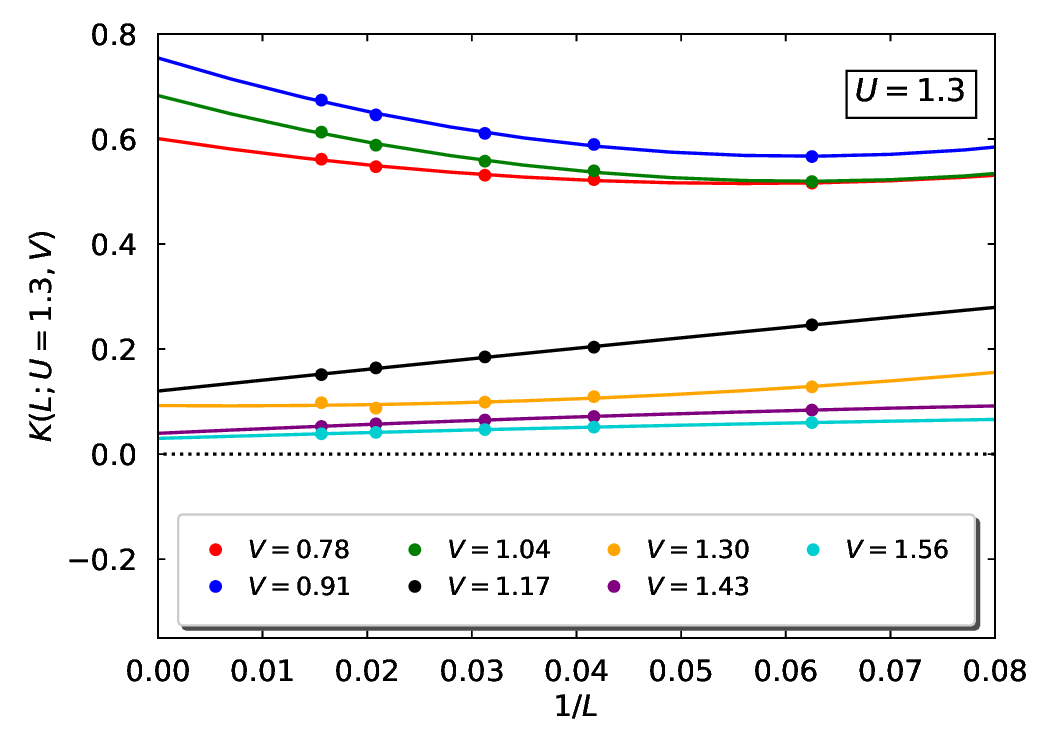}\\
   (b) \includegraphics[width=8cm]{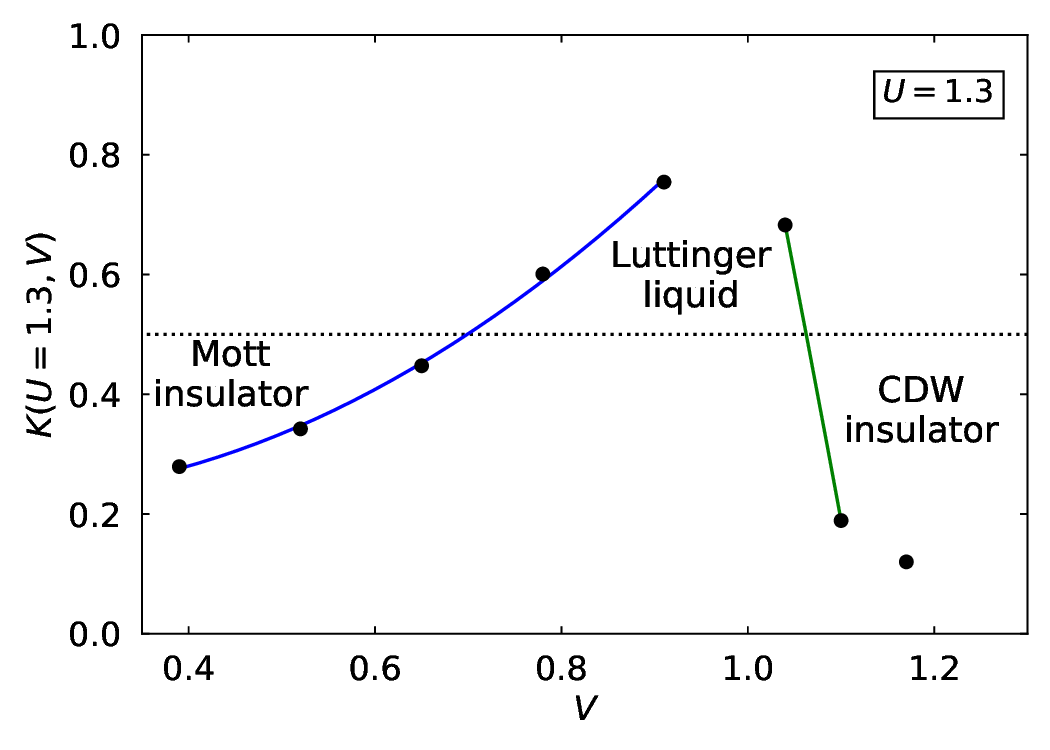}
  \end{center}
  \caption{Luttinger parameter $K(L;U,V)$ from DMRG
    for the $1/r$-Hubbard model with $1/r$-long-range interactions for fixed $U=1.3$; 
    (a) as a function of inverse system size for $L=16,24,32,48,64$ for various values for $V$ including a second order polynomial fit to the thermodynamic limit. 
    (b) as a function of $V$ in the thermodynamic limit. The lines correspond to a first-order (green) and second-order (blue) polynomial fit.}
    \label{LuttingerParamU13vertical}
\end{figure}

\subsection{Transitions at fixed Hubbard interaction}
\label{subsec:fixedU}

Since the Luttinger parameter characterizes electron correlations and explicitly identifies the Luttinger liquid, 
this parameter allows for the identification of two quantum phase transitions within a single plot, 
as we observe only the breakdown of the Luttinger liquid, see Fig.~\ref{LuttingerParamU13vertical}.
The system enters a Luttinger liquid phase once the Luttinger parameter exceeds $K > 1/2$. 

However, achieving the theoretical value of $K = 1$ within the metallic Luttinger liquid is hindered by finite-size effects. 
Note that the results from Fig.~\ref{LuttingerParamU13vertical} lead to the vertical lines for $U = 1.3$ 
in the phase diagram.

Of course, it is not possible to specify the nature of the insulating phases from the Luttinger parameter alone.  
One might argue that, due to finite-size effects, 
the phase transition into the CDW phase is more easily detected, i.e., the curve $K(U,V)$ for fixed $U$ is steeper
as a function of $V$ in the CDW phase than in the Mott insulating phase.

This ambiguity does not pose a problem because other physical quantities, e.g., the structure factor, permit to
identify the CDW phase, as we discuss next.

\subsection{Structure factor} 
\label{subsec:strucfactor}

\begin{figure}[t]
    \begin{center}
    \includegraphics[width=8cm]{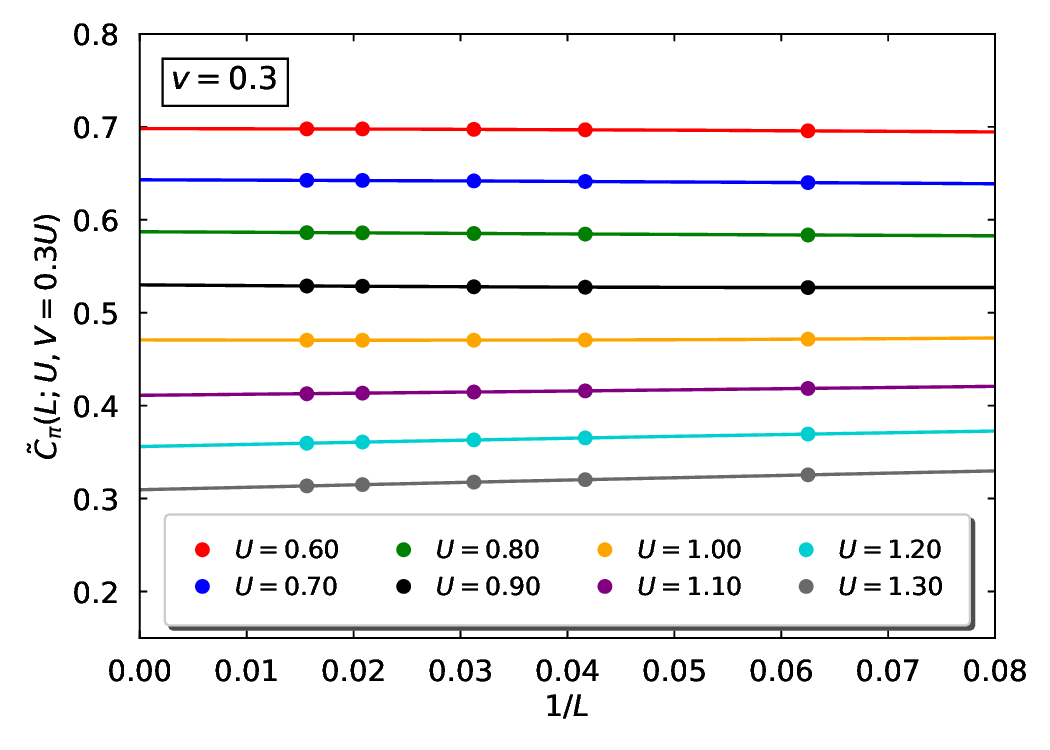}
    \end{center}
    \caption{Structure factor $\tilde{C}^{\rm NN}(\pi,L;U,V)$ as a function of $1/L$ for various values of $U$ 
    for the $1/r$-Hubbard model with $1/r$-long-range interaction $v=0.3$
    for system sizes $L=16,24,32,48,64$. Lines are a second-order polynomial extrapolation
    to the thermodynamic limit, see Eq.~(\ref{eq:fitstucturefactorinL}).\label{fig:structurefactoratpi03}}
\end{figure}

For the $1/r$-Hubbard model, the finite-size corrections to the structure factor $\tilde{C}_{\pi}(L;U,V)\equiv \tilde{C}^{\rm NN}(\pi,L;U,V)$,
\begin{equation}
    \tilde{C}_{\pi}(L;U,V)= \tilde{C}_{\pi}(U,V) +\frac{C_1(U,V)}{L} + \frac{C_2(U,V)}{L^2} \; ,
    \label{eq:fitstucturefactorinL}
\end{equation}
permit to locate the critical interaction strength. 
In Fig.~\ref{fig:structurefactoratpi03} we show the structure factor for $v=0.3$ and various values of $U$ as a function of inverse system size
for $L=16,24,32,48,64$. 

\begin{figure}[t]
  \begin{center}
  \includegraphics[width=8cm]{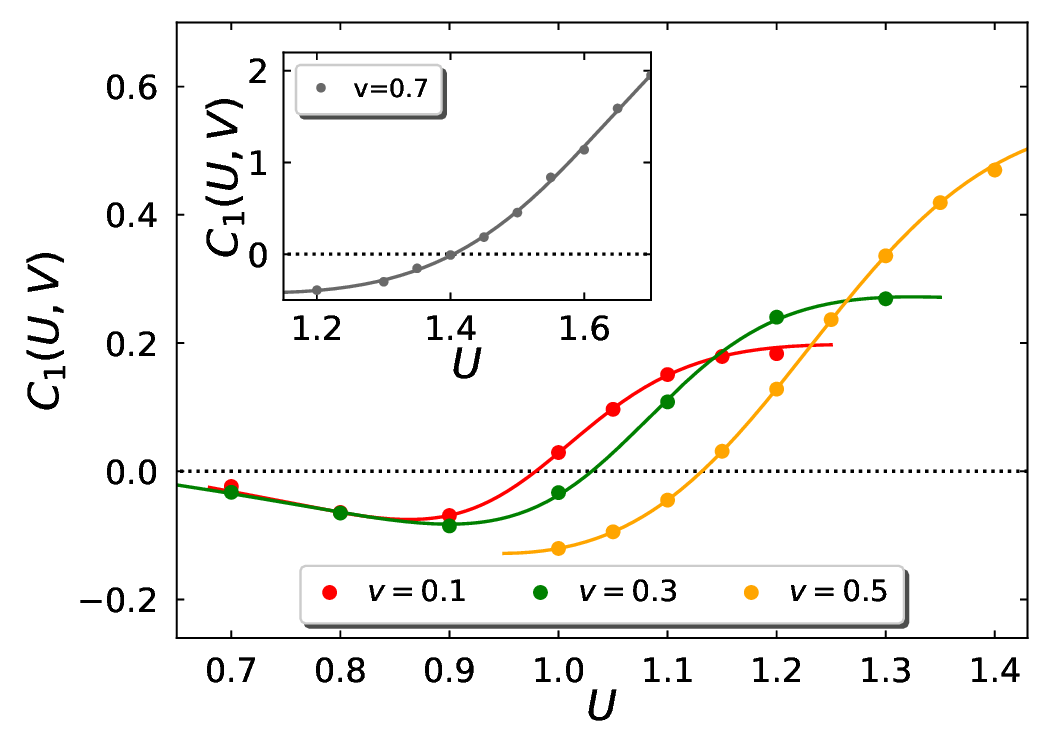}
  \end{center}
  \caption{Finite-size coefficient $C_1(U,V)$ of the structure factor as a function of $U$
  for the $1/r$-Hubbard model with $1/r$-long-range interactions for $v=0.1$, $v=0.3$, and $v=0.5$ (inset: $v=0.7$), see Eqs.~(\ref{eq:fitstucturefactorinL}) and~(\ref{eq:C1iszero}). 
  Lines are fitted Fano resonance curves~(\ref{eq:Fanoformula}).
  \label{fig:structurefactorC1}}
\end{figure}

As can be seen from the figure, 
the coefficient in $1/L$ changes its sign at the critical
interaction strength, 
\begin{equation}
C_1(U_{\rm c,sf}^{\rm Mott}(V),V)=0 \; .
\label{eq:C1iszero}
\end{equation}
To see this more clearly, in Fig.~\ref{fig:structurefactorC1} we show the coefficient $C_1(U,V)$
as a function of $U$ for $v=0.1$, $v=0.3$, $v=0.5$, and $v=0.7$ and fit the data to 
a Fano resonance,
\begin{equation}
     C_1^{\rm Fano}(U,V)=a(V)  + b(V)
  \frac{[q_{\rm F}(V)\Gamma(V) +U-U_{\rm c}(V)]^2}{[\Gamma(V)]^2+[U-U_{\rm c}(V)]^2}\; .
    \label{eq:Fanoformula}
\end{equation}
Analogously, we find the critical interaction strength $U_{\rm c}^{\rm CDW}(V)$ for the transition into the CDW phase from 
the $1/L$ corrections to the CDW order parameter~(\ref{eq:ourCDWorderparameterdef}), $N_1(U,V)$ in Eq.~(\ref{eq:CDWfit2nd}),
instead of $C_1(U,V)$.

As in our two previous studies of the $1/r$-Hubbard model~\cite{1overRHubbard,1overRHubbard-NN},
a bound state that interacts with the continuum manifests in observables, thereby contributing 
a Fano resonance to various physical quantities, with weight of the order $1/L$.
Using the Fano resonance formula and the conditions $C_1(U_{\rm c,sf}^{\rm Mott},V)=0$ and
$N_1(U_{\rm c,sf}^{\rm CDW},V)=0$, respectively, the $1/L$-corrections
of the structure factor and of the CDW order parameter
provide the estimate $U_{\rm c,sf}(V)$ for the critical interaction.
The resulting data for various~$v$ are listed in Table~\ref{tab:one}.

The critical interaction strength $U_{\rm c,sf}^{\rm Mott}(V)$ systematically underestimates the exact value for the
Mott transition by a few percent. Together with the critical interaction strength from the Luttinger parameter $U_{\rm c,LL}^{\rm Mott}(V)$
we thus can set tight limits to $U_{\rm c}^{\rm Mott}(V)$.

\subsection{Identification of the CDW}
\label{sec:CDWtransition}

Lastly, we discuss the phase transitions between the metallic Luttinger liquid and the charge-density wave insulator. 
For this case, the $\gamma$-parameters of the ground-state energy, $\gamma_0$, and of the two-particle gap, $\gamma_2$, 
are no longer suitable. Instead, we address the Luttinger parameter and the order parameter of the charge-density wave.
We qualitatively assess the CDW transition by observing the opening of the two-particle gap, as shown in Fig.~\ref{fig:Delta2}.
A quantitative analysis would require much larger system sizes and a wider range of data points to reduce finite-size effects. 
We solely use the two-particle gap to verify the consistency of our results.

\subsubsection{Luttinger parameter and CDW order parameter}

As shown in Fig.~\ref{LuttingerParamU13vertical}, the Luttinger parameter serves as an indicator for the CDW phase transition.
The missing logarithmic corrections induce an inaccuracy in the quantitative determination of the critical interaction. 
In any case, the Luttinger parameter permits to identify easily phase transitions to/from a Luttinger liquid phase.

To estimate the error in the Luttinger parameter for curves with fixed $U$, the Mott transition on the left side of Fig.~\ref{LuttingerParamU13vertical}b can be used as a reference. 
While the depicted phase transition is determined at $V_{\rm c,LL}^{\rm Mott}(U=1.3) \approx 0.699$, 
the fit of all Mott transitions from Section~\ref{FittingMottTransitions} gives a critical interaction $V_{\rm c}^{\rm Mott}(U=1.3)\approx 0.773$, 
corresponding to a deviation of approximately $10\%$.

The CDW order parameter can be evaluated as shown in Fig.~\ref{fig:CDWorderparameter}. 
Due to finite-size effects, the phase transition appears to be continuous, requiring a large number of data points beyond the transition for an accurate quantitative determination.

\subsubsection{Fano resonance curve}
\begin{figure}[ht]
  \begin{center}
  \includegraphics[width=8cm]{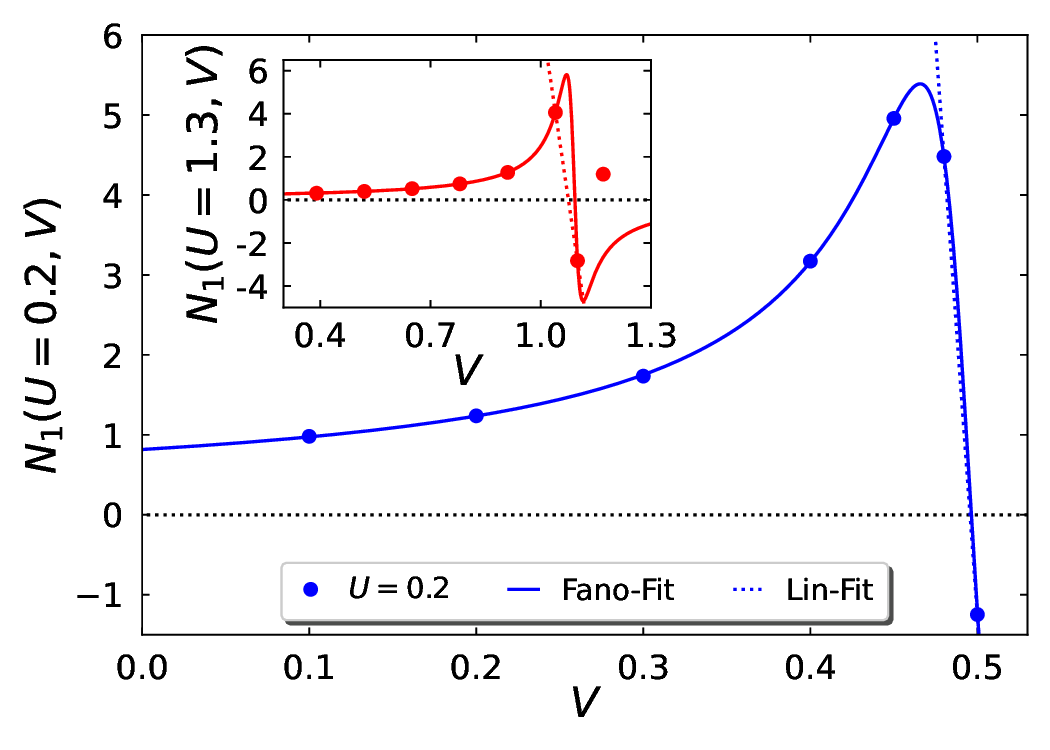}
  \end{center}
  \caption{$N_1(U,V)$ as a function of $V$ and fixed $U$
  for the extended $1/r$-Hubbard model with $1/r$-long-range interactions for $U=0.2$ (inset: $U=1.3$). 
  Lines are fitted Fano resonance curves~(\ref{eq:Fanoformula}) and dotted lines are linear fits. 
  To stabilize the fit for $U=1.3$, we did not include the data point for $V=1.17$.
  \label{fig:orderparameterN1}}
\end{figure}

As demonstrated in our previous works~\cite{1overRHubbard,1overRHubbard-NN} and in Section~\ref{subsec:strucfactor}, 
a bound state exists that couples to the continuum, thereby inducing Fano resonances in various physical quantities.
This behavior is also seen in the parameter $N_1(U,V)$ 
within the order parameter of the CDW, as defined in Eq.~(\ref{eq:CDWfit2nd}); see, e.g., 
Fig.~14b in~\cite{1overRHubbard-NN}. 
The phase transition can be clearly identified through the sign change of the parameter $N_1(U,V)$, 
see Fig.~\ref{fig:orderparameterN1} for fixed $U=0.2$ and $U=1.3$ in the inset. 

Due to the limited number of data points, particularly after the sign change, a full verification of the Fano resonance curve~(\ref{eq:Fanoformula}) is not possible. 
However, the available data points align well with the fitted curve. Moreover, once two data points are available, situated just before and after the transition, 
a simple linear fit function is sufficient to provide reasonable estimates for the transition. 

\begin{table}[t]
  \begin{ruledtabular}
    \begin{tabular}[t]{rrrrr}
      $U$ & $V_{\rm c,LL}(U)$ & $V_{{\rm c},N_1}^{\rm Fano}(U)$ & $V_{{\rm c},N_1}^{\rm Lin}(U)$ & $\overline{V}_{\rm c}^{\rm CDW}(U)$ \\[1pt]
      \hline \\[-7pt]
      $0.2$ & $0.493$ & $0.496$ & $0.496$ & $0.495$ \\
      $1.3$ & $1.062$ & $1.093$ & $1.075$ & $1.076$ \\
      \hline\\
      \hline\\[-7pt]
      $v=V/U$& $U_{\rm c,LL}(V)$ & $U_{{\rm c},N_1}^{\rm Fano}(V)$ & $U_{{\rm c},N_1}^{\rm Lin}(V)$ & $\overline{U}_{\rm c}^{\rm CDW}(V)$ \\[1pt]
      \hline\\[-7pt]
      $1.0$ & $0.762$ & $0.800 $ & $0.783$ & $0.782$ \\
    \end{tabular}
          \end{ruledtabular}
  \caption{Critical interaction strengths for the metal-to-CDW insulator transition in the $1/r$-Hubbard model with $1/r$-long-range interactions,
as obtained from the Luttinger parameter $V_{\rm c,LL}$
and the parameter $N_1(U,V)$ from Eq.~(\ref{eq:CDWfit2nd}) using a Fano resonance fit function $V_{\rm c,N_1}^{\rm Fano}(U)$, see Eq.~(\ref{eq:Fanoformula}), 
or a linear fit function $V_{\rm c,N_1}^{\rm Lin}(U)$; in DMRG, $16\leq L\leq 64$ lattice sites were addressed.
    \label{tab:two}}
\end{table}

In Table~\ref{tab:two} we collect the results for the CDW transitions at fixed $U=0.2,1.3$ and for $v=V/U=1$. 
It is seen that the estimate from the extrapolation of the order parameter in Fig.~\ref{fig:CDWorderparameter}, $V_{\rm c,op}(U=1.3)=1.07$,
is in good agreement with the result from the Luttinger and Fano extrapolations, $\overline{V}_{\rm c}(U=1.3)=1.076$, as collected in Table~\ref{tab:two}.

All approaches lead to consistent results. 
The extrapolations from the Luttinger parameter underestimate the critical interactions whereas the results
from the Fano resonance give an upper bound to $V_{\rm c}^{\rm CDW}(U)$. 
As for the Mott transition, this permits to set boundaries to the exact critical value of the metal-to-insulator transition.

\section{Renormalization of the phase diagram}
\label{sec:RenormalizationPhaseDiagram}

In this section, we discuss the quantum phase diagram for the $1/r$-Hubbard model
with $1/r$-long-range interactions and its mapping from the quantum phase diagram of the $1/r$-Hubbard model with nearest-neighbor interactions.

\subsection{Quantum phase diagrams}

\begin{figure}[t]
    \begin{center}
    (a) \includegraphics[width=8cm]{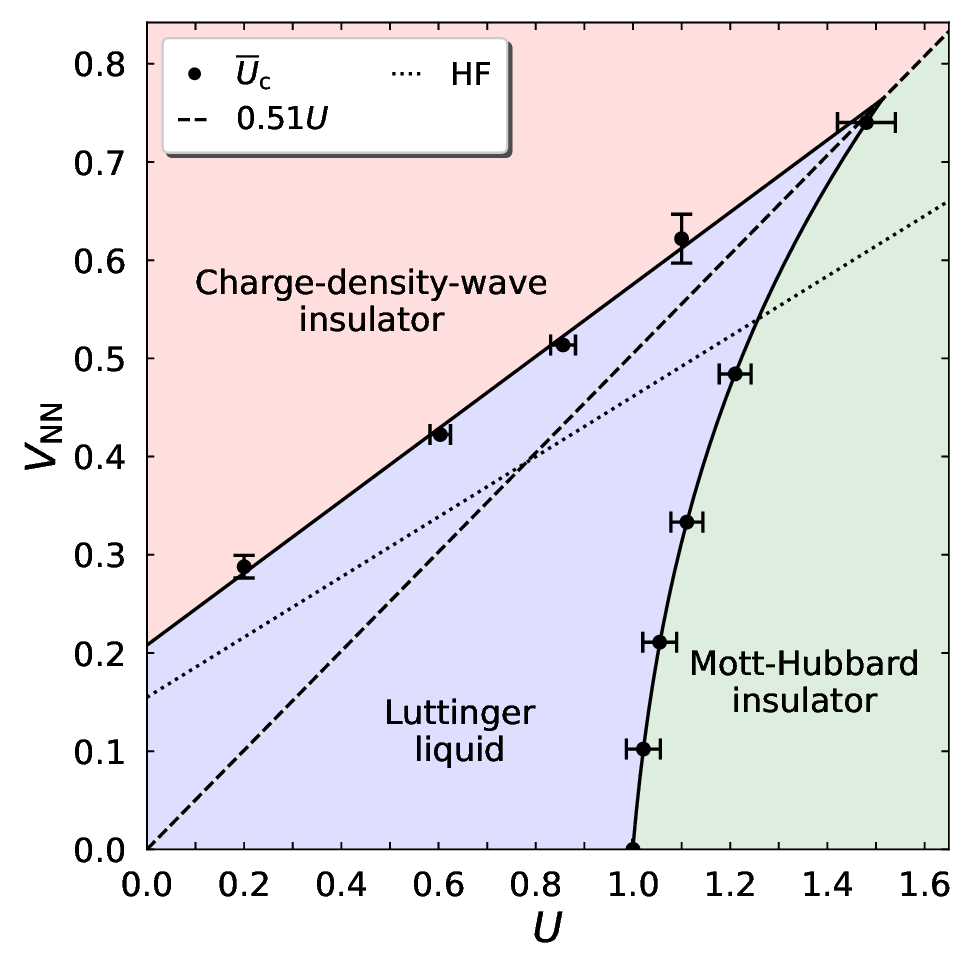}\\
    (b) \includegraphics[width=8cm]{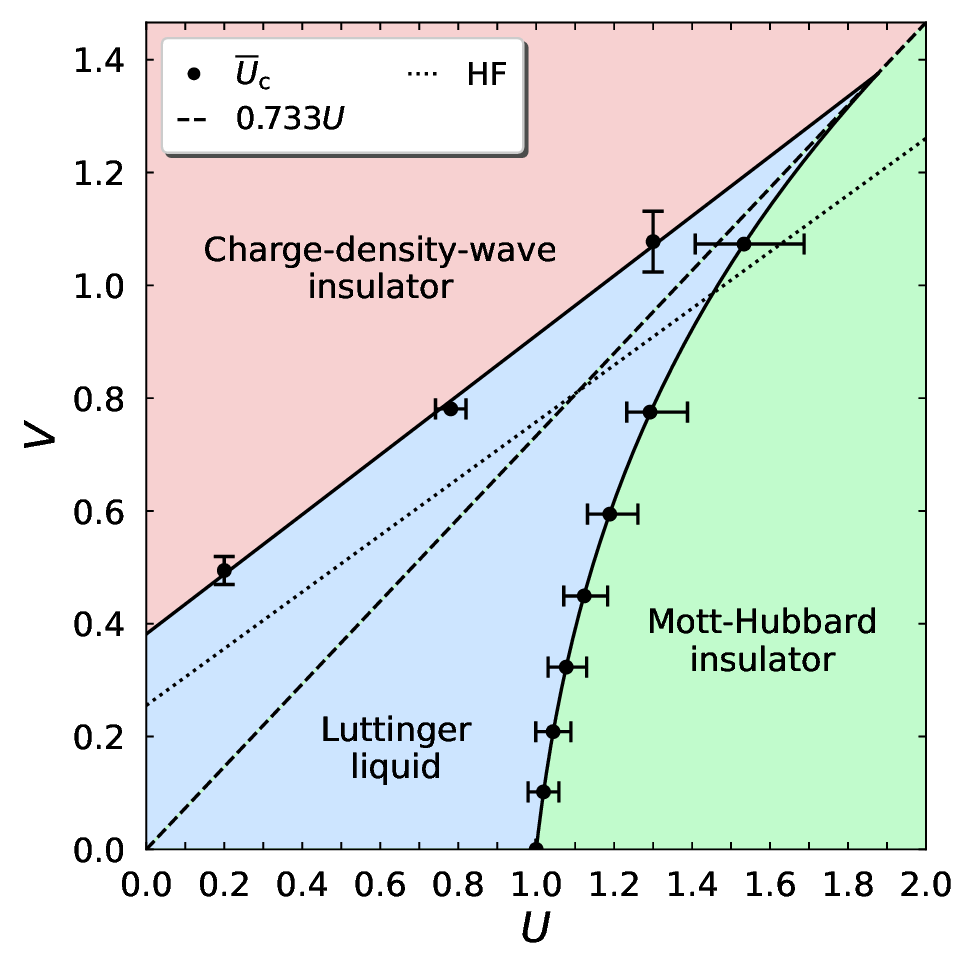}
    \end{center}
    \caption{(a) Phase diagram of the one-dimensional $1/r$-Hubbard model with nearest-neighbor interactions.
    (b) Phase diagram of the one-dimensional $1/r$-Hubbard model with $1/r$-long-range interactions. 
      Continuous lines: polynomial fits of the phase-separation lines,
      see Sects.~\ref{CDW-Transition-Fitting} and~\ref{FittingMottTransitions}; dotted line: Hartree-Fock (HF) result for
    the transition between metal and charge-density-wave insulator.\label{QPDiagramAndComparisonagain}}
\end{figure}

For convenience, we reproduce the quantum phase diagrams 
in Fig.~\ref{QPDiagramAndComparison} in Fig.~\ref{QPDiagramAndComparisonagain},  where we
  add the Hartree-Fock results for the transition between metal and CDW insulator as dotted lines
  and the strong-coupling limit for the separation lines between Mott-Hubbard and CDW insulators
  as dashed lines, respectively.
As in our previous work~\cite{1overRHubbard-NN},
we refrain from the analysis of the transition between Mott and CDW insulators.

As shown numerically for the standard extended Hubbard
model~\cite{Jeckelmann2002,Jeckelmannreply2005,Satoshis2007,MundNoackLegeza2009},  
the transition is discontinuous for large interactions, $U,V \gg W$, as also inferred from the atomic limit,
see appendix, where $V_{\rm NN,c}^{\rm al}=U/2$ for the extended Hubbard model.
  Since the kinetic energy slightly stabilizes the Mott-Hubbard insulator,
the phase separation line in Fig.~\ref{QPDiagramAndComparisonagain}a for large interactions is
approximately given by $V_{\rm c,NN}(U\gg W)\approx 0.51 U$ (dashed line).
The same line of reasoning gives
$V_{\rm c,LR}^{\rm al}=U/[2\ln(2)]$ in the atomic limit in the presence of $1/r$-long-range interactions, and
$V_{\rm c,LR}(U\gg W)\approx 0.733 U \approx 0.51 U/\ln(2)$ (dashed line).

For the standard extended Hubbard model and for moderate interactions, $U, V \approx W$, and fixed $U$,
a transition occurs from the Mott insulator to a bond-order-wave insulator 
as a function of $V$ before the CDW insulator dominates at large nearest-neighbor interactions.
Beyond the triple point where Luttinger liquid, Mott insulator, and CDW insulator meet in our phase diagrams, we also expect
to find a region with a bond-order wave in extended $1/r$-Hubbard models.
Therefore, the dashed line separating Mott-Hubbard insulator and charge-density-wave insulator 
should be taken as a guide to the eye only.
For this reason, we only show the phase diagram for $U\leq 1.9$, 
and the analysis of the triple point and the bond-order wave phase are beyond the scope of our present work.

\subsubsection{Parameterization of the CDW critical line}
\label{CDW-Transition-Fitting}

For $U\leq 1.9$, the transition from the Luttinger liquid to the CDW insulator is qualitatively correctly described by
Hartree-Fock theory. In very good approximation, the critical interactions $V_{\rm c}^{\rm CDW}(U)$ are linear functions of~$U$. 
As shown in the appendix,  the dotted Hartree-Fock lines
in Fig.~\ref{QPDiagramAndComparisonagain}
are given by
\begin{equation}
V_{\rm c,HF}^{\rm CDW}(U)=a^{\rm HF}+ b^{\rm HF} U +\mathcal{O}(U^2)
\label{eq:HFCDWlinear}
\end{equation}
with
\begin{eqnarray}
  a_{\rm NN}^{\rm HF} \approx 0.1548 &\quad ,\quad & b_{\rm NN}^{\rm HF}\approx 0.3064 \; , \nonumber \\
 a_{\rm LR}^{\rm HF} \approx 0.2553 &\quad ,\quad & b_{\rm LR}^{\rm HF}\approx 0.5027 \; ,
\end{eqnarray}
for the $1/r$-Hubbard models with nearest-neighbor and $1/r$-long-range interactions, respectively.

Hartree-Fock theory
overestimates the stability of the CDW phase and thus underestimates the critical interaction strength,
$V_{\rm c,HF}^{\rm CDW}(U)<V_{\rm c}^{\rm CDW}(U)$. Therefore, in 
Fig.~\ref{QPDiagramAndComparisonagain}, the dotted Hartree-Fock lines
systematically lie below the CDW transition lines derived from our DMRG data, see Table~\ref{tab:two} for the $1/r$-Hubbard model with $1/r$-long-range interactions
and Table~I in our previous work~\cite{1overRHubbard-NN} for the $1/r$-Hubbard model with nearest-neighbor interactions.

For a quantitative comparison, we use the mean values from Table~\ref{tab:two} and apply a linear fit to the three data points to avoid over-fitting and
for a direct comparison with the Hartree-Fock results.
We thus write
\begin{equation}
V_{\rm c,CDW}(U) = a^{\rm CDW} + b^{\rm CDW} U + \mathcal{O}(U^2)\;,
\label{eq:VcCDWlinear}
\end{equation}
and find
\begin{eqnarray}
    V_{\rm c,NN}^{\rm CDW}(U) &\approx & 0.2079 + 0.3675 U \nonumber \; , \\
    V_{\rm c,LR}^{\rm CDW}(U) &\approx & 0.3818 + 0.5293 U
    \label{eq:VcCDWlinearDATA}
\end{eqnarray}
for the models with nearest-neighbor and $1/r$-long-range interactions, respectively.
Equation~(\ref{eq:VcCDWlinearDATA}) defines the continuous phase separation lines between
Luttinger liquid and CDW insulator in Fig.~\ref{QPDiagramAndComparisonagain}.

\subsubsection{Parameterization of the Mott critical line}
\label{FittingMottTransitions}

For purely local interactions, 
the Mott-Hubbard transition is known to occur at $U_{\rm c}^{\rm Mott}(V=0)=1$~\cite{GebhardRuckenstein,Gebhardbook}, an analytical result that is well reproduced by DMRG~\cite{1overRHubbard}, 
see also Table~\ref{tab:one}. 
The inclusion of repulsive interactions beyond the purely local Hubbard interaction stabilizes the metallic phase, i.e., 
non-local interactions {\em increase\/} the critical interaction strength, $U_{\rm c}^{\rm Mott}(V>0)\geq 1$. 
Apparently, the additional repulsive interaction softens the two-particle scattering potential
in position space and thus makes it less effective. 

In the phase diagrams~\ref{QPDiagramAndComparisonagain} we show $V_{\rm c}^{\rm Mott}(U)$ which displays a fairly steep
dependence on~$U$ close to $U=1$. In turn, its inverse function, $U_{\rm c}^{\rm Mott}(V)$, is a fairly flat function of $V$
around $V=0$ so that a Taylor series of $U_{\rm c}^{\rm Mott}(V)$ in $V$ around the bare Hubbard limit converges much better
than  a Taylor series of $V_{\rm c}^{\rm Mott}(U)$ in $U$ around $U=1$.

\begin{figure}[t]
  \begin{center}
  \includegraphics[width=8cm]{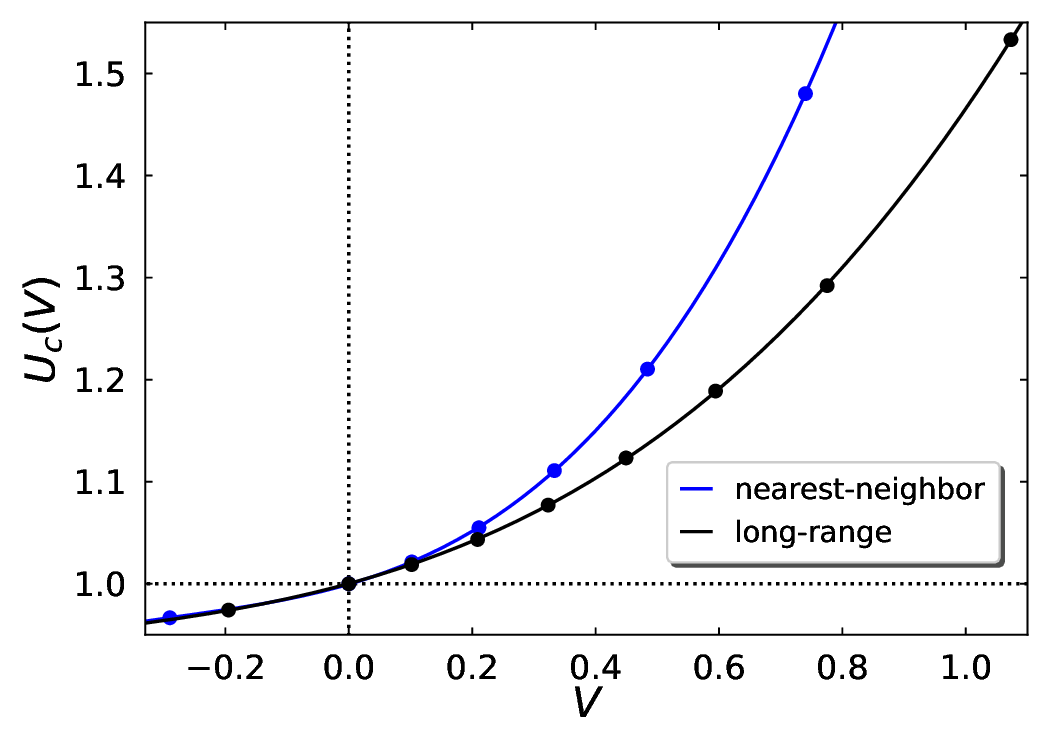}
  \end{center}
  \caption{Cubic fits, eq.~(\ref{Eq:MottFitFunctions}), for the critical interactions $U_{\rm c}^{\rm Mott}(V)$ for the Mott transitions in 
  the $1/r$-Hubbard models with nearest-neighbor interactions (blue solid line) and $1/r$-long-range interactions (black solid line).
  \label{fig:MottTransFit}}
\end{figure}

In Fig.~\ref{fig:MottTransFit} we show the data for $U_{\rm c}^{\rm Mott}(V)$ for the Mott transition
for the $1/r$-Hubbard models with nearest-neighbor and $1/r$-long-range interactions,
respectively. 

For the data points we use the mean values of Table~\ref{tab:one} in this work and in Ref.~\cite{1overRHubbard-NN}.
For the fit we employ a third-order polynomial,
\begin{eqnarray}
\label{Eq:MottFitFunctions}
U_{\rm c}^{\rm Mott}(V) &=& b_0 + b_1 V + b_2 V^2 +b_3 V^3\;,
\end{eqnarray}
where $b_i$ naturally differ for nearest-neighbor and $1/r$-long-range interactions. 
The corresponding critical lines, $V_{\rm c}^{\rm Mott}(U)$, are shown in the phase diagrams~\ref{QPDiagramAndComparisonagain}.
In the range $1\leq U\leq 1.9$ and due to the fact that $b_3$ is small enough,
we can approximate
\begin{equation}
V_{\rm c}^{\rm Mott}(U) \approx \frac{-b_1 + \sqrt{b_1^2 - 4 (b_0-1) b_2 + 4 b_2 (U-1)}}{2 b_2}
\end{equation}
with very good accuracy. Since $b_0\approx 1$ and $b_1\ll 1$, at first glance the critical curve $V_{\rm c}^{\rm Mott}(U)$ seems to display a square-root
singularity near $U=1$. However, this is not the case as the critical line does not display any singularities around $(U=1,V=0)$.

Nevertheless, the fits for $U_{\rm c}^{\rm Mott}(V)$ performed in Fig.~\ref{fig:MottTransFit} deserve special attention.
It is important to include at least one data point for attractive interactions, $v=V/U=-0.3$ for nearest-neighbor interactions
and $v = -0.2$ for $1/r$-long-range interactions.  
Otherwise, the fitted curve would exhibit a nonsensical minimum close to $V=0$ that would prohibit the inversion of the function. From a physical point of view,
an attractive nearest-neighbor interaction de-stabilizes the Luttinger liquid because the Hubbard scattering potential becomes   
more prominent in position space, jumping from $-|V|$ for electrons on neighboring sites to $+U$ for on-site pairs.

The (stabilized) fit results in the curves 
\begin{eqnarray}
U_{\rm c, LR}^{\rm Mott}(V) &\!=\!& 1.0002 + 0.1659 V + 0.1851 V^2 + 0.1145 V^3\,,\nonumber\\
U_{\rm c,NN}^{\rm Mott}(V) &\!=\!& 0.9996 + 0.1760 V + 0.3362 V^2 + 0.4104 V^3\,.\nonumber \\
\label{fittedFunctions}
\end{eqnarray}
Recall that $U_{\rm c}^{\rm Mott}(V=0)=1$ so that the observation $|b_0-1|\ll 1$ implies a very good quality of the fit.
The inverse of $U_{\rm c}^{\rm Mott}(V)$ is shown in the phase diagram~\ref{QPDiagramAndComparisonagain}.

\subsection{Renormalization}

Fig.~\ref{QPDiagramAndComparisonagain} shows that the two phase diagrams display the same qualitative features.
Moreover, as we shall argue now, it is possible to map the two phase diagrams onto each other using simple
renormalization factors for the phase transition lines for the CDW and Mott transitions, respectively.

\subsubsection{CDW transition}
\label{renormCDW}

\begin{figure}[t]
  \begin{center}
  \includegraphics[width=8cm]{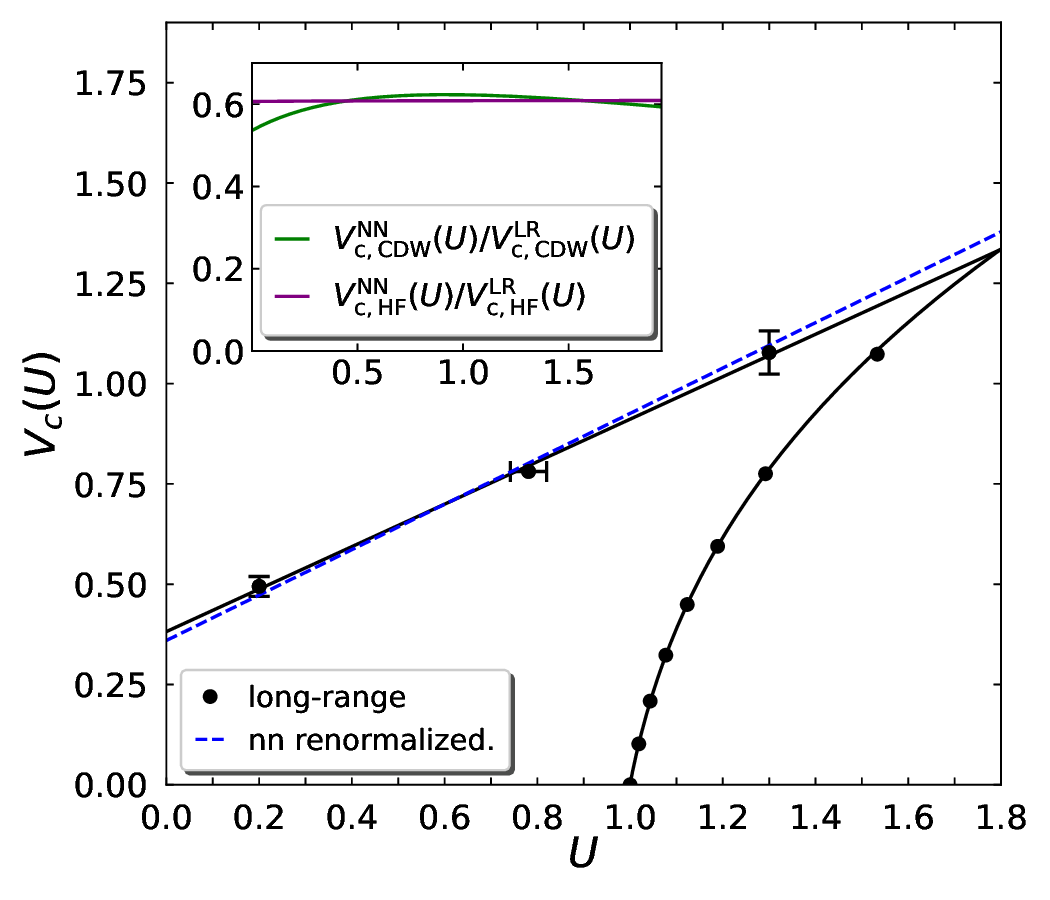}
  \end{center}
  \caption{Renormalization of the CDW phase transition line. 
  The phase transitions for the $1/r$-Hubbard model with $1/r$-long-range interactions are shown as black dots and lines, together with the renormalized CDW phase transition line
  from the $1/r$-Hubbard model with nearest-neighbor interactions (blue dashed line); here, we applied the renormalization factor $R_{\rm CDW} = 0.61$. 
  The inset displays the exact ratio $V_{\rm c,NN}^{\rm CDW}(U) / V_{\rm c,LR}^{\rm CDW}(U)$ (green solid line) 
  and Hartree-Fock ratio $V_{\rm c,HF}^{\rm NN}(U)/V_{\rm c,HF}^{\rm LR}(U)$ (purple solid line) that are close to $R_{\rm CDW}\approx 0.61$ for all $0\leq U\leq 1.9$.
  \label{CDWRenormalized}}
\end{figure}

As can be seen in Fig.~\ref{QPDiagramAndComparisonagain} and inferred quantitatively from Eqs.~(\ref{eq:HFCDWlinear})--(\ref{eq:VcCDWlinearDATA}),
the phase transition curves between the Luttinger liquid and CDW insulator are very well approximated by straight lines.
The corresponding renormalization is given by
\begin{eqnarray}
    R_{\rm exact}^{\rm CDW}(U)&=& \frac{V_{\rm c,NN}^{\rm CDW}(U)}{V_{\rm c,LR}^{\rm CDW}(U)}  \\ 
   &\approx& 
    \frac{0.2079 + 0.3675 U}{0.3818 + 0.5293 U}\approx 0.5445+0.2077 U \;.\nonumber
\end{eqnarray}
The corresponding result from Hartree-Fock theory is given by
\begin{eqnarray}
    R_{\rm HF}^{\rm CDW}(U)&=& \frac{V_{\rm c,NN}^{\rm HF}(U)}{V_{\rm c,LR}^{\rm HF}(U)} \\
    &\approx &
    \frac{0.1548+ 0.3064 U}{0.2553 + 0.5027 U}\approx 0.6063 +0.0062 U \;.\nonumber 
\end{eqnarray}
Both curves are shown in the inset of Fig.~\ref{CDWRenormalized}.
The comparison shows that the factors $R^{\rm CDW}_{\rm exact}(U)$
and  $R_{\rm HF}^{\rm CDW}(U)$ fairly agree with each other over the region of interest, $0\leq U \leq 1.9$.
Recall that the critical interactions from DMRG carry an error estimate.

Since $R_{\rm HF}^{\rm CDW}(U)$ is almost independent of~$U$, 
we choose 
\begin{equation}
R_{\rm CDW}=  R_{\rm HF}^{\rm CDW}(U=0)\approx 0.61
\end{equation}
as our overall renormalization factor for the CDW transition.
In Fig.~\ref{CDWRenormalized} we plot the phase separation lines for the $1/r$-Hubbard model with $1/r$-long-range interactions
together with the renormalized phase transition line for the CDW transition
\begin{eqnarray}
V_{\rm c, ren}^{\rm CDW}(U) &=& \frac{V_{\rm c, NN}^{\rm CDW}(U)}{R_{\rm CDW}} \nonumber \\
&\approx & \frac{0.2079 + 0.3675 U}{R_{\rm CDW}} \approx 0.34 +0.60 U \;.
\label{eq:VcrenCDW}
\end{eqnarray}
The comparison with Eq.~(\ref{eq:VcrenCDW}), 
\begin{eqnarray}
    V_{\rm c,LR}^{\rm CDW}(U) &\approx&  0.3818 + 0.5293 U \nonumber \\
&\approx& 0.34 +0.60 U\approx V_{\rm c, ren}^{\rm CDW}(U)\; ,
\end{eqnarray}
visualized in Fig.~\ref{CDWRenormalized}, shows
that the original and renormalized points for the CDW transition agree within error bars.

\begin{figure}[t]
  \begin{center}
  \includegraphics[width=6.5cm]{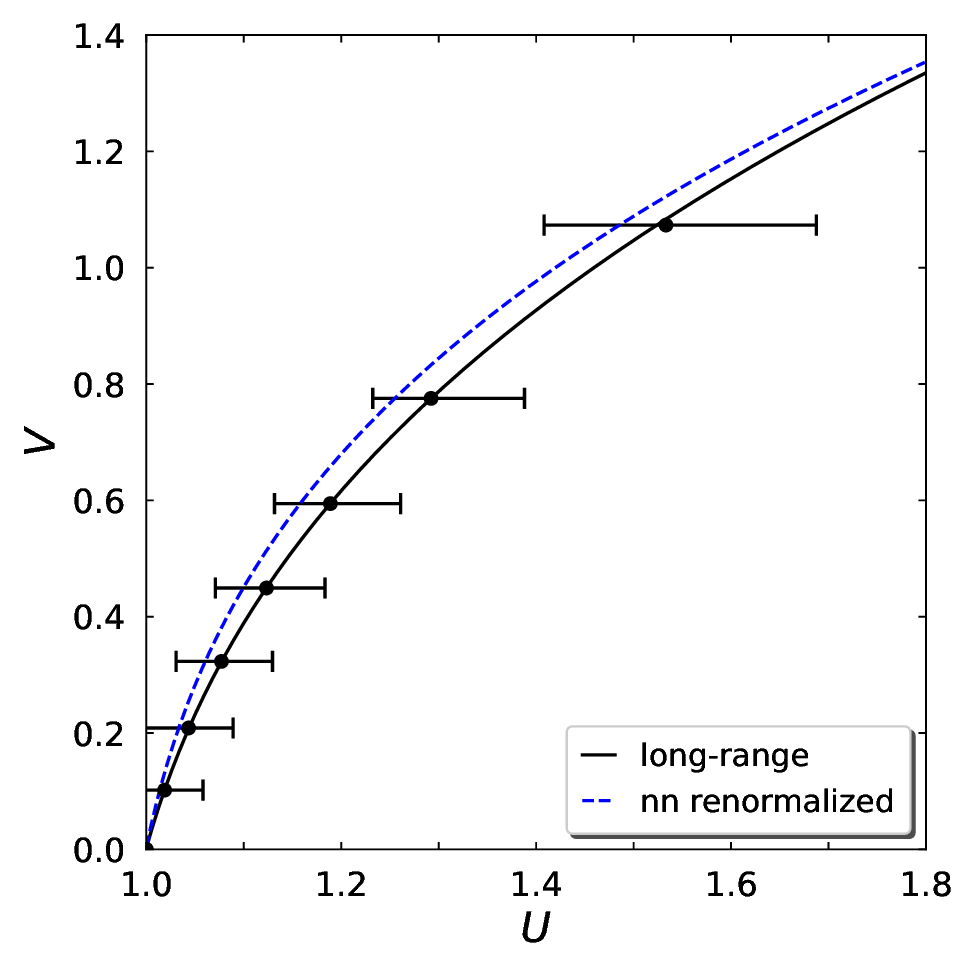}
  \end{center}
  \caption{Renormalization of the Mott-Hubbard phase transition line. 
  The Mott-Hubbard phase transition line for the $1/r$-Hubbard model with $1/r$-long-range interactions (black solid line) reasonably agrees
  with the Mott-Hubbard phase transition for the extended $1/r$-Hubbard model (blue dashed line) when renormalized using the atomic-limit factor $R^{\rm al}_{\rm Mott}=\ln 2\approx 0.693$.
  \label{PhaseDiagrammMottRenormierung}}
\end{figure}

\subsubsection{Mott-Hubbard transition}
\label{sec:renormMott}

For the renormalization of the Mott-Hubbard phase transition line, we employ the renormalization factor that occurs in the atomic limit (al), 
where the kinetic energy is neglected.
As shown in the appendix, we have
\begin{equation}
    R^{\rm al}_{\rm Mott} = \frac{V_{\rm c,NN}^{\rm al}(U)}{V_{\rm c,LR}^{\rm al}(U)} = \ln 2 \approx 0.693\;.
    \label{eq:RMOTT}
\end{equation}
Evidently, this choice selects the strong-coupling regime as starting point. 
Therefore, we do not expect perfect results for intermediate couplings, $U \approx W$, where the Mott-Hubbard transition actually occurs.

In Fig.~\ref{PhaseDiagrammMottRenormierung} we show the Mott-Hubbard phase transition line for the $1/r$-Hubbard model with $1/r$-long-range interactions in
comparison with the Mott-Hubbard phase transition for the extended $1/r$-Hubbard model, renormalized using the atomic-limit factor $R^{\rm al}_{\rm Mott}\approx 0.693$. 

The two curves agree within error bars, with a systematic deviation of the critical interactions of some $5\%$ for moderate interactions. 
This is not too surprising because we use only a single number as our renormalization factor. 
As in the case of the CDW transition, the renormalization is fairly small and very mildly dependent on the Hubbard parameter.

Close to $U=1$, a linear extrapolation for $V_{\rm c}^{\rm Mott}(U)$ according to Eq.~(\ref{Eq:MottFitFunctions}) gives
\begin{eqnarray}
  V_{\rm c,NN}^{\rm Mott}(U) = 0.0025 + 5.6269 (U-1) + \mathcal{O}((U-1)^2)\;, \nonumber \\
    V_{\rm c,LR}^{\rm Mott}(U) = 0.0022 + 6.0106 (U-1) + \mathcal{O}((U-1)^2)\;.\nonumber \\
\end{eqnarray}
In linear order, the renormalization factor should be $R^{\rm lin}_{\rm Mott} \approx 0.94$ for moderate couplings, instead of $R^{\rm al}_{\rm Mott}\approx 0.69$
for strong couplings. The agreement of these factors is still very reasonable, given the fact that moderate couplings $U\approx W$ 
are far from the strong-coupling limit $U \gg W$.

\section{Conclusions}
\label{sec:conclusions}

In our conclusions, we first summarize our main results.
Next, we discuss the relevance of our findings in a broader context.

\subsection{Summary}
\label{subsec:summary}

In this work, we used the SU(2) spin adapted density-matrix renormalization group (DMRG) method for up to $L=64$ lattice sites and antiperiodic boundary conditions
with bond dimension up to $D_{\rm SU(2)}^{\rm max}=6000$
to derive
the quantum phase diagram of the one-dimensional $1/r$-Hubbard model with $1/r$-long-range interactions at half band-filling.
As in our previous publication~\cite{1overRHubbard-NN} on the $1/r$-Hubbard model with nearest-neighbor interactions, we employed the
ground-state energy density, the two-particle gap, the structure factor, and the Luttinger parameter to
identify the quantum phase transitions between the metallic Luttinger liquid for weak interactions, the Mott-Hubbard insulator
for dominant on-site (Hubbard) interactions, and the charge-density wave (CDW) insulator for strong inter-site interactions.

The comparison with the quantum phase diagram for the $1/r$-Hubbard model
with nearest-neighbor interactions leads to two central results:
\begin{enumerate}
    \item Long-range interactions do not change the quantum phase diagram {\em qualitatively}.
    \item The quantum phase diagrams can {\em quantitatively\/} be mapped onto each other 
    by using constant renormalization factors for the quantum phase transition lines. 
\end{enumerate}
A model with only nearest-neighbor interactions mimics the long-range model
    when we put $V_{\rm NN}\approx R_{\rm qp} V_{\rm LR}$ with $R_{\rm qp}= 0.61 \ldots 0.69$.
At small interactions, the renormalization factor for the CDW transition, $R_{\rm CDW}\approx 0.61$, 
results from Hartree-Fock theory. For large interactions,  the atomic limit provides
the renormalization factor $R_{\rm Mott}\approx 0.69$ for the CDW and Mott-Hubbard transitions.

\subsection{Discussion}

In this last subsection we discuss the relevance of our findings for interacting many-particle systems. We come to our main conclusion 
that the inclusion of long-range interactions is not crucial for a proper description 
of interacting many-electron systems.  
Therefore, the screening problem is not severe, i.e., 
long-range interactions can be replaced 
by effective short-range interactions throughout the ground-state
phase diagram with metallic and (Mott) insulating phases. 

\subsubsection{Application to conjugated polymers}

Polymers such as poly-acetylene and poly-di\-acety\-lene~\cite{Sariciftci,SchottinLanzani}
are insulators where the $1/r$-long-range interaction is only statically screened with
a dielectric constant $\varepsilon_d\approx 2.3$. 
For large electron separations, the long-range interaction must equal the Coulomb interaction,
\begin{equation}
    \frac{V_{\rm LR}}{d/a} = \frac{e^2}{d \varepsilon_d} \; ,
\end{equation}
where $d$ is the distance between electrons, measured in physical units such as \AA ngstr\o m, 
$a\approx 1.4\, \text{\AA}$ is the average bond-length in trans poly-acetylene, and $e$ is the electronic charge.
With the Bohr radius $a_{\rm B}\approx 0.529\, \text{\AA}$ and $E_{\rm Ryd}=e^2/(2a_{\rm B})\approx 13.6\, \text{eV}$,
we have as an estimate for the effective interaction in the extended Hubbard model ($R\approx 0.65$)
\begin{eqnarray}
    V_{\rm NN}&=& R V_{\rm LR} = \frac{R}{\varepsilon_{d}} \frac{2a_{\rm B}}{a} \frac{e^2}{2a_{\rm B}} \nonumber \\
    &\approx & \frac{0.65}{2.3} \frac{1.06\text{\AA}}{1.4\text{\AA}} 13.6 \, \text{eV} = 2.9 \, \text{eV}\; .
\end{eqnarray}
When the actual zig-zag geometry of poly-acetylene is taken into account~\cite{PhysRevLett.51.1191,DUIJVESTIJN1985461},
the unit cell comprising two carbon atoms has a diameter of $2a= 2.45\, \text{\AA}$ which slightly increases
the effective nearest-neighbor interaction, $V_{\rm NN}\approx 3.3\, \text{eV}$. 

A value $V_{\rm NN}=3\, \text{eV}$ was successfully applied for the description of the optical properties of
poly-acetylene using the extended Hubbard model~\cite{Baeriswyl1992}.
It is thus seen that the factor $R\approx 2/3$ appropriately describes the renormalization of the long-range Coulomb interaction. Note, however, that nowadays long-range
Coulomb interactions in fairly long polymers can appropriately be treated 
using the DMRG method~\cite{Barford,PhysRevB.87.245116}.

\subsubsection{Advantages for numerical investigations}

For the same system size, 
the numerical effort to run DMRG calculations for $1/r$-long-range interactions 
scales with system size~$L$, i.e., it is typically
about a factor 100 or so higher than for nearest-neighbor interactions.
For other numerical approaches such as quantum Monte Carlo (QMC),
long-range interactions also pose a major obstacle.

Therefore, it is important to know that a numerical analysis of 
many-particle systems with short-range interactions is relevant
for real systems with long-range Coulomb interactions. 
Those studies and their results should not be dismissed as irrelevant just because
models with short-range interactions are employed.

Moreover, extensive parameter scans can be carried out more easily for models
with short-range interactions, e.g., when electron-phonon couplings are included
or in the presence of doping. When the region of interest is appropriately narrowed down,
the more costly inclusion of long-range Coulomb terms
is facilitated using the renormalization factor $R_{\rm qp}$.

\subsubsection{Correlation functions}
Models with algebraically decaying interactions may display algebraically
decaying correlation functions even in the presence of a gap, see, e.g.,
Ref.~\cite{correlationfunctionscommmathphys}. Therefore, it would be interesting
to study in detail the behavior of real-space correlation functions
close to the quantum phase transitions. 

Unfortunately, as seen from Figs.~\ref{fig:momdis} and \ref{fig:CNNq},
our systems sizes $L \leq 64$ 
are too small to address the asymptotic region $L \gg r \gg 1$
where short-range correlations have died out $(r \gg 1$) 
and finite-size effects are still negligible, $r \ll L$.
The Fourier transform of the single-particle density matrix,
the momentum distribution $n(k)$, looks discontinuous at first glance despite
the fact that it is continuous both in the Luttinger liquid
and in the Mott and CDW insulating phases. 
Likewise, the density-density correlation function $\tilde{C}^{\rm NN}(q)$
looks unspectacular and it is difficult, yet feasible, to extract the 
Luttinger parameter from it using eq.~(\ref{eq:KfromCNN}).
Much larger systems, of the order of ten thousand sites, are required for a 
detailed analysis of the asymptotic behavior of real-space correlation functions.
Fortunately, for our analysis of the quantum phase diagram, 
we do not have to rely on correlation functions.

\subsubsection{Remaining issues}

The $1/r$-Hubbard model is but a single example where the conceptual equivalence of models 
with short-range and long-range interactions could be tested. Therefore, the question arises:
how generic are our findings?

First of all, it should be stated that effective models are frequently 
and successfully employed in describing low-energy properties of physical systems. 
For example, $s$-wave scattering is sufficient for the description of the
interaction of slow neutrons with nuclei. 
A sufficiently strong short-range interaction binds optically excited electrons and holes into excitons
in condensed matter theory in the same way as the actual long-range Coulomb interaction.
In this sense, it does not come as an utter surprise that the long-range Coulomb interaction can
be replaced by an effective short-range interaction in the analysis of physical quantities.
The important insight gained from the example of the $1/r$-Hubbard model 
lies in the fact that the concept of short-range interactions apparently also applies to 
the ground-state properties of interacting {\em many}-particle systems.

Second, we point out that correlations are strongest in one dimension
because the particles cannot move around each other.
In other words, if screening works in one dimension, it should work even more so in higher dimensions.
This ties in with the fact that the Hartree approximation for inter-site interactions
becomes exact in the limit of infinite dimensions~\cite{Uhrig}, and 
it takes modifications to the dynamic mean-field theory (DMFT)~\cite{DMFT}
to re-introduce inter-site interactions in the limit of high dimensions (`extended' DMFT~\cite{EDMFT}). 
We remark in passing that the $1/r$-Hubbard model might be peculiar 
in one dimension where nesting is commonplace but
nesting of the band structure at half band-filling is rather the exception in higher dimensions.

Finally, we remark that our conclusions are based on a single non-trivial model.
More work is necessary to corroborate our findings for other models.
As our next testing case, we are currently investigating spinless fermions  
with $1/r$-long-range interaction on a half-filled chain for which
we recently investigated the CDW transition in the presence of nearest-neighbor 
interactions~\cite{PhysRevB.106.205133}.

\begin{acknowledgments}
  \"O.L. has been supported by the Hungarian National Research,
  Development, and Innovation Office (NKFIH) through Grant No.~K134983, and TKP2021-NVA
  by the Quantum Information National Laboratory of Hungary.
  \"O.L. also acknowledges financial support from the Hans Fischer Senior Fellowship program funded by the Technical University of Munich
  -- Institute for Advanced Study and from the Center for
  Scalable and Predictive methods for Excitation and Correlated phenomena
  (SPEC),
  which is funded as part of the Computational Chemical Sciences Program
  FWP 70942 by the U.S.\
  Department of Energy (DOE), Office of Science, Office of
  Basic Energy Sciences,
  Division of Chemical Sciences, Geosciences, and Biosciences
  at Pacific Northwest National Laboratory.
\end{acknowledgments}

\appendix*

%

\section{Hartree-Fock theory}
\label{HartreeFock}
\renewcommand{\thesection}{\Alph{section}}

In this appendix we derive the Hartree-Fock theory for the one-dimensional $1/r$-Hubbard model
with $1/r$-long-range interactions at half band-filling
when a Fermi-liquid phase competes with a charge-density-wave phase.

\subsection{CDW Hartree-Fock Hamiltonian}

In Hartree-Fock theory, we decouple the Hubbard interaction as follows,
\begin{equation}
\hat{D}^{\rm HF}=\hat{D}^{\rm H}= \sum_l \Bigl[\langle \hat{n}_{l,\uparrow} \rangle \hat{n}_{l,\downarrow}
+ \hat{n}_{l,\uparrow} \langle \hat{n}_{l,\downarrow} \rangle 
-\langle \hat{n}_{l,\uparrow} \rangle \langle \hat{n}_{l,\downarrow} \rangle \Bigr]\;.
\end{equation}
For the long-range terms, we have Hartree and Fock contributions,
\begin{equation}
  \hat{V}^{\rm HF} = \hat{V}^{\rm H} + \hat{V}^{\rm F} \; ,   
\end{equation}
with
\begin{eqnarray}
\hat{V}^{\rm H} &=& V \sum_{r=1}^{L/2} V(r) \sum_{l=1}^{L}\Bigl[ \left( \langle \hat{n}_l \rangle -1\right) 
\left( \hat{n}_{l+r}  -1\right)  \nonumber \\
&&\hphantom{\sum_{r=1}^{L/2} V(r) \sum_{l=1}^{L}\Bigl[} + \left( \hat{n}_l -1\right)  \left( \langle \hat{n}_{l+r} \rangle -1\right) 
\nonumber \\
&&\hphantom{\sum_{r=1}^{L/2} V(r) \sum_{l=1}^{L}\Bigl[} - \left( \langle \hat{n}_l \rangle -1\right)  \left( \langle \hat{n}_{l+r} \rangle -1\right) \Bigr]\;
\end{eqnarray}
and 
\begin{eqnarray}
\hat{V}^{\rm F}&=& V \sum_{r=1}^{L/2} V(r) \sum_{l=1,\sigma}^L \Bigl[ \langle \hat{c}_{l,\sigma}^+ \hat{c}_{l+r,\sigma}^{\vphantom{+}}\rangle
\hat{c}_{l,\sigma}^{\vphantom{+}} \hat{c}_{l+r,\sigma}^+ \nonumber \\
&&\hphantom{\sum_{r=1}^{L/2} V(r) \sum_{l=1,\sigma}^L\Bigl[} + 
\hat{c}_{l,\sigma}^+ \hat{c}_{l+r,\sigma}^{\vphantom{+}}
\langle \hat{c}_{l,\sigma}^{\vphantom{+}} \hat{c}_{l+r,\sigma}^+ \rangle \nonumber\\
&&\hphantom{\sum_{r=1}^{L/2} V(r) \sum_{l=1,\sigma}^L\Bigl[} - 
\langle \hat{c}_{l,\sigma}^+ \hat{c}_{l+r,\sigma}^{\vphantom{+}}\rangle 
\langle \hat{c}_{l,\sigma}^{\vphantom{+}} \hat{c}_{l+r,\sigma}^+ \rangle \Bigr] .\;
\end{eqnarray}
Here, $\langle \hat{A} \rangle$ denotes the ground-state expectation value 
of the operator $\hat{A}$,
 $   \langle \hat{A}\rangle \equiv \langle \Phi_0 | \hat{A} | \Phi_0\rangle  $,
with $|\Phi_0\rangle$ as the ground state of the Hartree-Fock Hamiltonian $\hat{H}^{\rm HF}$, see below.

We make the CDW Ansatz for the order parameter
\begin{equation}
  \langle \hat{n}_{l,\sigma} \rangle   = \frac{1}{2} \left( 1 + (-1)^l \Delta\right)
  \label{appeq:selfconstDelta}
\end{equation}
with the real CDW parameter $\Delta \geq 0$. This Ansatz implies that
the unit cell has doubled. 

Furthermore, we introduce the abbreviation ($r\neq 0$)
\begin{equation}
B_r(l)=   \langle \hat{c}_{l,\sigma}^+ \hat{c}_{l+r,\sigma}^{\vphantom{+}}\rangle = {\rm i}b_r(l)\; .
\label{appeq:selfconstB}
\end{equation}
Particle-hole symmetry implies that $B_r(l)$ is purely complex at half band-filling, 
\begin{equation}
  B_r^*(l)=\langle \hat{c}_{l+r,\sigma}^+ \hat{c}_{l,\sigma}^{\vphantom{+}}\rangle
= \langle \hat{c}_{l+r,\sigma}^{\vphantom{+}} \hat{c}_{l,\sigma}^+ \rangle
= -B_r(l) \; ,
\end{equation}
where we used particle-hole symmetry in the second step. 
Therefore, $b_r(l)$ is real.
Since the unit cell has doubled, we make the Ansatz that $b_r(l)$ alternates just as the 
local order parameter,
\begin{equation}
    b_r(l)= b_r +(-1)^l d_r \; ,
\end{equation}
allowing thus for a bond-order wave. 

With these abbreviations, we can rewrite the Hartree-Fock interaction at half band-filling as
\begin{eqnarray}
   \hat{D}^{\rm H}&=& \frac{L}{4}\left(1 -\Delta^2 \right)
   +\frac{\Delta}{2} \sum_{l,\sigma} (-1)^l \hat{n}_{l,\sigma} \; ,\\
\hat{V}^{\rm H} &=& - L\Delta^2 \overline{V}
     +2\Delta \overline{V}\sum_{l=1,\sigma}^L (-1)^l \hat{n}_{l,\sigma} \; , \\
     \overline{V} &=&  V \sum_{r=1}^{L/2}(-1)^r V(r) \; ,
\end{eqnarray}
and
\begin{eqnarray}
 \hat{V}^{\rm F} &=& 2L V\sum_{r=1}^{L/2}V(r)\left(b_r^2+d_r^2\right) \nonumber \\
 && +V \sum_{r=1}^{L/2}V(r) b_r \!\sum_{l=1,\sigma}^L  
   \rmi \bigl[\hat{c}_{l,\sigma}^+ \hat{c}_{l+r,\sigma}^{\vphantom{+}}
   - \hat{c}_{l+r,\sigma}^+ \hat{c}_{l,\sigma}^{\vphantom{+}} \bigr]  \nonumber \\
&&    +V\sum_{r=1}^{L/2}V(r) d_r \!\sum_{l=1,\sigma}^L (-1)^l  
   \rmi \bigl[\hat{c}_{l,\sigma}^+ \hat{c}_{l+r,\sigma}^{\vphantom{+}}
   - \hat{c}_{l+r,\sigma}^+ \hat{c}_{l,\sigma}^{\vphantom{+}} \bigr] .\nonumber \\
\end{eqnarray}
The resulting single-particle problem defines the Hartree-Fock Hamiltonian for a possible CDW ground state
\begin{equation}
    \hat{H}^{\rm HF}= \hat{T} + U \hat{D}^{\rm H} +  \hat{V}^{\rm H}+\hat{V}^{\rm F} \; .
\end{equation}
It has to be solved self-consistently, i.e., the parameters $\Delta$, $b_r$ and $d_r$ must be chosen such that the ground state
fulfills Eqs.~(\ref{appeq:selfconstDelta}) and~(\ref{appeq:selfconstB}).

\subsection{Diagonalization of the Hartree-Fock Hamiltonian}

In the CDW phase, the Hartree-Fock Hamiltonian is identical for both spin species,
$\hat{H}^{\rm HF}=\sum_{\sigma} \hat{H}_{\sigma}^{\rm HF}$.
Dropping the spin index we must diagonalize 
\begin{eqnarray}
\hat{H}_{\rm sf} &=&\sum_k \epsilon(k) \hat{C}_k^+\hat{C}_k^{\vphantom{+}}
+\left(\frac{U}{2}+2\overline{V}\right)\Delta \sum_l (-1)^l\hat{n}_{l} +C \nonumber \\
&& 
+ V\sum_{r=1}^{L/2} V(r) b_r \sum_{l=1}^L  
   {\rm i}\left[\hat{c}_l^+ \hat{c}_{l+r}^{\vphantom{+}}
   - \hat{c}_{l+r}^+ \hat{c}_l^{\vphantom{+}} \right] \nonumber \\
   && 
+ V\sum_{r=1}^{L/2} V(r) d_r \sum_{l=1}^L (-1)^l  
   {\rm i}\left[\hat{c}_l^+ \hat{c}_{l+r}^{\vphantom{+}}
   - \hat{c}_{l+r}^+ \hat{c}_l^{\vphantom{+}} \right]\nonumber \\
\end{eqnarray}
 for spinless fermions (`sf'), where 
 \begin{equation}
    C=L\frac{U}{8}(1-\Delta^2)-L \frac{\overline{V}\Delta^2}{2}+L V\sum_{r=1}^{L/2}V(r)\left(b_r^2+d_r^2\right) \; .
    \label{eq:defC}
 \end{equation}
 In momentum space, the Hamiltonian reads
\begin{eqnarray}
    \hat{H}_{\rm sf} &=& C+ \sideset{}{'}\sum_k     \Bigl[
    \left(\epsilon(k)+b_{\rm e}(k)\right) \hat{C}_k^+\hat{C}_{k}^{\vphantom{+}} \nonumber \\
    && \hphantom{C+ \sideset{}{'}\sum_k     \Bigl[}
    +\left(\epsilon(k+\pi)+b_{\rm o}(k)\right) \hat{C}_{k+\pi}^+\hat{C}_{k+\pi}^{\vphantom{+}} \Bigr] \nonumber \\
    &&    +\left(\frac{U}{2}+2\overline{V}\right)\Delta \sideset{}{'}\sum_k \left(\hat{C}_k^+\hat{C}_{k+\pi}^{\vphantom{+}}
    +\hat{C}_{k+\pi}^+\hat{C}_{k}^{\vphantom{+}}\right) \nonumber \\
    && +\sideset{}{'}\sum_k \left(d(k)\hat{C}_k^+\hat{C}_{k+\pi}^{\vphantom{+}}
    +d^*(k) \hat{C}_{k+\pi}^+\hat{C}_{k}^{\vphantom{+}}\right) \; ,
\end{eqnarray}
where the prime on the sum indicates the $k$-space region $-\pi<k<0$. 

Moreover, we
introduced the abbreviations
\begin{eqnarray}
 b_{\rm e}(k)&=& -2 V\sum_{r=1}^{L/2} V(r)b_r \sin(kr) \; ,\nonumber \\
  b_{\rm o}(k)&=& -2 V\sum_{r=1}^{L/2} (-1)^rV(r)b_r \sin(kr) \; ,
  \end{eqnarray}
and
\begin{eqnarray}  d(k)&=& V\sum_{r=1}^{L/2} \rmi V(r) d_r  \left( (-1)^r e^{\rmi k r} - e^{-\rmi kr} \right) \nonumber\\
  &=& d_{\rm R}(k)+\rmi d_{\rm I}(k) \; , \nonumber \\
  d_{\rm R}(k) &=& - V\sum_{r=1}^{L/2} V(r) d_r \left(1+(-1)^r\right) \sin(kr) \; , \nonumber \\
  d_{\rm I}(k) &=& - V\sum_{r=1}^{L/2} V(r) d_r \left(1-(-1)^r\right) \cos(kr) \; . \nonumber \\
  \label{eq:defbsandds}
\end{eqnarray}
We diagonalize $\hat{H}_{\rm sf}$ with the help of the linear transformation
\begin{eqnarray}
\hat{C}_k &=& c_k \hat{\alpha}_k -s_ke^{\rmi \psi_k} \hat{\beta}_k \; , \nonumber \\
\hat{C}_{k+\pi} &=& s_k e^{-\rmi \psi_k} \hat{\alpha}_k +c_k \hat{\beta}_k  \; ,
\end{eqnarray}
where we abbreviate $c_k \equiv \cos(\varphi_k)$ and $s_k=\sin(\varphi_k)$,
and $\varphi_k$ and $\psi_k$ are real phases.
For each $-\pi < k<0 $, $\hat{h}_k$ must be diagonal in the new operators where
\begin{eqnarray}
\hat{h}_k&=& \left(\epsilon(k)+b_{\rm e}(k)\right) \nonumber \\
&& \times
\left( c_k\hat{\alpha}_k^+ - s_k e^{-\rmi \psi_k} \hat{\beta}_k^+ \right) 
\left( c_k\hat{\alpha}_k^{\vphantom{+}} - s_k e^{\rmi \psi_k} \hat{\beta}_k^{\vphantom{+}} \right)  \nonumber \\[6pt]
&& + \left(\epsilon(k+\pi)+b_{\rm o}(k)\right) \nonumber \\
&& \times
\left( s_ke^{\rmi \psi_k}\hat{\alpha}_k^+ + c_k \hat{\beta}_k^+ \right) 
\left( s_k e^{-\rmi \psi_k}\hat{\alpha}_k^{\vphantom{+}} + c_k \hat{\beta}_k^{\vphantom{+}} \right)  \nonumber \\[6pt]
&& + \left(\frac{U\Delta}{2}+2\overline{V}\Delta + d(k) \right)\nonumber\\
&& \times \left( c_k\hat{\alpha}_k^+ - s_k e^{-\rmi \psi_k} \hat{\beta}_k^+ \right) 
\left( s_k e^{-\rmi \psi_k}\hat{\alpha}_k^{\vphantom{+}} + c_k \hat{\beta}_k^{\vphantom{+}} \right) \nonumber \\[6pt]
&& +  \left(\frac{U\Delta}{2}+2\overline{V}\Delta + d^*(k) \right)\nonumber\\
&& \times \left( s_ke^{\rmi \psi_k}\hat{\alpha}_k^+ + c_k \hat{\beta}_k^+ \right) 
\left( c_k\hat{\alpha}_k^{\vphantom{+}} - s_k e^{\rmi \psi_k} \hat{\beta}_k^{\vphantom{+}} \right) \; .\nonumber \\
\end{eqnarray}
The mixed terms proportional to $\hat{\alpha}_k^+\hat{\beta}_k^{\vphantom{+}}$ in $\hat{h}_k$ read
\begin{eqnarray}
 M_k&=&   \left(\epsilon(k)+b_{\rm e}(k)\right) (-c_k s_ke^{\rmi \psi_k}) \nonumber \\
&&  + \left(\epsilon(k+\pi)+b_{\rm o}(k)\right)(c_k s_ke^{\rmi \psi_k}) \nonumber \\
 && + \left(\frac{U\Delta}{2}+2\overline{V}\Delta + d(k) \right) c_k^2  \nonumber \\
 && +  \left(\frac{U\Delta}{2}+2\overline{V}\Delta + d^*(k) \right) (-s_k^2 e^{2\rmi \psi_k}) \, .\,\nonumber 
 \\
\end{eqnarray}
We demand that the mixing term vanishes, $M_k=0=M_k^*$,
which implies
\begin{eqnarray}
\text{lhs} & =& \text{rhs} \;, \nonumber \\
\text{lhs} &=& \sin(2\varphi_k) \left( \epsilon(k) -\epsilon(k+\pi) +b_{\rm e}(k) -b_{\rm o}(k) \right) \; , \nonumber \\
\text{rhs}&=&  \left(\frac{U\Delta}{2}+2\overline{V}\Delta + d(k) \right)e^{-\rmi \psi_k}\cos(2\varphi_k) \nonumber \\
&& + \left(\frac{U\Delta}{2}+2\overline{V}\Delta + d^*(k) \right)e^{\rmi \psi_k}\cos(2\varphi_k) \nonumber \\
&&+ \left(\frac{U\Delta}{2}+2\overline{V}\Delta + d(k) \right)e^{-\rmi \psi_k} \nonumber \\
&& - \left(\frac{U\Delta}{2}+2\overline{V}\Delta + d^*(k) \right)e^{\rmi \psi_k} \nonumber \\
&=& 2\cos(2\varphi_k) \left( \frac{U\Delta}{2}+2\overline{V}\Delta + d_{\rm R}(k) \right) \cos(\psi_k) 
\nonumber \\
&& +2\cos(2\varphi_k) d_{\rm I}(k) \sin(\psi_k) \nonumber \\
&& -2\rmi  \left(\frac{U\Delta}{2}+2\overline{V}\Delta + d_{\rm R}(k) \right)\sin(\psi_k) \nonumber \\
&& +2\rmi d_{\rm I}(k) \cos(\psi_k) \; ,
\label{eq:lhsisrhs}
\end{eqnarray}
where we used Eq.~(\ref{eq:defbsandds}) and 
\begin{eqnarray}
    c_k^2 &=& \frac{1}{2} \left( 1+\cos(2\varphi_k) \right) \; , \nonumber \\
    s_k^2 &=& \frac{1}{2} \left( 1-\cos(2\varphi_k)\right) \; , \nonumber \\
    c_ks_k &=& \frac{1}{2} \sin(2\varphi_k) \; .
\end{eqnarray}
Apparently, the imaginary terms of the right-hand-side of eq.~(\ref{eq:lhsisrhs}) must vanish, which leads to
\begin{equation}
\tan(\psi_k) = \frac{2d_{\rm I}(k)}{U\Delta +4\overline{V}\Delta +2d_{\rm R}(k)}
\end{equation}
for the phase $\psi_k$. 
Moreover, the real part of the eq.~(\ref{eq:lhsisrhs}) leads to
\begin{eqnarray}
    \tan(2\varphi_k) &=& \frac{[-Z(k)]}{\epsilon(k) -\epsilon(k+\pi) +b_{\rm e}(k) -b_{\rm o}(k)}\geq 0 \; , \nonumber \\
    Z(k) &=& -(U\Delta +4 \overline{V}\Delta+2 d_{\rm R}(k))\cos(\psi_k)\nonumber \\
    && -2d_{\rm I}(k)\sin(\psi_k) \nonumber \\
    &=& \sqrt{\left(U\Delta +4\overline{V}\Delta +2d_{\rm R}(k)\right)^2 + 4 [d_{\rm I}(k)]^2} \; ,\nonumber \\
\end{eqnarray}
where we assumed that $U\Delta +4\overline{V}\Delta +2d_{\rm R}(k)< 0$.
The diagonal terms give
\begin{equation}
    h_k= E_{\alpha}(k) \hat{\alpha}_k^+\hat{\alpha}_k^{\vphantom{+}} + E_{\beta}(k) \hat{\beta}_k^+\hat{\beta}_k^{\vphantom{+}}
\end{equation}
with
\begin{eqnarray}
    E_{\alpha}(k) &=& (\epsilon(k)+b_{\rm e}(k)) c_k^2 +
    (\epsilon(k+\pi)+b_{\rm o}(k)) s_k^2\nonumber \\
    && + \left(\frac{U\Delta}{2}+2\overline{V}\Delta + d(k) \right)e^{-\rmi \psi_k}c_ks_k \nonumber \\
    && + \left(\frac{U\Delta}{2}+2\overline{V}\Delta + d^*(k) \right)e^{\rmi \psi_k}c_ks_k \nonumber \\
    &=& \frac{1}{2}\left[ \epsilon(k)+\epsilon(k+\pi)+b_{\rm e}(k)+b_{\rm o}(k)\right] -s(k) \; , \nonumber \\
     E_{\beta}(k) &=&\frac{1}{2}\left[\epsilon(k)+\epsilon(k+\pi)+b_{\rm e}(k)+b_{\rm o}(k)\right] +s(k)\nonumber\; , \\
     \label{eq:EalphaEbeta}
\end{eqnarray}
where
\begin{eqnarray}
    s(k) &=& \frac{1}{2} \sqrt{[Z(k)]^2+ \left[ \epsilon(k)-\epsilon(k+\pi)+b_{\rm e}(k)-b_{\rm o}(k)\right]^2} \nonumber \\
    &=& \frac{1}{2} \Bigl[
\left(U\Delta +4\overline{V}\Delta +2d_{\rm R}(k)\right)^2 + 4 [d_{\rm I}(k)]^2 \nonumber \\
&& \hphantom{\biggl[}
+ \left[ \epsilon(k)-\epsilon(k+\pi)+b_{\rm e}(k)-b_{\rm o}(k)\right]^2
    \Bigr]^{1/2} . \label{eq:sk}
\end{eqnarray}
The Hartree-Fock Hamiltonian becomes diagonal with the quasi-particle dispersion $E_{\alpha,\beta}(k)$,
\begin{equation}
    \hat{H}^{\rm HF} =2C + \sideset{}{'}\sum_{k,\sigma} \left[E_{\alpha}(k) \hat{\alpha}_{k,\sigma}^+\hat{\alpha}_{k,\sigma}^{\vphantom{+}}
    + E_{\beta}(k) \hat{\beta}_{k,\sigma}^+\hat{\beta}_{k,\sigma}^{\vphantom{+}}\right] \; ,
\end{equation}
where we re-introduced the spin index. 
Since $E_{\alpha}(k)< E_{\beta}(k)$ for all $-\pi<k<0$,
the ground state at half band-filling contains only $\alpha$-particles,
\begin{equation}
    |\Phi_0\rangle= \prod_{-\pi<k<0,\sigma} \hat{\alpha}_{k,\sigma}^+ |{\rm vac}\rangle \; .
\end{equation}

\subsection{Self-consistency equations and CDW transition}

The self-consistency equation~(\ref{appeq:selfconstDelta}) 
becomes
\begin{eqnarray}
    \Delta &=&  \frac{1}{L} \sideset{}{'}\sum_{k} \langle \hat{C}_{k,\sigma}^+ \hat{C}_{k+\pi,\sigma}^{\vphantom{+}}
    + \hat{C}_{k+\pi,\sigma}^+ \hat{C}_{k,\sigma}^{\vphantom{+}}\rangle \nonumber \\
&=&  \frac{1}{L} \sideset{}{'}\sum_{k} \sin(2\varphi_k)\cos(\psi_k) \nonumber \\
&=& - \frac{1}{L} \sideset{}{'}\sum_{k} \frac{\left[U\Delta/2+2\overline{V}\Delta +d_{\rm R}(k)\right]}{s(k)} \nonumber\\
&=& -\int_{-\pi}^0 \frac{{\rm d}k}{2\pi} \frac{\left[U\Delta/2+2\overline{V}\Delta +d_{\rm R}(k)\right]}{s(k)}
    \end{eqnarray}
in the thermodynamic limit. 
The self-consistency equation~(\ref{appeq:selfconstB}) for the bond dimerization
becomes
   \begin{eqnarray}
    \rmi b_r  +(-1)^l \rmi d_r &=& \frac{1}{L} \sum_k e^{-\rmi k l} e^{\rmi k(l+r)} \langle \hat{C}_{k,\sigma}^+ \hat{C}_{k,\sigma}^{\vphantom{+}}\rangle
    \nonumber \\
    && +\frac{1}{L} \sum_k e^{-\rmi k l} e^{\rmi (k+\pi)(l+r)} \langle \hat{C}_{k,\sigma}^+ \hat{C}_{k+\pi,\sigma}^{\vphantom{+}}\rangle \nonumber \\
\end{eqnarray}
so that
\begin{equation}
    \rmi b_r = \frac{1}{L} \sideset{}{'}\sum_{k}e^{\rmi kr}\left[\langle \hat{C}_{k,\sigma}^+ \hat{C}_{k,\sigma}^{\vphantom{+}}\rangle
 +  (-1)^r  \langle \hat{C}_{k+\pi,\sigma}^+ \hat{C}_{k+\pi,\sigma}^{\vphantom{+}}\rangle\right]
\end{equation}
and
\begin{equation}
    \rmi d_r = \frac{1}{L} \sideset{}{'}\sum_{k}e^{\rmi kr}\left[(-1)^r\langle \hat{C}_{k,\sigma}^+ \hat{C}_{k+\pi,\sigma}^{\vphantom{+}}\rangle
 +  \langle \hat{C}_{k+\pi,\sigma}^+ \hat{C}_{k,\sigma}^{\vphantom{+}}\rangle\right]\; .
 \label{eq:dralmost}
\end{equation}
This results in
\begin{eqnarray}
    \rmi b_r &=& \frac{(1-(-1)^r)}{2} \rmi \tilde{b}_r \nonumber\; ,  \\
    \rmi \tilde{b}_r &=&-\frac{1}{L} \sideset{}{'}\sum_{k}e^{\rmi kr} \frac{\left[ \epsilon(k)-\epsilon(k+\pi)+b_{\rm e}(k)-b_{\rm o}(k)\right]}{2s(k)}\nonumber \; ,\\
    \tilde{b}_r&=& -\int_{-\pi}^0 \frac{\rmd k}{2\pi} \sin(kr) \frac{\left[ \epsilon(k)-\epsilon(k+\pi)+b_{\rm e}(k)-b_{\rm o}(k)\right]}{2s(k)} \nonumber \\
\end{eqnarray}
in the thermodynamic limit. Here, we used that $b_r$ is real. Note that $b_r\neq 0$ only for odd distances, $r=2m+1$. From eq.~(\ref{eq:defbsandds})
it thus follows that $b(k)\equiv b_{\rm e}(k)=-b_{\rm o}(k)$ with
\begin{eqnarray}
    b(k)&=&-2V\sum_{r=1}^{L/2} V(r) \tilde{b}_r\frac{(1-(-1)^r)}{2} \sin(kr) \;  , \\
    \tilde{b}_r &=&  -\int_{-\pi}^0 \frac{\rmd k}{2\pi} \sin(kr) \frac{\left[ \epsilon(k)-\epsilon(k+\pi)+2b(k)\right]}{2s(k)} \nonumber 
\end{eqnarray}
with 
\begin{eqnarray}
   s(k) &=& \frac{1}{2} \Bigl[
\left(U\Delta +4\overline{V}\Delta +2d_{\rm R}(k)\right)^2 + 4 [d_{\rm I}(k)]^2 \nonumber \\
&& \hphantom{\frac{1}{2} \biggl[}
+ \left[ \epsilon(k)-\epsilon(k+\pi)+2b(k)\right]^2
    \Bigr]^{1/2} \; .  
    \label{eq:skagain}
\end{eqnarray}
The dispersion relation thus simplifies to
\begin{eqnarray}
    E_{\alpha}(k) 
    &=& \frac{1}{2}\left[ \epsilon(k)+\epsilon(k+\pi)\right] -s(k) \; , \nonumber \\
     E_{\beta}(k) &=&\frac{1}{2}\left[\epsilon(k)+\epsilon(k+\pi)\right] +s(k)\; .
     \label{eq:EalphaEbetaagain}
\end{eqnarray}
Furthermore, eq.~(\ref{eq:dralmost}) leads to
\begin{eqnarray}
    \rmi d_r &=& \frac{1}{L} \sideset{}{'}\sum_{k}e^{\rmi kr} c_ks_k \left[(-1)^r e^{-\rmi \psi_k} + e^{\rm i\psi_k}\right]   \label{eq:drodd}\\
   d_r &=&   \frac{1-(-1)^r}{2} d_{r,{\rm o}}+  \frac{1+(-1)^r}{2} d_{r,{\rm e}} \; , \nonumber \\
   d_{r,{\rm o}} &=& \frac{1}{L} \sideset{}{'}\sum_{k}2c_ks_k \cos(kr) \sin\psi_k \nonumber \\
   &=& -\int_{-\pi}^0 \frac{\rmd k}{2\pi} \cos(kr) \frac{d_{\rm I}(k)}{s(k)}  \; , \nonumber \\
   d_{r,{\rm e}} &=& \frac{1}{L} \sideset{}{'}\sum_{k}2c_ks_k \sin(kr) \cos\psi_k \nonumber \\
   &=&  -\int_{-\pi}^0 \frac{\rmd k}{2\pi} \sin(kr) \frac{U\Delta/2 +2\overline{V}\Delta + d_{\rm R}(k)}{s(k)} \; . \nonumber
\end{eqnarray}
Recall that
\begin{eqnarray}
  d_{\rm R}(k) &=& - V\sum_{r=1}^{L/2} V(r) d_{r,{\rm e}} \left(1+(-1)^r\right) \sin(kr) \; , \nonumber \\
  d_{\rm I}(k) &=& - V\sum_{r=1}^{L/2} V(r) d_{r,{\rm o}} \left(1-(-1)^r\right) \cos(kr) \; . \nonumber\\\label{eq:defbsanddsagain}
\end{eqnarray}
When we only have a nearest-neighbor interaction, i.e., $V(r)=\delta_{r,1}$,
the set $\{d_{\rm R}(k)=0,d_{r,{\rm o}}=0\}$ provides a self-consistent solution
and we recover the equations given in Ref.~\cite{1overRHubbard-NN}.

\subsection{Purely real dimerization function}

As a convenient simplification, we will search solutions for a purely real dimerization,
$d_{r, {\rm o}}=0$. Eq.~(\ref{eq:defbsanddsagain}) shows that the Ansatz $d_{r,{\rm o}}=0$
leads to $d_{\rm I}(k)=0$. When inserted in eq.~(\ref{eq:drodd}), we recover $d_{r,{\rm o}}=0$.
Self-consistency is achieved, irrespective of all other variational parameters.
It implies that
\begin{equation}
    \psi_k =0
\end{equation}
and
\begin{equation}
    Z(k) = -[U\Delta +4 \overline{V}\Delta+2 d(k)] \; .
\end{equation}
The self-consistency equations simplify to
\begin{eqnarray}\label{SelfConEquations}
\Delta &=& -\int_{-\pi}^0 \frac{{\rm d}k}{2\pi} \frac{\left[U\Delta/2+2\overline{V}\Delta +d(k)\right]}{s(k)} \nonumber \; ,\\
    b(k)&=&-2V\sum_{r=1}^{L/2} V(r) \tilde{b}_r\frac{(1-(-1)^r)}{2} \sin(kr) \;  , \nonumber \\
    \tilde{b}_r &=&  -\int_{-\pi}^0 \frac{\rmd k}{2\pi} \sin(kr) \frac{\left[ \epsilon(k)-\epsilon(k+\pi)+2b(k)\right]}{2s(k)} \nonumber \; , \\
 d(k) &=& - V\sum_{r=1}^{L/2} V(r) d_r \left(1+(-1)^r\right) \sin(kr) \nonumber \; ,\\
  d_r   &=&  -\int_{-\pi}^0 \frac{\rmd k}{2\pi} \sin(kr) \frac{U\Delta/2 +2\overline{V}\Delta + d(k)}{s(k)} \; ,
\end{eqnarray}
where
\begin{eqnarray}
   s(k) &=& -\frac{1}{2} \Bigl[
\left(U\Delta +4\overline{V}\Delta +2d(k)\right)^2  \nonumber \\
&& \hphantom{\frac{1}{2} \biggl[}
+ \left[ \epsilon(k)-\epsilon(k+\pi)+2b(k)\right]^2
    \Bigr]^{1/2} \; .  
    \label{eq:skfinal}
\end{eqnarray}
Recall that the parameters $\tilde{b}_r$ are defined only for odd~$r$, 
and the parameters $d _r$ are defined only for even $r$.

The self-consistency equations show that, (i), the Coulomb interaction at odd particle distances influences the charge-density-wave order, and, (ii),
the Coulomb interaction at even distances can trigger a bond-order-wave instability.

\subsection{Minimization of the Hartree-Fock ground-state energy}

As an alternative to the solution of the self-consistency equations,
we can also address the Hartree-Fock ground-state energy per spin species,
\begin{equation}
    \frac{e_0^{\rm HF}(U,V)}{2} = \frac{C}{L}  + \int_{-\pi}^0 \frac{\rmd k}{2\pi} \left[\frac{1}{2}\left[ \epsilon(k)+\epsilon(k+\pi)\right] -s(k)\right]
    \end{equation}
    with 
\begin{eqnarray}
\frac{C}{L} &=& \frac{U}{8}(1-\Delta^2)- \frac{\overline{V}\Delta^2}{2}\\
    && + V\sum_{r=1}^{L/2}V(r)\left[ \frac{(1-(-1)^r)}{2} \tilde{b}_r^2  + \frac{(1+(-1)^r)}{2} d_r^2\right] \; , \nonumber 
\end{eqnarray}
where we used eqs.~(\ref{eq:defC}) and~(\ref{eq:EalphaEbetaagain}) for a purely real dimerization function. From eq.~(\ref{eq:skfinal}) we have
\begin{eqnarray}
   s(k) &=& \frac{1}{2} \Bigl[
\left(U\Delta +4\overline{V}\Delta +2d(k)\right)^2  \nonumber \\
&& \hphantom{\frac{1}{2} \biggl[}
+ \left[ \epsilon(k)-\epsilon(k+\pi)+2b(k)\right]^2
    \Bigr]^{1/2} , 
    \label{eq:skfinalagain}
\end{eqnarray}
where 
\begin{eqnarray}
    b(k)&=&-V\sum_{r=1}^{L/2} V(r) \tilde{b}_r(1-(-1)^r) \sin(kr) \;  , \nonumber \\
 d(k) &=& - V\sum_{r=1}^{L/2} V(r) d_r \left(1+(-1)^r\right) \sin(kr) \; .
\end{eqnarray}
Recall that 
\begin{equation}
    \overline{V} =  V \sum_{r=1}^{L/2}(-1)^r V(r) <0 \; .
\end{equation}
The task is to minimize the Hartree-Fock ground-state energy $e_0^{\rm HF}(U,V)$ with respect to the
variational parameters $\Delta$, $\tilde{b}_r$ ($r$: odd), and $d_r$; $r$ is even.

\subsection{CDW critical interaction for weak coupling}

In the absence of bond order, $d(k)=0$ and $d_r=0$, the critical interaction strength is obtained from
putting $\Delta=0^+$ in the above equations. We thereby assume a continuous transition, which is the case for $U,V\lesssim W$. We then have
\begin{equation}
    s(k) = -\frac{1}{2}\left(\epsilon(k)-\epsilon(k+\pi)+2b_{\rm c}(k)\right)
\end{equation}
for $-\pi<k<0$. Thus, with $\Delta$ from Eq.~(\ref{SelfConEquations}),
\begin{equation}
    1=\int_{-\pi}^0\frac{\rmd k}{\pi} \frac{U/2+2\overline{V}_{\rm c}}{1/2\left[\epsilon(k)-\epsilon(k+\pi)+2b_{\rm c}(k)\right]}\; ,
\end{equation}
where $\epsilon(k)-\epsilon(k+\pi)=-\pi t$ and
\begin{eqnarray}
    b_{\rm c}(k) &=& -2V_{\rm c,LR}\sum_{r=1} ^\infty V(r)\tilde{b}_r \left(\frac{1-(-1)^r}{2}\right)\sin(kr) \nonumber\\
    &=& \frac{2V_{\rm c, LR}}{\pi}B(k) \; , \nonumber \\
B(k)&=&\sum_{m=0}^\infty \frac{\sin((2m+1)k)}{(2m+1)^2} \; ,
\end{eqnarray}
because
\begin{eqnarray}
    \tilde{b}_r &=& -\int_{-\pi}^0 \frac{\rmd k}{2\pi} \sin(kr)\frac{\epsilon(k)-\epsilon(k+\pi)+2 b(k)}{2 s(k)} \nonumber\\
    &=& \int_{-\pi}^0\frac{\rmd k}{2\pi} \sin(kr)
    \,\,=\,\, -\frac{1}{2\pi r}(1-(-1)^r)
\end{eqnarray}
in the thermodynamic limit. 
With $\overline{V}=-V_{\rm c,LR}\ln 2$ from
\begin{equation}
    \sum_{r=1}^\infty \frac{(-1)^r}{r} = \ln 2\;,
    \label{eq:log2sum}
\end{equation}
$V_{\rm c,LR}$ results from
\begin{equation}
    1 = \int_0^\pi \frac{\rmd k}{\pi} \frac{4 V_{\rm c, LR} \ln 2  -U   }{W/2+4V_{\rm c, LR}B(k)/\pi}\;,
\end{equation}
where $B(k)\geq 0$ for $0\leq k\leq \pi$. Thus, $V_{\rm c, LR}>U/(4\ln 2)$ is a condition.

For small $U$ and with $V_{\rm c,LR}=a_{\rm LR}+U b_{\rm LR}$, we find
\begin{equation}
    1=\int_0^\pi\rmd k\,\frac{8a_{\rm LR}\ln 2}{8a_{\rm LR}B(k)+\pi}\;,
\end{equation}
which leads to $a_{\rm LR}\approx 0.255298$ to leading order. To first order, we have
\begin{equation}
    \frac{1}{b_{\rm LR}} = 32\pi(\ln 2)^2a_{\rm LR} \int_0^\pi\,\left(\frac{1}{8a_{\rm LR}B(k)+\pi}\right)^2\;,
\end{equation}
so that we find $b_{\rm LR}\approx 0.502707$.

For the case of nearest-neighbor interactions, the same analysis gives, see Eq.~(A22) of Ref.~\cite{1overRHubbard-NN},
\begin{equation}
     \frac{1}{\alpha_{\rm c}-2U/\pi} = \frac{\pi}{\sqrt{1-\alpha_{\rm c}^2}}
     -\frac{2}{\sqrt{1-\alpha_{\rm c}^2}}\arctan\left(\frac{\alpha_{\rm c}}{\sqrt{1-\alpha_{\rm c}^2}}\right)
\end{equation}
with the abbreviation $\alpha_{\rm c}=8V_{\rm c,NN}/\pi$. 
This equation must be solved numerically for given~$U$. 

For weak Hubbard interactions, we set 
\begin{equation}
    \alpha_{\rm c}(U)=\alpha_0+\alpha_1 U
\end{equation} 
to find
$\alpha_0\approx 0.394235$ so that $V_{\rm c,NN}^{\rm HF}(U=0)\approx 0.154816 =a_{\rm NN}^{\rm HF}$.
Moreover, 
\begin{equation}
    \alpha_1=2(1-\alpha_0^2)/[\pi(1-2\alpha_0^2)] \; ,
\end{equation}
so that $\alpha_1\approx 0.780192$ and $b_{\rm NN}^{\rm HF}\approx 0.306381$
for the CDW transition for the extended $1/r$-Hubbard model in Hartree-Fock theory.

\vspace{6pt}

\subsection{CDW critical interaction in the atomic limit}

Since the transition is discontinuous for strong interactions, 
we must compare the ground-state energies of the CDW and non-ordered phases directly.

In the atomic limit, we can ignore the contribution of the kinetic energy and may focus on the disordered state with one electron
per site that has the energy $E_{\rm dis}=0$. In comparison, the CDW state, with double occupancies on even lattice sites
and holes on odd sites, has the energy
\begin{eqnarray}
    E_{\rm NN}^{\rm CDW} &=& \frac{L}{2} U +L (-V) \; , \nonumber \\
E_{\rm LR}^{\rm CDW} &=& \frac{L}{2} U +L \sum_{r=1}^{L/2} V \frac{(-1)^r}{d(r)} 
\end{eqnarray}
for nearest-neighbor interactions and $1/r$-long-range interactions.
Using eq.~(\ref{eq:log2sum})  we thus find 
 for the extended Hubbard model 
\begin{equation}
     e_{\rm NN}^{\rm CDW} = \frac{U}{2} -V
     \end{equation}
    and
     \begin{equation}
      e_{\rm LR}^{\rm CDW} =\frac{U}{2} -V\ln(2)
\end{equation}
for the Hubbard model with $1/r$-long-range interactions, respectively.

In the atomic limit, $\hat{T}\equiv 0$, the CDW transitions
occur at $V_{\rm c,NN}^{\rm al}(U)=U/2$
for the extended Hubbard model and
at $V_{\rm c,LR}^{\rm al}(U)=U/(2\ln(2))$ for the Hubbard model with $1/r$-long-range interactions,
respectively. These considerations justify the phase separation lines in Fig.~\ref{QPDiagramAndComparison}.

Moreover, the renormalization factor that maps the ground-state energies onto each other is given by
\begin{equation}
 R^{\rm Mott}(U,V\gg W)  = \frac{V_{\rm c, NN}^{\rm al}(U)}{ V_{\rm c, LR}^{\rm al}(U)} =\ln(2) \;.
\label{eq:VcrenMOTT}
\end{equation}
This factor is used in the renormalization of the critical line for the Mott transition.

\bibliography{LRHM}

\end{document}